\documentclass[12pt,twoside,a4paper]{book}

\usepackage[latin1]{inputenc}
\usepackage{a4}         
\usepackage{fancyhdr}   
\usepackage[english]{babel}
\usepackage{makeidx}    
\usepackage{nomencl}		
\usepackage{color}			
\usepackage{t1enc}			
\usepackage{latexsym}		
\usepackage{wrapfig}		
\usepackage{amsthm}			
\usepackage{amssymb}   

\usepackage{graphicx}
\usepackage{pslatex}
\usepackage{ifthen}

\usepackage[T1]{fontenc}
\usepackage{pslatex}

\usepackage{psfrag}
\usepackage{subfigure}

\usepackage{listings} 

\usepackage{amsmath}

\newcommand{\blankpage}
{
  \clearpage{\pagestyle{empty}\cleardoublepage}
}

\graphicspath{
	{figures/}
}

\newlength{\gw}

\lstdefinelanguage{pseudo}
{
morekeywords={and, not, or, for, if},
sensitive=true, 
morecomment=[l]{//}, 
morecomment=[s]{/*}{*/}, 
morestring=[b]"
}

\lstdefinelanguage{glsl} 
{
morekeywords={uniform, float, return, vec2},
sensitive=true, 
morecomment=[l]{//}, 
morecomment=[s]{/*}{*/}, 
morestring=[b]"
}
\definecolor{black_color}{rgb}{0.0,0.0,0.0}
\definecolor{listingbgr_color}{rgb}{0.72,0.93,0.52}
\theoremstyle{definition}

\makeglossary

\begin{document}

\frontmatter


\begin{titlepage}

\begin{center}

\parbox{8cm}{\includegraphics[height=1.6cm]{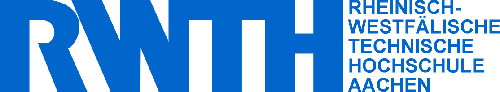}}
\hfill
\parbox{2.5cm}{\includegraphics[width=2.3cm]{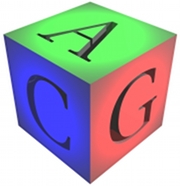}}

\vspace{0.5cm}

\vspace{0.75cm}

\textsf
{
Fakultät für Mathematik, Informatik und Naturwissenschaften\\
Lehrstuhl für Informatik VIII (Computergraphik und Multimedia)\\
Prof. Dr. Leif Kobbelt
}

\rule{\linewidth}{1pt}

\vspace{1.75cm}
\LARGE
\textbf{Bachelor Thesis}

\vspace{1.7cm}
\huge
Virtual Texturing

\vspace{3.0cm}
\Large
Andreas Neu\\
\large
Matrikelnummer: 271784

\vspace{0.5cm}
April 2010

\vspace{1.05cm}
\rule{\linewidth}{1pt}

\vspace{0.5cm}
\textsf{\textbf{
\normalsize
\begin{tabular}{ll}
Erstgutachter:  & Prof. Dr. Leif Kobbelt\\
Zweitgutachter: & Prof. Dr. Bastian Leibe\\
\end{tabular}
}}
\end{center}

\end{titlepage}


\chapter*{Acknowledgement}
First of all I would like to thank Prof. Dr. Leif Kobbelt and Prof. Dr. Bastian Leibe at the RWTH Aachen University 
for reviewing this thesis and giving me the opportunity to write it in the field of Computer Graphics, which 
fascinates me since my first steps on the Commodore 128D.
\\\\
A special thanks goes to my advisor Volker Sch\"onefeld, who got me interested in \emph{Megatexturing}
and \emph{Virtual Texturing}. He always took some of his spare time to read my long emails and had
enough patience to give me advice when I got stuck during the course of this thesis.
\\\\
Furthermore I want to thank Sebastian Raubach for reading the thesis and giving constructive feedback 
and Andreas T\"onnis for not falling asleep when I talked about Virtual Texturing (which was quite often in
the past months).
\\\\
Last but not least I want to thank the talented people at \emph{map-factory.org}, whose maps we used in order
to test our implementation.

\blankpage


\pagestyle{fancyplain}

\begin{figure}
I hereby affirm that I composed this work independently and used no other than the specified sources and tools and
that I marked all quotes as such.

\medskip

\begin{flushright}
Aachen, \today

\vspace{1.5cm}
\small
(Andreas Neu)
\end{flushright}
\end{figure}

\blankpage

{
\pagestyle{fancyplain}
\tableofcontents
\newpage
}

\mainmatter

\chapter{Introduction}
	\section{Motivation}\label{01_motivation}
		Texturing is an important technique to add fine details to geometrical surfaces.
		Due to the limited size of texture memory on current graphic cards, artists have to work within those
		limits.
		Hence, they have to work with a fixed texture budget and reuse this limited set quite often throughout a scene.
		Unfortunately this introduces visually disturbing patterns. Once these patterns are spotted, they remind
		the viewer of seeing something artificial.
		To avoid this from happening, artists traditionally employ a technique called \emph{multi texturing}, where multiple
		textures are blended together for the purpose of making the scene less repetitive.
		Although this technique worked quite well so far, there exist some downsides: 
		Firstly it is computationally expensive to use multiple texture look-ups per fragment, because it stresses the memory throughput, which is often the main bottleneck.
		Secondly it is not a very intuitive approach and thus
		impacts the workflow of artists negatively. \\\\
		\emph{Virtual Texturing} is a technique that can be used to texture every spot in the scene uniquely.
		It supports virtually unlimited detail by dividing the texture intro a hierarchy of smaller parts, so
		called \emph{pages}, and keeping only those parts in the memory that are currently needed. Because
		of that, it is possible to provide areas in the texture exclusively for every surface, so that they
		really can be textured in a unique fashion, even if that would exceed the video cards memory. 
		This approach suits the work of artists, because they
		can concentrate on the details in every area individually to provide the anticipated looks without
		worrying about the texture budget.
		\\\\
		Since Virtual Texturing can only keep a limited set of texture parts in the memory, it has to progressively load 
		new parts as soon as they are needed.
		For example, changes to the viewpoint can cause pages to be become visible that are not currently available in the memory.
		As long as they stay unavailable, the viewer will see visual artifacts, which can be very disturbing.
		The obvious way to reduce these artifacts is to optimize the streaming itself. 
		But this does not work in cases where the latency is high, or the bandwidth is limited, such as internet streaming.
		In these cases page misses will be inevitable, hence it is important to decide in what order required pages are streamed, to minimize the artifacts 
		and optimize the visual experience for the viewer.	

	\section{Contributions of this Thesis} \label{01_contribution}
		This thesis deals from a practical point of view with the implementation of a renderer that employs
		\emph{Virtual Texturing} in order to use textures of such large sizes that every spot of a scene can be textured individually.\\
		Furthermore we introduce a accompanying tool chain, which allows its user to create textures of multiple gigabytes and
		to process existing data in a way such that they can be used with the generated textures.\\
		\\
		From a more analytical viewpoint we investigated several ideas that aim to improve the visual quality as soon as latency is
		inevitable. This includes an analysis of several heuristics that give an indication on which pages have to be streamed next
		and a technique that tries to predict the future need.
		In addition to that we developed a method that can be utilized to measure the performance of these techniques.
		Provided with the results of this study, we present a combination of these ideas yielding better results compared to basic heuristics. 

	\section{Organization of the Thesis} \label{01_organization}
		The present thesis is divided into 5 chapters. During the course of the second chapter we shortly
		review related works in this field. 
		The third chapter gives an overview over the basic concepts behind Virtual Texturing, so that we can discuss several details of 
		our rendering system and its accompanying tool chain throughout the rest of the chapter.
		In chapter four we present and discuss our results of the previously mentioned analysis. 
		The last chapter closes the thesis with a conclusion and a look on some subjects that could be worthwhile to investigate in the future.\\
		\\
		Furthermore there is an appendix added at the end to make the  overall thesis more readable and give
		the reader the chance to review certain mathematical formulas, that we employ within our discussion.

\blankpage

\chapter{Related work}
The related work section is split into two sections.
In the first part we review various publications on  Virtual Texturing itself. But before that, we also discuss two older approaches that
use huge amounts of texture data for the purpose of rendering interactive scenes and hence are directly related to the topic of this
thesis.
During the second section we review existing literature on \emph{Sparse Voxel Octrees}, a technique that extends the fundamental idea of
Virtual Texturing to the visualization of voxels in order to provide a rendering system with unique and highly detailed geometry.

\section{Virtual Texturing}
		Tanner et al.~\cite{Tanner98} described the idea of a \emph{Clipmap}, which was  
		an early attempt to emulate the residence of a single large texture within the limited memory space.
		It is basically a mipmap pyramid in which each level is clipped to a certain region that is defined by a so called 
		\emph{clipcenter}. 
		The clipcenter is calculated based on the viewpoint, so that only the visible parts of the texture data have to be stored in memory.
		Each clipcenter represents a specific region within its mip level. Due to this fact it works well in the case of a terrain, whose parametrization 
		exploits this property. 
		But in the case of an arbitrary textured geometry it can happen that parts become visible that are not necessarily near to each other within the texture.
		This means, that in this case the clipcenter would cover a large region that contains unused texture parts.
		\\\\
		In 2004 Darbon et al.~\cite{Darbon04} proposed in a technical report a system that can be employed to use 
		huge amounts of texture data on arbitrary textured meshes. 
		In contrast to using one large texture it works with a set of several texture files that were used to texture different meshes.
		For each mesh a so called \emph{Texture Load Map} is used to identify texture tiles that are visible from the current viewpoint. 
		In order to do so it renders the processed geometry into texture space. The output of this rendering process is
		a texture whose size matches the number of tiles within the texture. This means that if a texel in the texture has been
		rendered then the corresponding tile is needed.
		For each tile that is not currently available in the memory a load request is sent to a \emph{Texture Producer}, which
		has access to all the texture files on the hard drive. As soon as the requested tiles become available, the system
		will update a corresponding \emph{Tile Pool}, which can be used in combination with a dynamic \emph{indirection table} to 
		render a mesh. 
		\\\\
		Using a single large texture to render arbitrary textured meshes in an efficient manner was pioneered by id Software
		with a technique called \emph{Megatexturing}. It has been used to texture the landscape environment of their commercial
		product \emph{Enemy Territory: Quake Wars}.
		For their next generation engine \emph{id Tech 5}, which is used in the upcoming game Rage, id Software
		extended the idea further to uniquely texture every object in the game, not only the terrain. 
		While implementation details are not publicly available, van Waveren addressed some of the issues of this technique 
		in 2009~\cite{Waveren09}.\\
		\noindent Sean Barrett gave a presentation about his implementation of Virtual Texturing at the 
		game developers conference 2008~\cite{Barrett08}.
		It features a detailed description of his fragment shader that can be used to identify the needed texture parts within the 
		screen space. Furthermore it describes how to render the currently visible part of the scene with just the available set
		of texture parts.\\
		\noindent Mittring~\cite{Mittring08} discussed the advantages and disadvantages of several solutions for different Virtual Texturing aspects 
		like rendering, content creation and data streaming.
		Instead of describing a complete solution, the report is a collection of several ideas that have been investigated by
		Crytek GmbH. 

\section{Sparse Voxel Octrees}
		Sparse Voxel Octrees is a technique that replaces the currently 
		used triangle rasterization pipeline by employing raycasting in combination with an octree of voxels.
		The octree represents a geometry set that, like the texture data in case of Virtual Texturing, is far too large to be contained completely within the memory.
		It can be considered as an extension to Virtual Texturing, since it uses the same underlying principles and suffers from similar 
		problems, because its visual quality also depends on how fast the data can be streamed from the backing storage.
		Due to this fact the results of this thesis could also be relevant for future research on Sparse Voxel Octrees.\\ 
		\noindent In 2008 Jon Olick described the idea of Sparse Voxel Octrees and the possibilities it provides for the game development 
		process~\cite{Olick08}.
		In the same year Crassin et al. proposed their implementation called \emph{Interactive GigaVoxels}~\cite{Crassin08}.\\
		\noindent In 2010 Laine and Karras published a technical report on their Sparse Voxel Octree implementation~\cite{Laine10}.
		It contains a discussion on several drawbacks of using voxels instead of triangles and features a analysis of the
		underlying memory usage.		

\blankpage

\chapter{Virtual Texturing}
	
	\section{Basic concepts} \label{03_basicConcepts}
\begin{figure}[h!]
 		\centering
  		\includegraphics[scale=1.0]{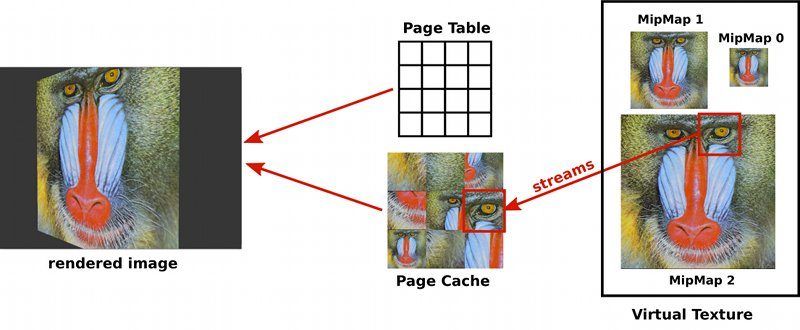}
  		\caption{High level overview of a Virtual Texturing system. The system streams the pages of a Virtual Texture into a much smaller
  				 page cache, which is then used in combination with a page table to render the scene with the currently available set.}
  		\label{fig_3_ideaOverview}
		\end{figure}

		Before we delve into the details, of our implementation in Section~\ref{03_renderer}, let us start with a discussion on basic
		concepts behind Virtual Texturing and introduce different keywords that we will encounter quite often 
		in the rest of this chapter.\\
		\\
		The basic idea of Virtual Texturing stems from the Design of Operating Systems, where a similar
		memory management technique called \emph{Paging} is used to give a running process the impression
		of having one large address space continuously available in the random access memory, although its
		parts may be scattered within the memory incoherently. They may not even be available currently and hence have to be loaded as soon as they are needed.\\
		\\
		Virtual Texturing transfers this idea to the process of texturing for the reason of providing
		a rendering system with a texture that is far too large to fit into the available memory of the
		graphics hardware.
		It emulates the residence of the large texture by actively managing a small working set within the sparse
		texture memory.\\
		\\
		We can see an abstract overview of this idea in Figure~\ref{fig_3_ideaOverview}.
		The shown system streams parts of a large texture set, called \emph{Virtual Texture}, into the faster but also smaller graphics memory, 
		so that it can use just the available set to render the scene in the best possible fashion.
		With \emph{best possible fashion}, we refer to the fact, that in order to guarantee interactive framerates, the
		system has to render the scene although some of the needed parts are still not available in the sparse memory, called
		\emph{Page Cache}.
		\subsection{Virtual Texture}\label{03_virtualTexture}
			The Virtual Texture, whose parts are progressively loaded by the rendering system, resides on a slower backing storage
			(e.g. a hard disc drive or an internet server) that has enough capacity to store it completely.
			It consists of a very large texture that is accompanied by all its mipmaps.
			The size of the highest resolution is chosen to be quadratic and a power of two (e.g. 32768 x 32768 Pixel). 
			Every mipmap has exactly double the size in width and height as its next lower level of detail. Doing so ensures that all mipmaps within 
			the Virtual Texture conform to the same properties as the highest resolution.\\
			We enumerate every mip level, starting with 0 at the lowest resolution in ascending order.
			
			\subsubsection{Pages} \label{03_pages}
				All the mipmaps of a Virtual Texture are divided up into equal sized parts in order to make the task of
				selecting and managing the set of texture parts as easy and efficient as possible.				
				\begin{figure}[h!]
  					\centering
  					\includegraphics[scale=1.0]{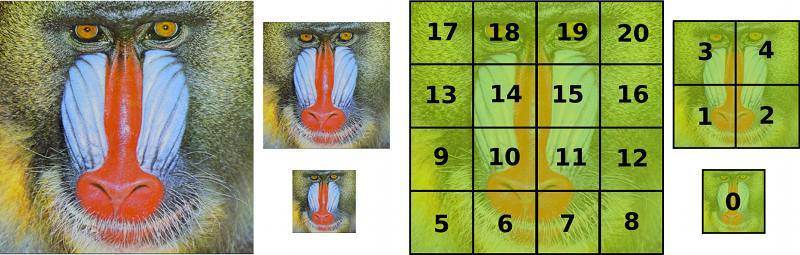}
  					\caption{All mipmaps of a Virtual Texture are split into equal sized parts, called pages. Starting with one
  					page on the lowest mipmap, the number of pages within a mipmap will increase by a factor of 4 for every mip level.}
  					\label{fig_3_splitting}
				\end{figure}	
				The size of each mipmap is, because of the assumptions we made above, quadratic and a power of two.
				Due to this reason we choose a page size $w_{page}, h_{page}$ that exhibits the same properties (e.g. 128 x 128 Pixel), so 
				that we get a discrete number of pages within every level of detail.
				We define the number $n_i$ of pages within mipmap $i$ to be
				\begin{center}
				\begin{math}
					\displaystyle
					n_i = 2^i \cdot 2^i
				\end{math}
				\end{center}
				This automatically implies that the dimension of mipmap 0 coincides with the chosen page size and that there is exactly one
				page within this level of detail. Furthermore it guarantees that the number of pages will increase by a factor of 4 for every
				mip level.\\
				\\
				For the reason of identifying each page, we can use both, a relative- and an absolute enumeration scheme.
				The example in Figure~\ref{fig_3_splitting} shows a absolute identification system. In the relative scheme, we 
				would identify for example page 6 as page 1 on mip level 2.
				By knowing the number of pages within every mipmap we can transform the page index from the relative into the absolute scheme
				quite easily. Let $p_{rel}^m$ denote the relative index of a page within mipmap $m$. Then the absolute index $p_{abs}$
				will be
				\begin{center}
				\begin{math}
					\displaystyle
					p_{abs} = p_{rel}^m + \sum\limits_{i=0}^{m-1} n_i = p_{rel}^m + \sum\limits_{i=0}^{m-1}(2^i \cdot 2^i)
				\end{math}
				\end{center}
				For the relative enumeration scheme we can furthermore split the page index into two coordinates $p_x^m , p_y^m$ that can
				be calculated by
				\begin{center}
				\begin{math}
					\displaystyle
					p_x^m = p_{rel}^m \mod 2^m \hspace{1cm} p_y^m = \left\lfloor \frac{p_{rel}^m}{2^m} \right\rfloor  
				\end{math}
				\end{center}
				For the other direction we can use
				\begin{center}
				\begin{math}
					\displaystyle
						p_{rel}^m = p_x^m + p_y^m \cdot 2^m
				\end{math}
				\end{center}
				
			\subsubsection{Page Hierarchy} \label{03_pageHierarchy}
\begin{figure}[h!]
  					\centering
  					\includegraphics[scale=1.0]{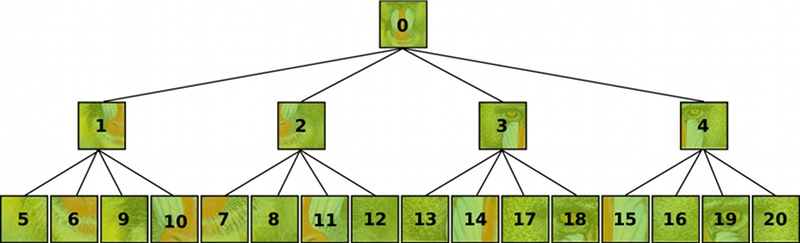}
  					\caption{Page Hierarchy for three mipmaps of a Virtual Texture. Every node, except a leaf, is connected 
  							 to exactly those four pages within the next mipmap, which contain the same texture information just in a
  							 higher resolution.}
  					\label{fig_3_hierarchy}
				\end{figure}

				By splitting all mipmaps in the way we described above, we get a nice property that
				can be exploited during rendering in order to have a fallback in the case that a needed page
				is not currently available in the cache.\\
				\\
				Looking at Figure~\ref{fig_3_splitting} reveals, that there exists a relation between the pages of 
				succeeding mipmaps.
				Every page within a mipmap can be connected with exactly four pages of the next higher level of
				detail. Doing so for all the mipmaps will provide a quadtree like the one shown in Figure~\ref{fig_3_hierarchy}.			
				Pages that are connected by an edge are called \emph{parents} and \emph{children}.
				All the pages that are on a path between a page and the root of the quadtree are called \emph{ancestors} of
				the considered page.\\
				\\
				As stated above a system can exploit this relationship during the course of rendering in order to have a fallback.
				This actually means, that it can use the lower resolution quality of any ancestor to texture the scene, while
				it waits for the needed page to be loaded.
				
		\subsection{Page Cache} \label{03_pageCache}
			Since Virtual Textures are in most cases too large to be contained completely within
			the texture memory of the graphics card, a rendering system has to keep only those pages
			in the memory that are currently needed.
			This active management of texture data takes place on a reserved area within the graphics
			cards memory called \emph{Page Cache}.  
			Like the  mipmaps of a Virtual Texture, the cache is split into equal sized parts, called \emph{frames}.
			These frames have exactly the same size as the pages, so that each page of the texture fits into any frame
			and could simply be exchanged by another page.\\
			\\
			All frames together form a grid as shown in Figure~\ref{fig_3_virtualPhysical} (a). We identify each frame within the grid by 
			using a pair of size-independent coordinates $f_x$ and $f_y$.
			Since every frame has exactly the same size as every page, we can calculate the offset $x_{offset}, y_{offset}$ of a frame within
			the cache by
			\begin{center}
			\begin{math}
				\displaystyle
				x_{offset} = f_x \cdot w_{page} \hspace{1cm} y_{offset} = f_y \cdot h_{page}
			\end{math}
			\end{center}
			with $w_{page} = h_{page}$ due to the quadratic nature of the pages.
			
		\subsection{Virtual- \& Physical coordinates} \label{03_virtualPhysicalAddr}
			All the geometry that is processed by a rendering system will use texture coordinates that have been
			created as if the complete texture on its highest resolution could be available. 
			We call these coordinates \emph{virtual coordinates}, since they stay in contrast to the reality that is
			dictated by the current state of the page cache.
			\begin{figure}[h!]
  				\centering
  				  \subfigure[]
  				{
  					\includegraphics[scale=0.8]{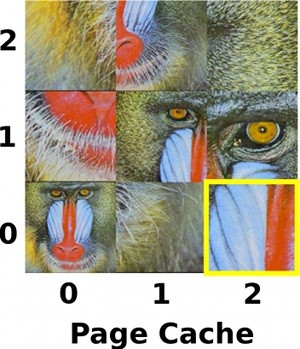}
  				}
  				\hspace{1cm}
  				\subfigure[]
  				{
  					\includegraphics[scale=1.0]{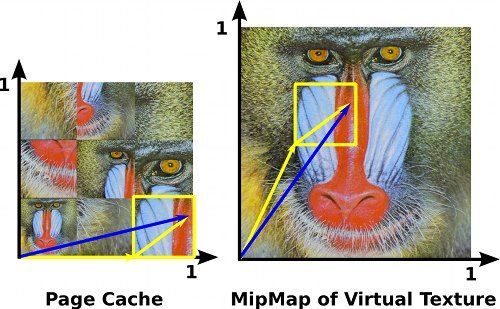}
  				}
  				
  			\caption{(a) Every frame in the cache is identified by two size independent coordinates $f_x, f_y$. 
  			The highlighted frame for example is identified as $f_x = 2, f_y = 0$.
  			(b) A page can reside at any possible position within the page cache, so we need different texture coordinates
  			in order to sample the texture data correctly.}
  			\label{fig_3_virtualPhysical}
			\end{figure}
			Due to the reason that a page can reside at any possible position within the page cache, we need, as shown in
			Figure~\ref{fig_3_virtualPhysical} (b), different coordinates to sample from the cache.
			Obviously a rendering system has to translate the given virtual texture coordinates into these so called \emph{physical coordinates}
			or otherwise will fail to render the scene correctly.
		\subsection{Page Table} \label{03_pageTable}
			In order to transform virtual coordinates into physical ones, the rendering system needs to know in which frame
			a specific page resides or which ancestor could be used as a fallback.
			\begin{figure}[h!]
  				\centering
  				\includegraphics[scale=1.0]{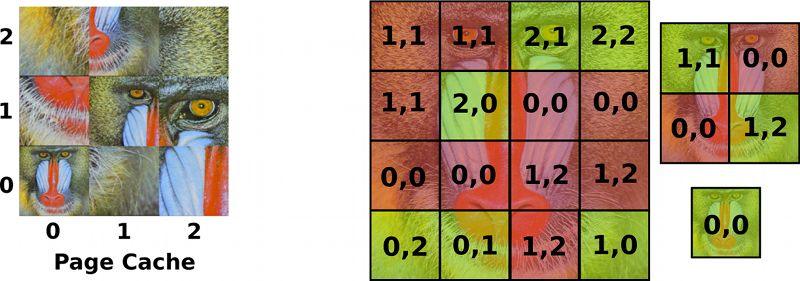}
  			\caption{Example for the data stored within the page table. 
  					If a page is available (green) then its entry within the page table contains the coordinates of the frame in which the page
  				     resides. Otherwise (red) it contains the coordinates of the frame that keeps the best available fallback.}
  			\label{fig_3_pageTableIdea}
			\end{figure}			
			For the reason of keeping track of these informations, the system maintains a so called \emph{Page Table}.
			It provides an entry for each page, which can be retrieved during rendering in order to get the coordinates of the
			frame from which it can sample the appropriate texture data. An example for the data that is stored within the page table is shown
			in Figure~\ref{fig_3_pageTableIdea}.
			\\\\
			Since the page cache is modified often, it is necessary to update this table in an efficient manner.

	\section{Renderer} \label{03_renderer}
		After we described the basic concepts of Virtual Texturing in a more high level style, we are now
		ready to take a closer look at specific details of our rendering system.\\
		\\
		Our rendering system is part of a small engine we developed and which runs on Mac OS X and Linux with OpenGL as a Graphics API.
		Since we can not discuss the complete engine in every detail, we will concentrate on implementation details that are relevant for Virtual Texturing.
		
		\subsection{Overview} \label{03_rendererOverview}		
\begin{figure}[h!]
  			\centering
  			\includegraphics[scale=2.0]{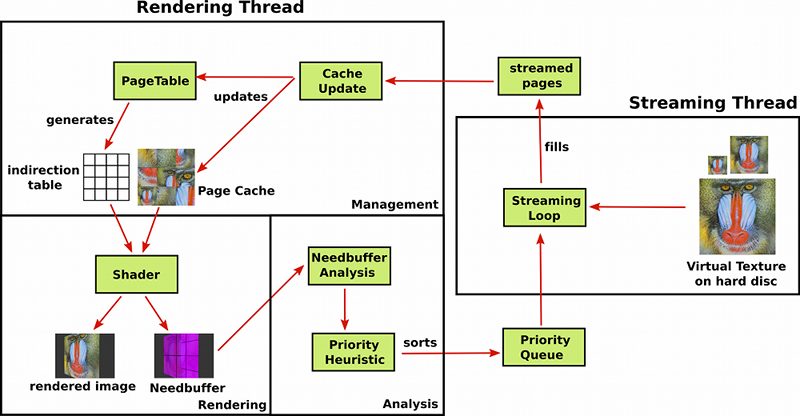}
  			\caption{Schematic overview of our Renderer. The system is divided into a rendering- and a streaming thread. Both
  					 communicate over two datasets. While the streaming thread is simply used to load the needed pages from the
  					 backing storage, the rendering thread fulfills three different tasks: management, rendering and analysis.}
  			\label{fig_3_renderer}
			\end{figure}

			Figure~\ref{fig_3_renderer} shows a schematic overview of our Virtual Texturing implementation.
			Let us discuss the complete system in a high level fashion, before we delve into the specific subparts individually.\\
			\\
			Our system is divided into two parts: rendering and streaming.	Both are implemented as separated threads for
			the reason of preventing the rendering part of getting stalled by waiting for unavailable pages to be loaded.\\
			\\
			The rendering thread is based on the ideas that have been proposed by Sean Barrett~\cite{Barrett08} and consists
			of three subparts:
			\begin{enumerate}
				\item \textbf{Management} Inserts newly streamed pages into the page cache and updates the page table accordingly. 
					Furthermore an indirection table is generated, which represents a compact version of the page table 
					that can be accessed by the shader during rendering.
				\item \textbf{Rendering} A shader that identifies the set of needed pages and renders the scene with the help of the indirection table
				    and those texture parts that are currently available in the page cache. It outputs a standard color image that will
				    be shown to the user and a second render target, called \emph{Needbuffer}, which contains the indices of the
				    needed pages.
				\item \textbf{Analysis} Analyzes the needbuffer to collect informations on the pages that have been marked as needed.
					These informations are further used to calculate priorities that determine the order in which the pages will be streamed.
				\end{enumerate}
				
				\noindent The streaming thread on the other side simply consists of a \emph{streaming loop} that has access to the Virtual Texture on 
				the hard disc. Its sole task is to load the pages that have been identified as needed by the rendering part in the
				appropriate order.\\
				\\
				The threads communicate through two datasets that are available to both of them:
				\begin{enumerate}
					\item \textbf{Priority Queue} Represents the list of pages that should be streamed. Its order
					is based on the priority values that are calculated for each page individually during the 
					analysis part of the rendering thread. 	
					\item \textbf{Streamed Pages} After a page has been loaded by the streaming thread, it will be stored in a list
					within the main memory. As soon as the rendering thread reaches the cache update, its management subpart will take 
					all pages within the list and insert them into the page cache.
				\end{enumerate}
		\subsection{Management}
			Before each frame is rendered we check for newly streamed pages within the \emph{streamedPages} dataset. 
			If this is the case, we modify the page cache and update the page table in order to keep track of the current cache status.
			\subsubsection{Modifying the page cache}
				The page cache is simply a two dimensional rgb texture that can be accessed by the shader during rendering.
				For the color depth we use 8 bit per channel, since it coincides with the color data stored in our Virtual Texture.\\
				\\
				We divide and organize the cache as discussed in Section~\ref{03_pageCache}.
				But in contrast to the theory explained there we use a frame size that is slightly larger than the one we use to
				divide the Virtual Texture into pages. Why we exactly have to do this will be explained in Section~\ref{03_filtering}
				and~\ref{03_pageBorderCreation}.\\
				\\
				Updating the cache is very simple. For each new page that is available, we find the next least recently used frame within 
				the cache and copy the texture data with the help of the OpenGL function \emph{glTexSubImage2D} into the respective 
				position within the texture.
				
			\subsubsection{Updating the page table}\label{03_rendererPageTable}
				We use a quadtree for the representation of the page table, in order to exploit the page hierarchy, which has been described
				in Section~\ref{03_pageHierarchy}.
				It is basically an array that provides exactly one entry for each page of the Virtual Texture. Every entry, except
				page 0, has a reference to its parent page, so that we can follow the references to get data that belongs 
				to the ancestors.\\
				\\
				In order to keep track of the current cache status, we store in each page entry the coordinates $f_x, f_y$ of the frame that contains 
				the page itself or the best available ancestor.
				As soon as something in the cache has been modified, we have to update the page table accordingly.				
				The above mentioned description of our quadtree has the nice property, that we can do this by just looping
				over the array that represents it.
\begin{center}
\lstset{language=pseudo}
\begin{lstlisting}
for(page = 0; page < MaxPageEntry; ++page)
{
	p = pageTable[page];
	if(p is not in cache and p has a parent)
		modify p to use the frame coordinates stored in parent
}
\end{lstlisting}			
\end{center}
			\subsubsection{Generating the indirection table}\label{03_indirectionTable}
				Since we can not use our page table as it is within the shader, we generate a so called \emph{indirection table}. This is
				simply a two dimensional texture, in which each page is represented by a texel. We generate and upload it to the graphics 
				cards memory directly after the page table has been updated as described above.
				For each page we retrieve the frame coordinates that are stored in the page table entry and encode them into the color channels
				of the appropriate texel. Furthermore we store the mip level $m$ of the page that resides within the respective frame.
				\begin{center}
				\begin{math}
					\displaystyle
					r = f_x \hspace{1cm} g = f_y \hspace{1cm} b = m
				\end{math}	
				\end{center}
				The shader can later use this information to transform the virtual coordinates into physical ones.
			
		\subsection{Rendering} \label{03_shader}
			A very important part of our implementation is the shader.
			Its tasks are
			\begin{enumerate}
				\item Identifying the currently needed set of pages
				\item Rendering the scene with the set that is available in the page cache
			\end{enumerate}
			Both tasks are achieved in the screen space for every fragment that gets rasterized.
			We started out with two shaders, so that each was handled separately.
			Doing so ended up being redundant, since we have to know which page is needed in order
			to retrieve the information that can be used to render the scene with the available set.
			So we switched to a unified shader with the help of \emph{multiple rendering targets}.\\
			\\
			The fragment shader has the following data as inputs available
			\begin{enumerate}
				\item The virtual texture coordinates $s,t \in [0,1]$ that have been interpolated across the face.
				\item The texture that represents the indirection table.
				\item The texture that represents the page cache.
			\end{enumerate}
								
			\subsubsection{Identification of needed pages} \label{03_identifyingNeededPages} 
\begin{figure}[h!]
 					\centering
 					\subfigure[]
					{
						\includegraphics[scale=1.0]{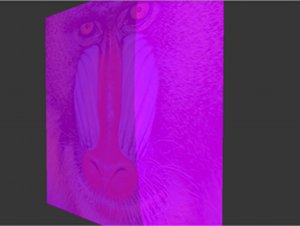}
					}	
 					\subfigure[]
					{
						\includegraphics[scale=1.0]{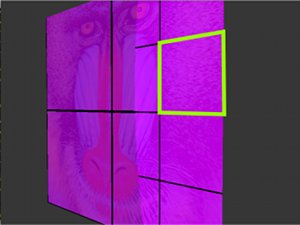}
					}		
				 	\caption{
  			 			 (a) Calculation of the mip level for every fragment
  			        	 (b) Identifying the page within the respective mip level
  			        	 }
  					\label{fig_3_ideaPageIdentification}
				\end{figure}

				The process of identifying the needed page for a fragment can be split into two steps 
				(see Figure~\ref{fig_3_ideaPageIdentification})
				\begin{enumerate}
					\item Calculation of the mip level
					\item Calculation of the page index within the mip level
				\end{enumerate}
				We calculate the mip level $m$ of the considered fragment by employing the edge-compression
				method, which is described in Appendix~\ref{app_lodCalculation}.\\
				\\
				We know from Section~\ref{03_pages} that the number of pages within mipmap $m$ is $2^m \cdot 2^m$.
				This means, that we have $2^m$ pages per row and $2^m$ pages per column.
				By scaling $s$ and $t$ accordingly, we can get the relative page coordinates $p_x^m$ and $p_y^m$ quite easily.
				\begin{center}
				\begin{math}
					p_x^m = floor(s \cdot 2^m) \hspace{1cm} p_y^m = floor(t \cdot 2^m)
				\end{math}
				\end{center}
				
			\subsubsection{Needbuffer bit depth}
				After we identified the page, we can store $p_x^m, p_y^m$ and $m$ into the color channels of the needbuffer. 
				How this is done in detail depends on the bit depth we are using, since it impacts the maximal number of pages 
				we can encode. 				
			
				\paragraph{32 Bit Per Channel}
					Using 32 Bit per color channel allows us to encode more pages than needed in most cases. 
					So we simply store the data as it is.
					\begin{center}
					\begin{math}
						r = p_x^m \hspace{1cm} g = p_y^m \hspace{1cm} b = m
					\end{math}
					\end{center}
					Although the output encoding stays easy, we have to keep in mind, that we later read the needbuffer into
					the random access memory, in order to analyze it.
					Using such a high resolution is quite inefficient and will slow down the performance.
				\paragraph{8 Bit Per Channel}
					With 8 Bit per channel we can only encode 256 symbols per color. 
					This means, that we could only address $256 \cdot 256 = 65536$
					pages per mipmap, if we would use the same encoding scheme as in the case of 32 Bit.
					This equates to an upper bound of textures that have a size of 32k x 32k.
					So in order to use textures of larger sizes, we exploit the usage of a alpha channel.
					\begin{center}
					\begin{math}
						\displaystyle
						r = p_x^m \mod 256 \hspace{1cm} g = p_y^m \mod 256 \hspace{1cm} b = m \hspace{1cm} a = \frac{p_x^m}{256} + \frac{p_y^m}{256} \cdot 16
					\end{math}
					\end{center}
					By doing so, we can encode $256^3 = 16777216$ pages per mipmap. This means that we actually could represent
					textures of the size 524k x 524k.				
				
			\subsubsection{Rendering with the available set}
\begin{figure}[h!]
 					\centering

 					\subfigure[]
					{
						\includegraphics[scale=0.7]{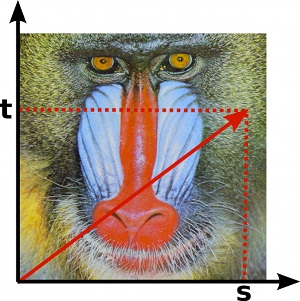}
					}
					\subfigure[]
					{
						\includegraphics[scale=0.7]{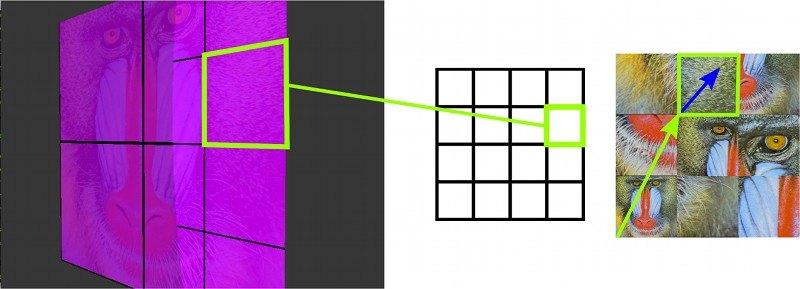}
					}
					\subfigure[]
					{
						\includegraphics[scale=0.7]{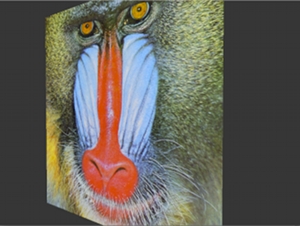}
					}	
					\hspace{1cm}
  					\subfigure[]
  					{
  						\includegraphics[scale=0.7]{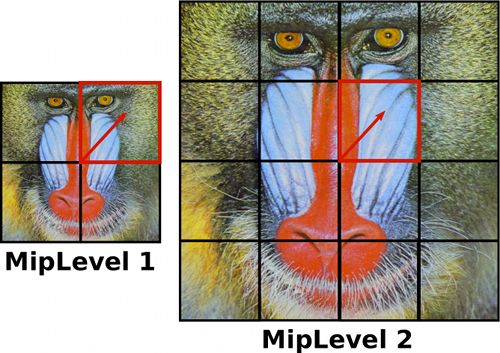}
  					}					
				 	\caption{
  			        	 (a) virtual coordinates $s,t$ 
  			        	 (b) we make a page table look up to get the external offset (green) and calculate the internal offset (blue) 
  			        	 (c) doing so for every fragment will yield the final image 
  			        	 (d) Example on why the internal offset depends on the mip level: We want to sample from a page within 
  			        	 mipmap 2, but the needed texture data is only available as a page from mip level 1. Using the same internal offset would yield incorrect results.}
  					\label{fig_3_ideaRendering}
				\end{figure}

				In order to sample the appropriate texel from the page cache, we have to transform the virtual coordinates $s$ and $t$ into physical ones, which
				as stated in Section~\ref{03_virtualPhysicalAddr} point to the appropriate texture data.
				To do so we have to answer the following questions:				
				\begin{enumerate}
					\item Which cache frame contains the needed texture data (external offset)?
					\item Which position within the frame should be sampled (internal offset)?
				\end{enumerate}
				Both questions can be answered by using the data we stored in the indirection table. So we use the page index we calculated
				in Section~\ref{03_identifyingNeededPages} to retrieve $f_x, f_y$ and mip level $i$ from the texel that represents the needed
				page within the indirection table.\\
				\\
				The external offset $x_{e}, y_{e}$ is simply the start position of the respective frame within the cache.
				We can calculate it from $f_x$ and $f_y$ as described in Section~\ref{03_pageCache}.
				\\\\
				Calculating the internal offset $x_{i}, y_{i}$ is a little trickier, since we have to take the mip level of the
				available page into account. Figure~\ref{fig_3_ideaRendering} (d) shows that using the same internal offset within different
				mipmaps would yield incorrect results.
				We know from Section~\ref{03_identifyingNeededPages} how we can calculate the page within a mip level. When we do the same for
				the available mip level $i$ and use the $fract$ operation instead of $floor$, we get the correct result
				\begin{center}
				\begin{math}
					\displaystyle
					x_{i} = fract(s \cdot 2^i) \hspace{1cm} y_{i} = fract(t \cdot 2^i)				
				\end{math}
				\end{center}
				Provided with both the external- and the internal offset we can use $s' = x_{e} + x_{i}$ and $t' = y_{e} + y_{i}$ to sample 
				the appropriate texel from the page cache.
		
			\subsubsection{Filtering} \label{03_filtering}	
				\begin{figure}[h!]
  				\centering
  					\subfigure[]
  					{
  						\includegraphics[scale=0.25]{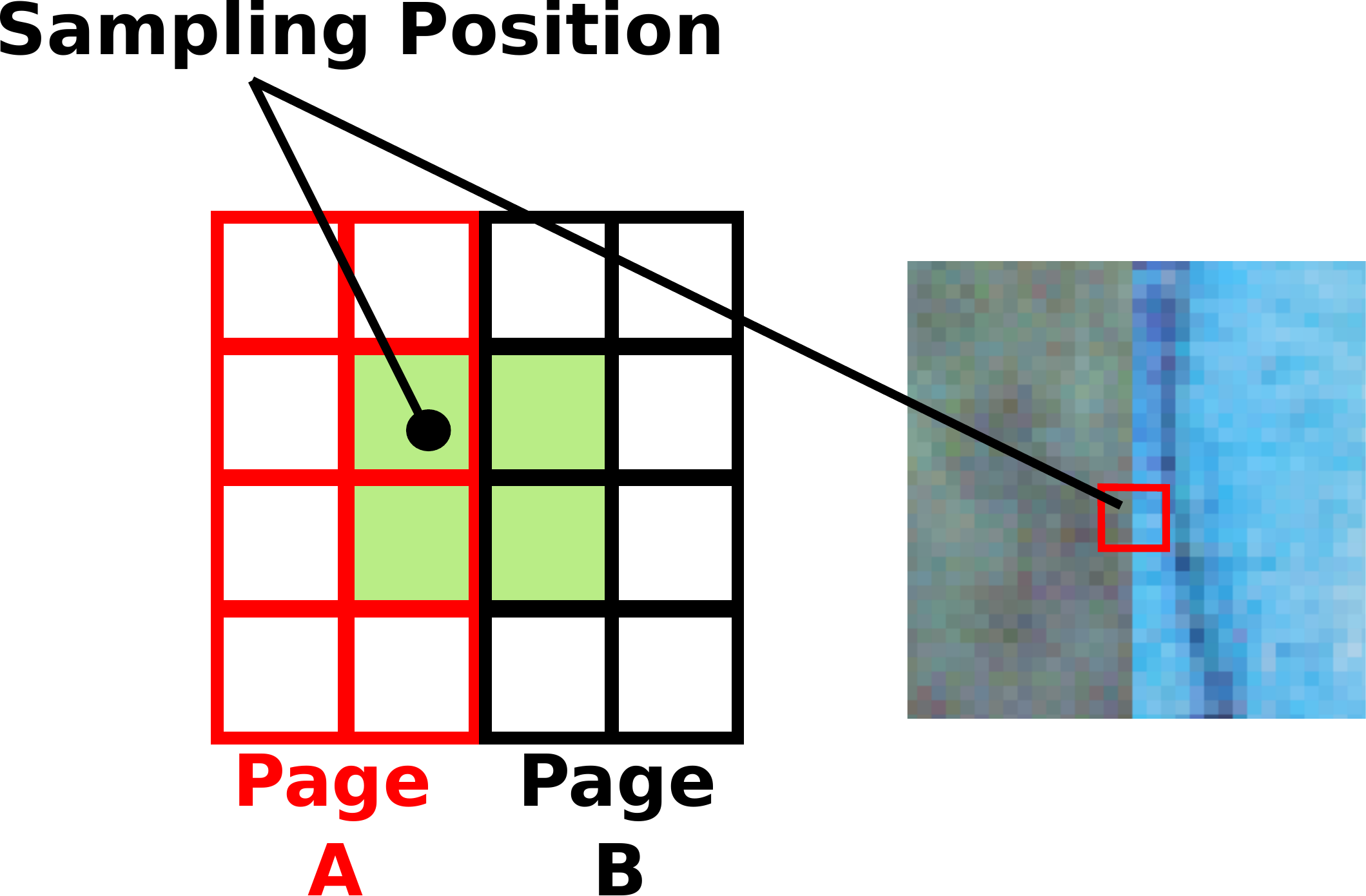}
					}
					\hspace{1cm}
  					\subfigure[]
  					{
  						\includegraphics[scale=0.7]{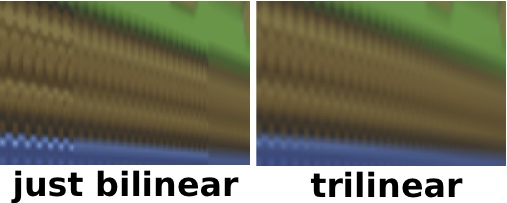}
					}
  				\hspace{1cm}
  				\caption{
  				(a) The problem of bilinear filtering. Sampling texels at the border of two pages within the page cache can lead
  				to a interpolation of quite different colors and hence introduce visual seams at page borders within the rendered image.
  				(b) Visual edges at the border between two neighbouring mip levels. Trilinear filtering diminishes this effect.} 
  				
  				\label{fig_3_bilinearProblem}
				\end{figure}
				
				\paragraph{Bilinear}
					We tried two ways to incorporate bilinear filtering. We started with a manually implementation
					within our shader, which ended up being prohibitively slow.					
					The better way to get bilinear filtering in a shader is to use the implementation that is
					provided by the chosen Rendering API, so that the interpolation of the texel data is done automatically during sampling.
					But doing so is not without its problems. Looking at Figure~\ref{fig_3_bilinearProblem} (a) shows that it can lead to incorrect
					interpolation results and hence to visual seams when a texel is sampled that lies at the border of two neighbouring 
					pages within the cache.
					We can circumvent this problem, when we provide a small pixel border around every page within the cache.
					This border contains texel data from the neighbouring pages within the Virtual Texture, so that we can sample
					the data as if the complete Virtual Texture would be available. Due to this fact, we have to store slightly larger pages within the
					the page cache. This is no contradiction to the theory explained in Section~\ref{03_pages}, because we still operate logically on pages
					of the chosen page size and just use the border as a workaround.
					How we get this border is described in Section~\ref{03_pageBorderCreation}.
					
				\paragraph{Trilinear}
					Barrett~\cite{Barrett08} proposes an implementation of trilinear filtering, that relies on two
					caches, so that it can make use of the Rendering APIs implementation. Mittring~\cite{Mittring08} states, 
					that this approach is really complicated and a waste of memory, since it stores identical texture data
					twice.
					We decided to use the obvious route that can be manually implemented within the shader.
					Since we have bilinear filtering already available, we can be sure, that every time
					we sample from the cache, we will get a correctly bilinear filtered result.
					So in order to use trilinear filtering, we simply sample for two adjacent mipmaps
					and use a linear interpolation to mix both results.
					Using trilinear filtering leads to visually better results, as shown in Figure~\ref{fig_3_bilinearProblem} (b),
					since it diminishes visible edges at the border of neighbouring mip levels.
		
		\subsection{Analysis}
			After the scene has been rendered, we update the screen and analyze the needbuffer in order to to find out which page needs to be loaded and 
			how important it is for the currently visible scene.

			\subsubsection{Evaluating the needbuffer}\label{03_needBufferEvaluation}
				In order to analyze the needbuffer, we read the corresponding image data from the graphics card into the
				main memory. This is the moment in which it can become inefficient to use 32 Bit per color channel,
				due to the fact that a lot more data has to be downloaded from the texture memory.\\
				\\
				As soon as a copy of the needbuffer resides in the working memory, we start to loop over all the pixels.
				For each pixel we transform the extracted relative page coordinates $p_x^m, p_y^m$ into the absolute index $p_{abs}$
				and mark the respective page within the page table as needed.
				Furthermore we collect different informations, that can be classified into two sets
				\begin{enumerate}
					\item Global information like cache hits and misses.
					\item Local information regarding the currently considered page, like the number of pixels that need it.
				\end{enumerate}		
				While the first set can be used to track the performance of our system during runtime, we store the page specific
				informations within the appropriate page table entries.\\
				\\
				After the complete needbuffer has been analyzed we employ a \emph{page priority heuristic} for the purpose of measuring
				the importance of each needed page.
			\subsubsection{Page Priority Heuristic}
				We implemented a set of different heuristics that calculate a priority for each needed page based on the information
				we gain within the analysis.
				The assigned priority of a page determines its position within the priority queue, in order to ensure that the most important 
				pages get streamed first.
				Since we stored the page specific informations within their page table entries, we can further use the heuristics in combination with
				exploited page hierarchy to estimate the priority of the ancestor pages, if we choose to implicitly stream them as well.		
				A more detailed description of the used heuristics will be given in Chapter~\ref{04_priorityHeuristics}, where we will
				discuss their impact on the rendering quality.
			
		\subsection{Streaming}\label{03_streamingOfPages}
			Although it would be the starting point for practical performance optimizations, it was out of scope to tweak the streaming thread and hence we stayed 
			with a very simple implementation for our system. 
			Furthermore we did not incorporate any form of realtime texture decompression methods~\cite{Waveren06}, which certainly would influence
			the design of the streaming part.\\
			\\
			Our streaming thread is simply a loop that accesses the Virtual Texture on the hard disc. Every cycle run it fetches
			the next needed page with the highest priority from the priority queue.
			The fetched texture data is then stored in the list called \emph{streamed pages}, so that the management part of the rendering thread
			can update the cache as soon as possible.
			How we exactly access the Virtual Texture on the hard disc and extract the needed texture data depends on the file format we chose.			
	
			\subsubsection{Virtual Texture file format} \label{03_fileFormat}
				Instead of using a traditional image layout for the resulting texture, we designed
				a simple file format that seemed more practical for the task of streaming the 
				pages. 
				\begin{figure}[h!]
 				\centering
  				\includegraphics[scale=1.0]{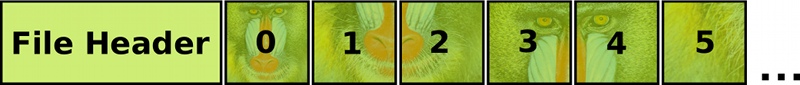}
  				\caption{All pages of all mipmaps are stored in a sequential order.}
  				\label{fig_3_fileFormat}
				\end{figure}
				As you can see in Figure~\ref{fig_3_fileFormat} we store the pages of all mipmaps in a sequential order
				inside the same file.
				This approach is beneficial, since every page represents a standalone unity within the file and
				could be accompanied by a magnitude of other useful information that could be exploited during rendering.
				Another nice property is the simplicity, with which we can reach every page within the file.\\
				\\
				Let $b$ denote the number of bytes per pixel. By knowing the absolute index $p_{abs}$ of the page we want to read,
				we can simply calculate the offset $o$ within the file as
				\begin{center}
				\begin{math}
					\displaystyle
			 		o = size_{fileheader} + (w_{page} \cdot h_{page} \cdot b) p_{abs}
				\end{math}
				\end{center}
				In contrast to a traditional image layout, where we would have to extract multiple pixel rows,
				we end up using fewer file operations in order to read a page.

	\section{Tool chain} \label{03_toolchain}
		During the course of this thesis we had to develop a set of purpose-built tools, 
		because existing image- and geometry processing software cannot operate on texture sizes of multiple gigabytes. 
		Due to the fact that the development of an interactive modeling application could not be
		accomplished in the scope of this thesis, we concentrated on a solution, that allows us
		to compile the output of traditional tools into a dataset, that can be used within the rendering
		system we discussed in Section~\ref{03_renderer}.
		
		\subsection{Texture Creation : vtmtc} \label{03_textureCreation}
			To create textures of the anticipated sizes, we developed a tool called \emph{vtmtc}. It has been
			designed to construct the resulting texture in a series of smaller steps and do as much
			preprocessing as needed for the purpose of simplifying the rendering system.
			\begin{figure}[h!]
 			\centering
  			\includegraphics[scale=0.8]{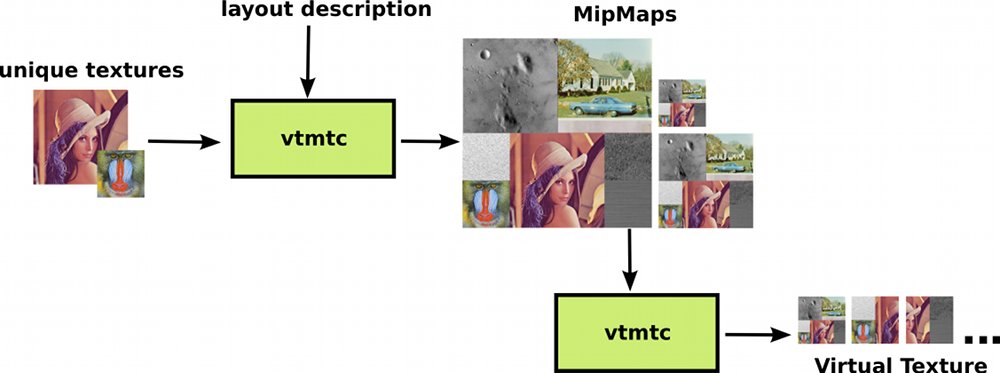}
  			\caption{The vtmtc pipeline for creating Virtual Textures}
  			\label{fig_3_vtmtc}
			\end{figure}
			
			\noindent As you can see in Figure~\ref{fig_3_vtmtc} everything starts out with a set of unique images and a
			layout that contains an exact position for each of them.
			Provided with this information, vtmtc creates a traditional two dimensional texture that represents
			the highest resolution mipmap of the Virtual Texture.
			This resulting texture is then scaled down to the next smaller mipmap size by employing bilinear
			filtering. We repeat this step with every mipmap until we reach the one that matches exactly
			the size of one page.\\ 
			The last processing step is the construction of a file that conforms to the format we discussed in Section~\ref{03_fileFormat}.
			We do so by cutting out one page at a time from the created set of mipmaps and storing them sequential in
			the filestream.

			\subsubsection{Page-Border Creation} \label{03_pageBorderCreation}
				As we discussed in Section~\ref{03_filtering} a pixel border around the pages is needed in order
				to prevent the rendering APIs implementation of bilinear filtering from filtering across the
				borders of neighbouring pages within the page cache.
				Looking again at the file format we introduced in Section~\ref{03_fileFormat} it seems
				obvious to add these borders during the creation of the texture and simply save pages of a slightly greater size so that they
				also contain a small amount of the surrounding texel data that belongs to their neighbouring pages.\\
				We get the borders more or less for free, when we simply resize the rectangle that determines the area that will 
				be cut out of a mipmap. 
				
			\subsubsection{NoiseValue Calculation} \label{03_noiseValueCalculation}
\begin{figure}[h!]
  				\centering
  				\includegraphics[scale=0.6]{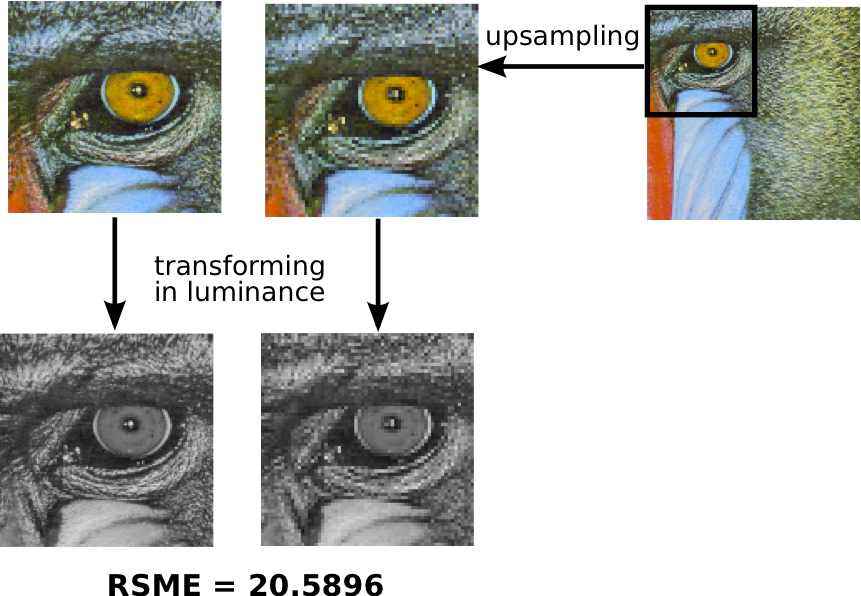}
  				\caption{Example for estimating the quality improvement the system would gain, if it would use the child instead
  				         of its parent page.}
  				\label{fig4_noiseCalculation}
				\end{figure}

				During our investigation on page priority heuristics, which will be discussed in Section~\ref{04_priorityHeuristics},
				we had the idea to provide the rendering system with a metric that gives an indication on how the quality will
				increase, if it uses the child page instead of its parent during rendering.\\
				\\
				The page hierarchy in Section~\ref{03_pageHierarchy} shows that every child represents exactly one fourth of its parent page 
				in a higher resolution. As a consequence of this we can only guarantee meaningful results if we compare the currently
				considered child with the correct quarter.\\
				As shown in Figure~\ref{fig4_noiseCalculation} we achieve this by determining the correct part within the parent and
				have it scaled up to the size of one page.
				This upsampled quarter and the child are then transformed into luminance in order to calculate
				the \emph{rooted mean squard error} (see Appendix~\ref{app_rmse}), which we chose as the metric to estimate the quality impact.
				\noindent Due to the fact that the rendering system needs these, as we call them, \emph{NoiseValues} to decide which page will be streamed next,
				we can not provide them alongside the page data within our texture file format.
				So an extra file is generated that can be read out during the initialization of the Virtual Texture.
		
		\subsection{Geometry retexturing : vtgeo} \label{03_retexturing}
				Due to the large number of projects that have been textured in a more traditional sense,
				it seems worthwhile to find a method that can automatically transform the texture coordinates of the
				provided geometry in a way that each of its faces can be textured unique.
				One of our tools, called \emph{vtgeo} does just that: it takes multiple already textured polymeshes
				and embeds them into one of our Virtual Textures and provides an individual area for each face.\\
				Although a little bit outdated, we chose the bsp file format from \emph{Quake 3 Arena} as a source for complex geometry, because
				there exists a plethora of freely available maps on the internet.
				\begin{figure}[h!]
  				\centering
  				\includegraphics[scale=0.8]{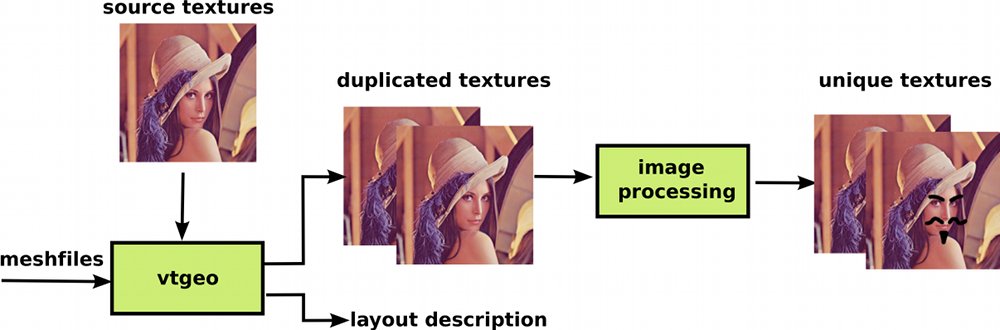}
  				\caption{Overview of the vtgeo pipeline.}
  				\label{fig_3_vtgeo}
				\end{figure}
				A schematic overview on vtgeo is shown in Figure~\ref{fig_3_vtgeo}. The starting point is a set of meshfiles that
				are textured by images from a given pool of source textures.
				During runtime vtgeo will process each mesh one by one. For each given polymesh it loops through the set of
				faces and fetches the texture that the triangles of the face are textured with.
				Since we want to texture every face uniquely, the texture is duplicated and saved on the hard disc. These 
				copies can be modified with traditional image processing tools in order to make each face look unique.\\
				\\
				Special care has to be taken of multiple texture repetitions across a face. The bsp file format for example
				allows texture coordinates that are not contained in the range of $[0,1]$. Quake 3 Arena
				uses these coordinates during rendering by simply repeating the texture in order to sample the appropriate texel.
				The problem is, that such a repetition does not make sense in the case of Virtual Texturing, where we want to
				texture every part of a face in a unique fashion. We avoided this problem by calculating the number
				of repetitions across a face and simply copying the source texture multiple times into the face specific duplicate and
				normalizing the texture coordinates of the respective vertices accordingly.\\
				\\
				While it fetches and duplicates the textures, vtgeo will create a layout file that can later be used
				with vtmtc in order to compile a Virtual Texture from the pool of unique textures.
				We will now describe how such a layout is generated and in which way the geometry has to be modified, so that
				it can be used with a texture that stems from such a layout.
				
			\subsubsection{Layout Creation \& Geometry Embedding} \label{03_geometryEmbedding}
\begin{figure}[h!]
 				\centering
				\subfigure[]
				{
					\includegraphics[scale=0.8]{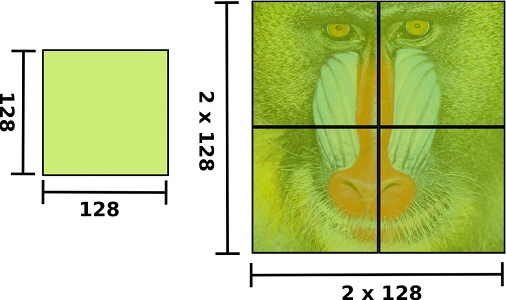}
				}
				\subfigure[]
				{
					\includegraphics[scale=0.8]{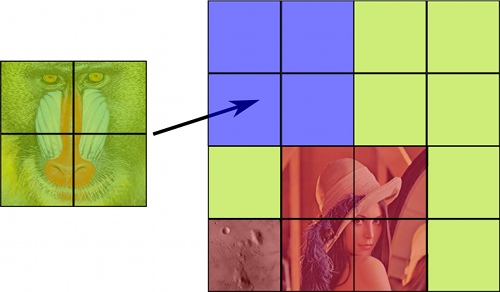}
				}
  				\caption{(a) Every grid entry has the size of one page. We estimate the number of entries a unique texture will fill in
  						 the layout. (b) We employ a first fit algorithm to find the next free space that is big enough to contain the
  						 unique texture.}
  				\label{fig_3_layout}
				\end{figure}

				After the system generated copies of all the referenced source textures, it has to find a layout
				for the highest resolution mipmap so that each face is provided with an individual area.		
				In order to determine these individual areas, it starts out with an empty layout that represents
				the texture to be created.
				The layout consists of a two dimensional grid of entries, which all have exactly the size of one page within the texture.
				\\\\
                For each face vtgeo estimates, as shown in Figure~\ref{fig_3_layout} (a), the number of grid entries that the corresponding unique 
                texture will fill horizontally and vertically.
				By knowing this size, measured in entries, a first fit algorithm is employed to find the next free area that is
				big enough to contain the texture. If there is no appropriate area available, vtgeo resizes the layout so that it represents
				a texture of the next higher resolution.
				After vtgeo found an area for each face in the geometry, the layout is ready to be used for creating the Virtual Texture.
        		But in order to use the texture in combination with the given meshes, a further processing task has to be done: transforming
				the texture coordinates of the vertices so that they coincide with the areas within the texture.\\
				\\
				With the starting point of the area within the destination texture and the sizes of both, the individual-
				and the destination texture, we can transform the texture coordinates.
				Let $x$ and $y$ denote the starting point of the area, while the width and height of the source- and destination textures
				are represented by $w_i, h_i$ and $w_d, h_d$. Let us further assume that these values are measured in pixels and so can
				be outside of $[0,1]$.
        		We can then transform the texture coordinates $s, t \in [0,1]$ of a vertex by using
				\begin{center}
				\begin{math}
					\displaystyle
						s' = \frac{(s \cdot w_i + x)}{w_d} \hspace{1.0cm} t' = \frac{(t \cdot h_i + y)}{h_d}
				\end{math}
				\end{center}
			
			\subsubsection{Problems of this approach} \label{03_embeddingProblems}			
				Although the taken approach worked quite well for our needs, we identified two problems, that can occur.
				\begin{figure}[h!]
 				\centering
				\subfigure[]
				{
					\includegraphics[scale=0.6]{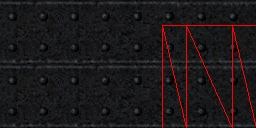}
				}
				\subfigure[]
				{
					\includegraphics[scale=0.4]{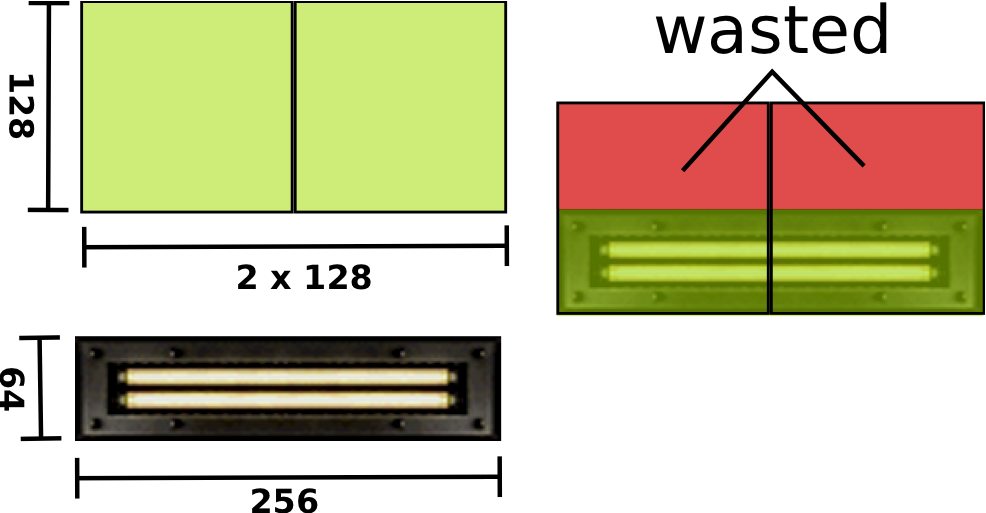}
				}
  				\caption{Problems of our approach. (a) We waste space by copying the complete texture, although the face needs just
  						 a small area. (b) Using textures that are not a multiple of the page size will lead to unused areas within
  						 the layout.}
  				\label{fig_3_embeddingProblems}
				\end{figure}
				\paragraph{No coherence}
					Faces that are close to each other within the geometry can be placed at totally different 
					places within the texture layout. Because of this lack of coherence, we can end up wasting time on loading
        			more pages than required in the optimal case.
					
				\paragraph{Wasted space}
					Our approach can waste quite a lot of texture space by simply inserting texture data that will never be
					used during rendering.			
					We identified two possible situations that can lead to this and visualized them in Figure~\ref{fig_3_embeddingProblems}.
					
					\begin{enumerate}
						\item Some faces may reference only a small part of the texture. Since our algorithm simply duplicates the original 
						      image, it can happen that it reserves far too much space for a single face. 
						\item As described above, the algorithm tries to construct a layout by using a grid of entries
							  in which each cell has exactly the size of one page. If textures are used, that have a width or height that is not
							  a multiple of this page size, we end up wasting space, due to the fact that an complete entry is marked as needed, 
							  although it is not completely filled.
					\end{enumerate}
					Since this is a common and hard problem in graphics, and since our approach worked sufficiently well, we did not invest more time into optimizing the algorithm used for layout generation.
	\section{Summary \& Result}
\begin{figure}[h!]
  			\centering
  			\includegraphics[scale=0.8]{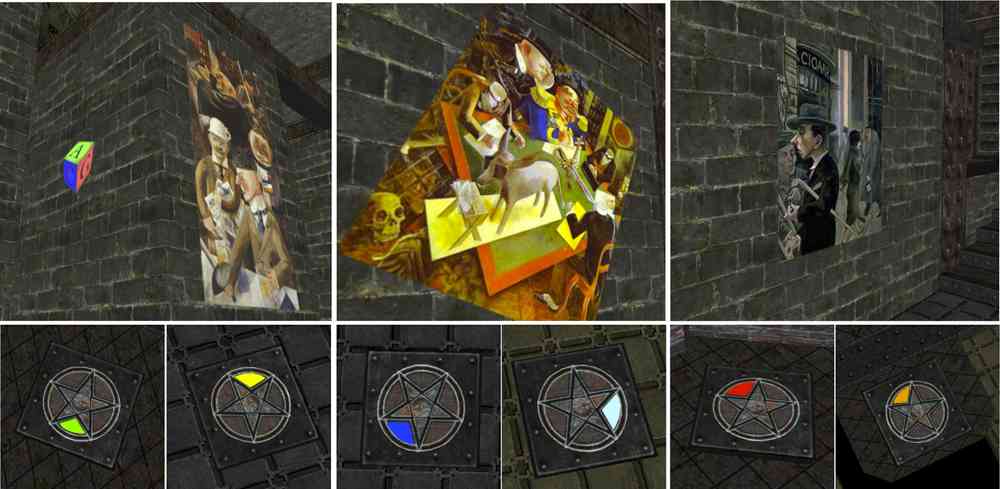}
  			\caption{A Quake 3 Arena map whose unique textures were modified to show paintings by George Grosz. The pentagram textures show that
  			we can really give each instance a unique look.}
  			\label{fig_3_result}
		\end{figure}

		In the course of this chapter we discussed our implementation of a renderer and the accompanying tool chain.
		As a result, we are able to take existing geometry and texture every of its faces uniquely.
		These textures are compiled into a Virtual Texture that can be used by our rendering system.
		Figure~\ref{fig_3_result} shows some example screenshots of a map that was modified to contain unique features.
		The remaining problem of our overall texturing process is, that due to the fact that our tool chain is non-interactive, we can not exactly
		say which copied texture belongs to which face in the geometry.
		
\blankpage

\chapter{Analysis}
\begin{figure}[h!]
	\centering
	\includegraphics[scale=1.2]{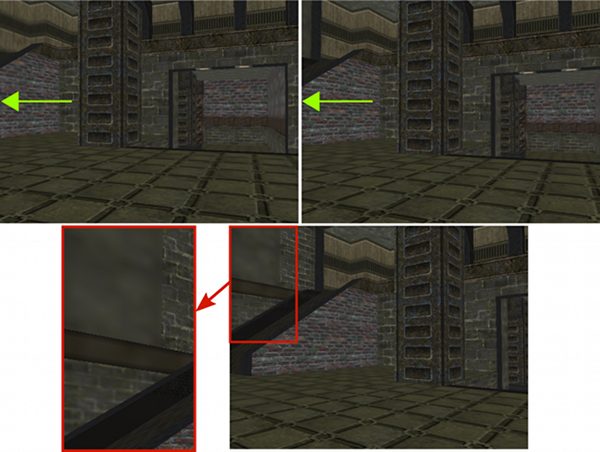}
	\caption{Camera motions like rotations will modify the current point of view. As a result pages will appear at the borders
			 of the screen, which have not been visible until now. The system will use lower resolutions as a fallback, until
			 the needed pages become available.}
	\label{fig_04_motionartifact}
	\end{figure}

\begin{figure}[h!]
	 	\centering
  		\includegraphics[scale=0.5]{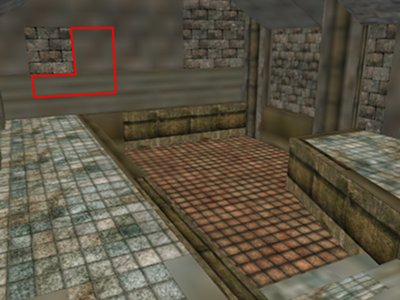}
  		\includegraphics[scale=0.5]{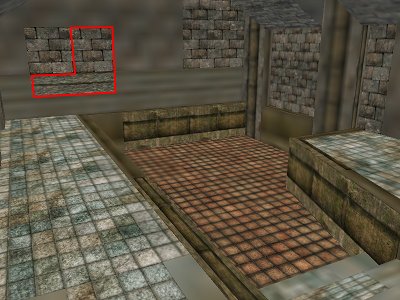}
  		\caption{Example of a so called \emph{LOD Snap}. A low resolution page is exchanged by its high quality counterpart within the course of one frame.}
  		\label{fig_04_lodSnapsExample}
	\end{figure}

	During the implementation we came across several problems that seemed worthwhile to investigate.\\
	\\
	Because of camera motions like rotations and translations, pages will become visible that are not
	currently in the memory. Figure~\ref{fig_04_motionartifact} shows an example of this.
	This effect becomes quite disturbing, especially on systems that suffer from high latencies, like
	internet applications or mobile devices, since we can load only a small number of pages per frame. 
	Due to this reason, it seems obvious to collect information about the importance of pages for the purpose of selecting
	those that have to be streamed first. We investigated this idea and discuss some of our results in Section~\ref{04_priorityHeuristics}.
	\\\\
	But even if we could improve the quality by choosing the pages cleverly, we will still experience an effect
	called~\emph{LOD Snap}~\cite{Waveren09}, which is exemplified in Figure~\ref{fig_04_lodSnapsExample}.
	It basically means, that a low resolution page that is shown in the current frame, will be exchanged by its high
	resolution counterpart within the next one.
	During the implementation it became obvious to us, that we can weaken the disturbance of this effect by
	streaming the ancestors of a page in advance. We tried several ideas and discuss them in Section~\ref{04_ancestorStreaming}.
	\\\\	
	\noindent While we use a particular indoor scene to discuss the results during Section~\ref{04_priorityHeuristics}, we also made tests on terrain scenes.
	Due to the different nature of both scene types, we have to take their unique properties into account. We discuss some of these differences
	and their impact on the quality in Section~\ref{04_pageCoherence}.
	
	\section{Evaluation method} \label{04_evaluationMethod}
		To evaluate the performance of several system configurations, we started with traditional metrics, like measuring \emph{cache hits}.
		While useful in the case of operating systems, it became evident, that these can not be used in our
		case, since we get no feedback on the visual quality. 
	
		\subsection{Quality measurement} \label{04_qualityBasedMetric}
			We settled on the idea to use quality assessments in order to see if one configuration leads to visually better results than another.
			The basic idea is to measure the difference between a sequence of reference images that are rendered with the 
			complete set of needed pages available, and one that is the result of using a specific system configuration.
			In order to get comparable results that were not aliased by the unpredictable behaviour of the streaming
			thread, we give every configuration a fixed limit of pages that it can load per frame.
			This is based on the assumption, that a good configuration will select the pages within this contingent
			more cleverly and thus will lead to better results.
			\\\\
			We started with \emph{rooted mean squared error} (RMSE) as a quality indicator, but added \emph{Structural Similarity} (SSIM)~\cite{Wang04} short time later, because
			it takes the human image perception into account and hence could lead to more interesting results.
			Both of these \emph{full reference} methods are further discussed in Appendix~\ref{app_frqa}. 
			
		\subsection{Testcases} \label{04_testCases}
			We tested different configurations of our system in fly-through scenes of indoor levels and terrains.
			This seemed interesting, since we could analyze their performance in more or less practical situations, where
			motion and complex geometry are predominant.\\
			\\	
			In order to understand the behaviour and influence of ancestor streaming strategies, we do not have to take motion into account.
			Due to this fact we chose different spots within a scene and fixed the camera. Then we allowed the system
			to stream one page per frame, so that we could take a close look on the influence of each single page that has
			been loaded.

	\section{Page Priority Heuristics} \label{04_priorityHeuristics}
		As mentioned in Section~\ref{03_needBufferEvaluation} we analyze the needbuffer in order to get the set of needed pages and
		estimate which page would be the best to stream. The complete set is ordered in a priority queue, which sorts
		the pages based on their priority values. In order to calculate this priority for each page, we employ a set of different
		heuristics that base their decision on different informations.
		
		\subsection{Discussed Dataset}
\begin{figure}[h!]
				\centering
				\includegraphics[scale=1.0]{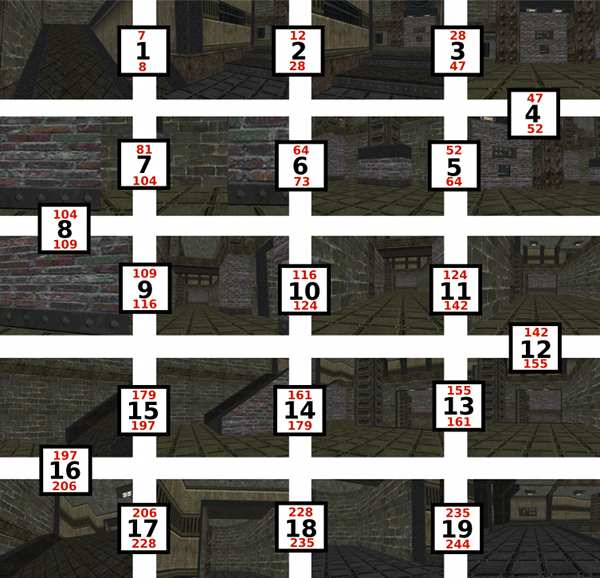}
				\caption{Keyframes of the dataset we want to discuss. The black numbers in the boxes are used to refer to a specific
						 motion, while the red ones identify starting- and ending frame of a motion.}
				\label{fig_04_storyboard}
			\end{figure}

			As an example during this discussion we will use a specific flythrough that we made in an indoor level. 
			Figure~\ref{fig_04_storyboard} shows keyframes of the taken reference images in order to make the discussion more comprehensible.
			Two neighbouring keyframes represent a specific motion that is identified by the black number within the box between
			them.
			The upper red number represents the frame within our flythrough at which the motion starts. The lower red number on the
			other hand marks the resulting keyframe.
			If the lower red number does not coincide with the upper red number of the next box, then the camera has not been moved
			in the mean time.
			\\\\
			We will concentrate in our discussion on two specific sections separately.
			
			\paragraph{Motion 1 up to 6} will be used in the discussion of the LookAhead Camera. The interesting point is motion \textbf{1}, which
			represents a very fast rotation that immediately stops after just one frame. We can interpret this as kind of a worst case, because
			we can not handle it properly at all. \textbf{2 + 3:} After a short stop we slightly rotate around the y-axis and translate in z direction. 
			\textbf{3 + 4 + 5:} We rotate slowly around the y-axis. \textbf{6:} We move to the wall that is now in front of the camera.
			
			\paragraph{Motion 7 up to 15} will be used to discuss the performance of several heuristics. It contains camera motions that are likely 
			to appear in an real-world application. \textbf{7:} We translate in the negative x direction along a wall that is in front of the camera.
			\textbf{8 + 9 + 10:} We rotate quite fast around the y-axis and very slowly around the x-axis in order to see the
			corridor. \textbf{11 + 12:} We move in z-direction through the corridor. 
			\textbf{13 + 14:} We rotate slowly around the y-axis. \textbf{15:}We move to the ramp in front of the camera.
			\\\\
			Every configuration we tested had the same set of pages at the start within its cache: the first three mip levels and all the pages
			that are visible from the starting position of the flythrough (see first keyframe).
			
		\subsection{Basic Heuristics} \label{04_BasicHeuristics}		
			We started out with a set of three basic heuristics. 
			\paragraph{Random}
				We assign each page a priority that is generated uniformly at random. Strictly speaking Random is no heuristic at all. 
				It simply represents the case that no heuristic is used.
			\paragraph{PixelSum}
				During the analysis of the needbuffer we count the pixels on screen that need the same page. A large number of pixel
				yield a higher priority.
          	
			\paragraph{Distance} 
				Alongside the needbuffer we read out the depth-buffer and compute the mean distance for a visible page. 
				A high priority is represented by a small distance.\\
				\begin{figure}[h!]
					\hspace*{-1.5cm}
					\subfigure[5 Pages per frame]
					{
						\includegraphics[scale=0.23]{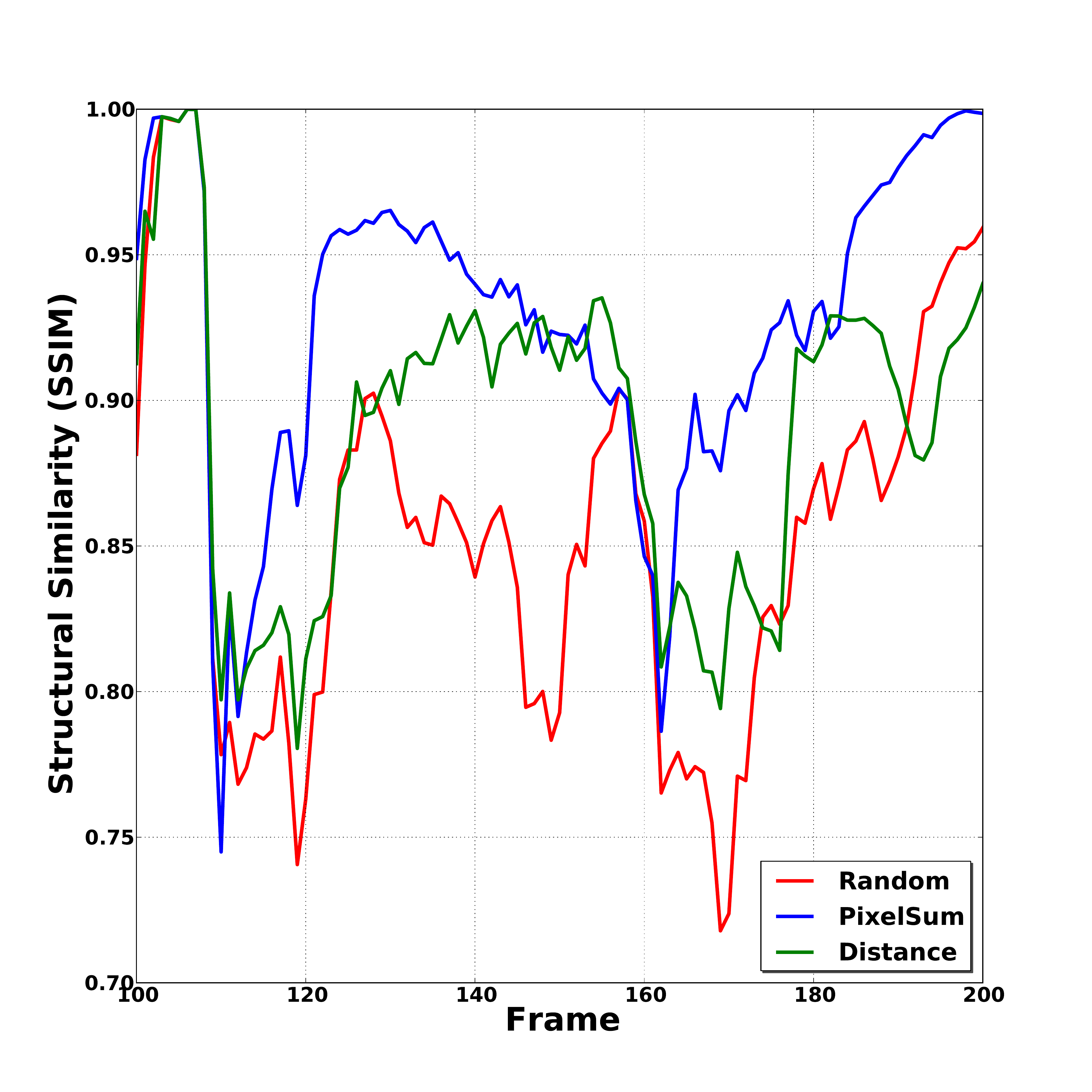}
					}
					\hspace*{-1.0cm}
					\subfigure[10 Pages per frame]
					{
						\includegraphics[scale=0.23]{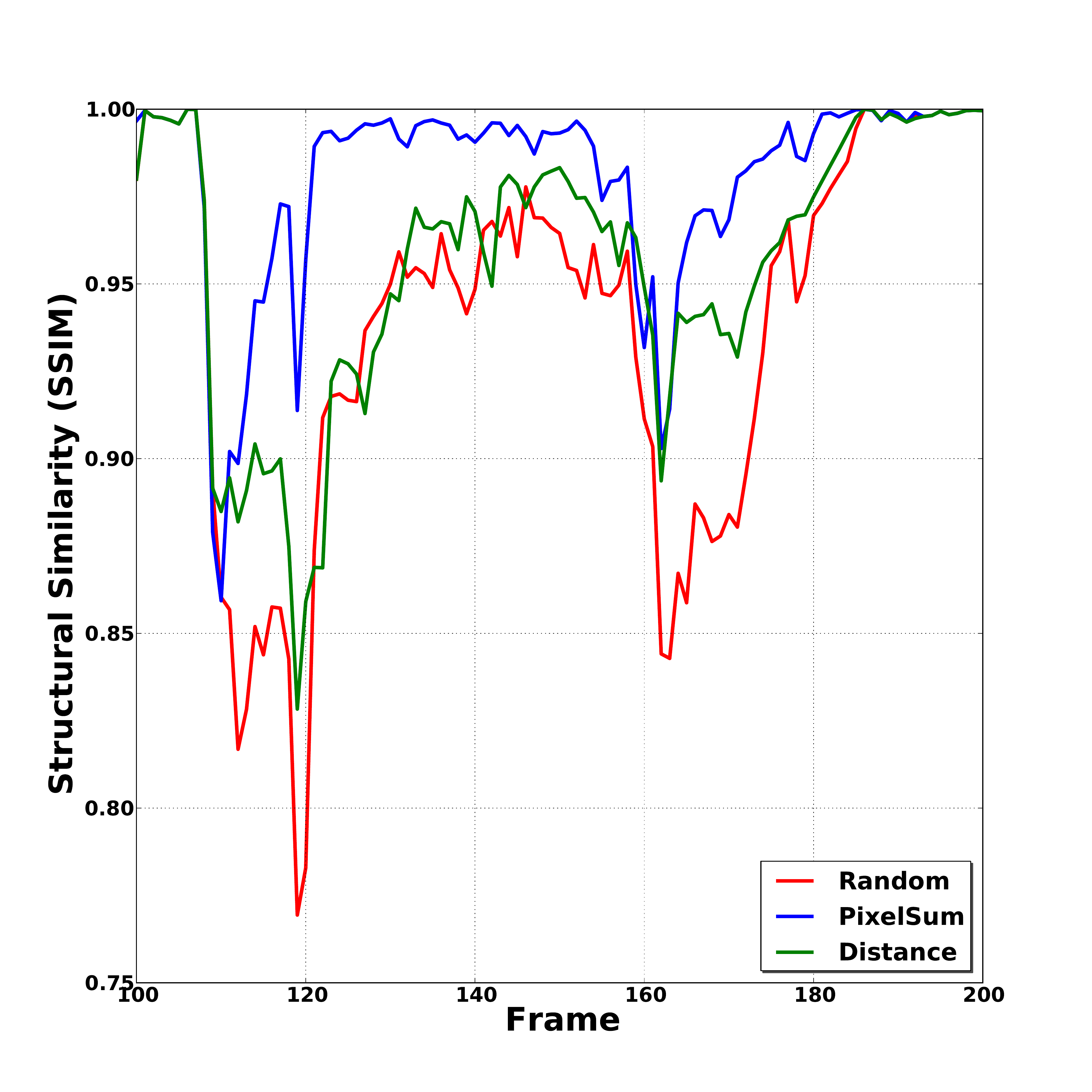}
					}

 					\caption{Comparison of the basic Heuristics within motion set 7 up to 15. 
 					Streaming pages in a specific order yields better results than selecting them randomly. 
 					PixelSum works pretty well.}
  					\label{fig_4_basicHeuristicsComp}
				\end{figure}
				
				\begin{figure}[h!]
					\centering
						\includegraphics[scale=0.23]{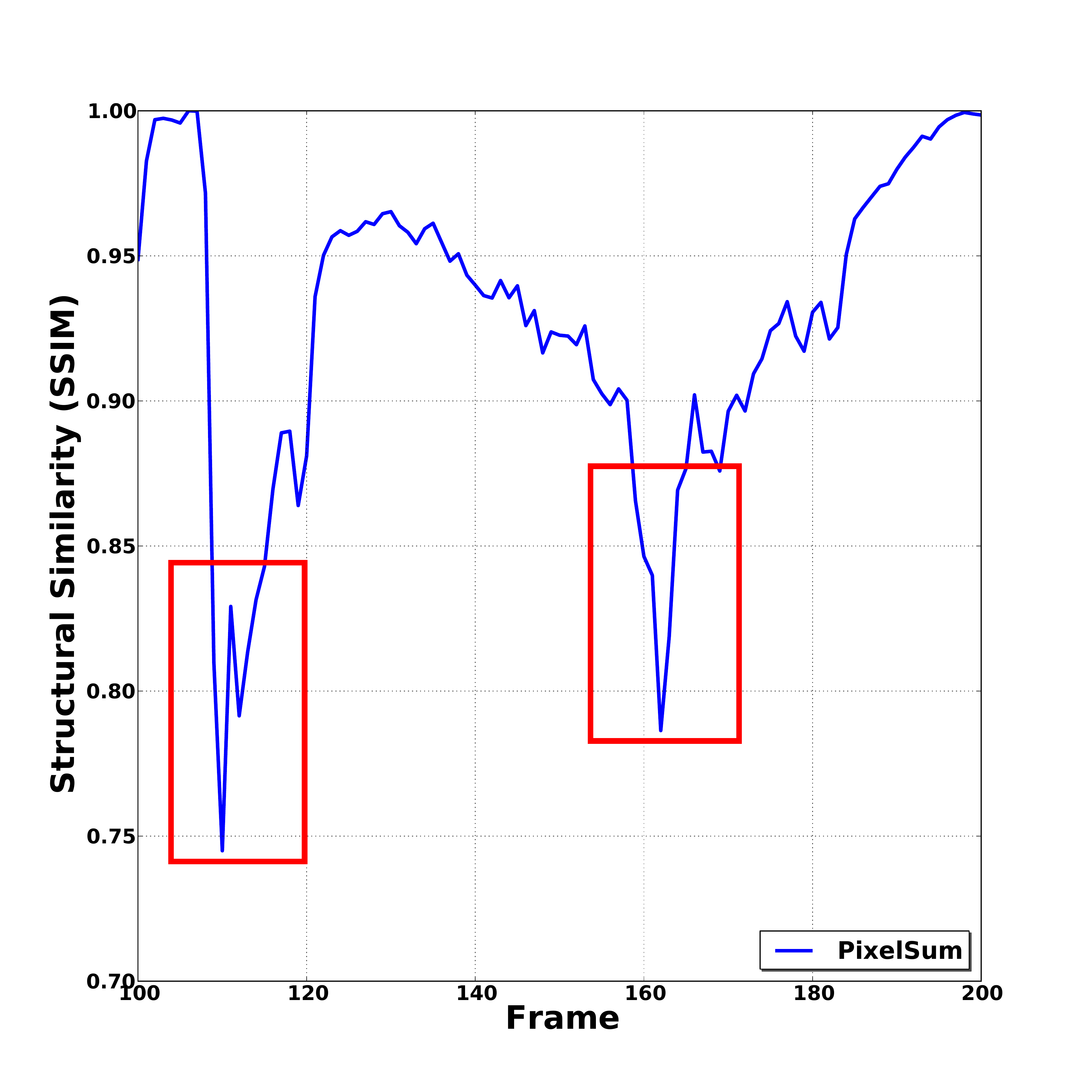}
					
 					\caption{Results of PixelSum. The red boxes mark two large peaks that represent huge quality losses.
 							These performance hits are caused by rotations.}
  					\label{fig_4_pixelSumRotationHighlighted}
				\end{figure}		
			\noindent Figure~\ref{fig_4_basicHeuristicsComp} shows a comparison of the three heuristics within motion set 7 up to 15. 
			Due to the fact that Random behaves badly compared to PixelSum and Distance we can conclude that streaming the pages in a
			non randomly order actually leads to better results and hence has an impact the visual quality. 
			Furthermore we see that PixelSum yields the best results in a setup where only a small number of pages can be loaded and that
			this trend holds on if we allow every configuration to stream twice as much.
			Figure~\ref{fig_4_pixelSumRotationHighlighted} shows again the results of PixelSum, where we used two red boxes to
			highlight large peaks. These peaks represent huge quality losses that are caused by rotations like the motions 9 and 14. 
			This means that it is highly relevant to diminish the effect of these rotations in order to improve the quality.

		\subsection{Advanced Heuristics}
			Since PixelSum yielded the best results we used it as a baseline for more advanced heuristics.
			\subsubsection{Weighted PixelSum}		
\begin{figure}[h!]
					\centering
					\includegraphics[scale=0.2]{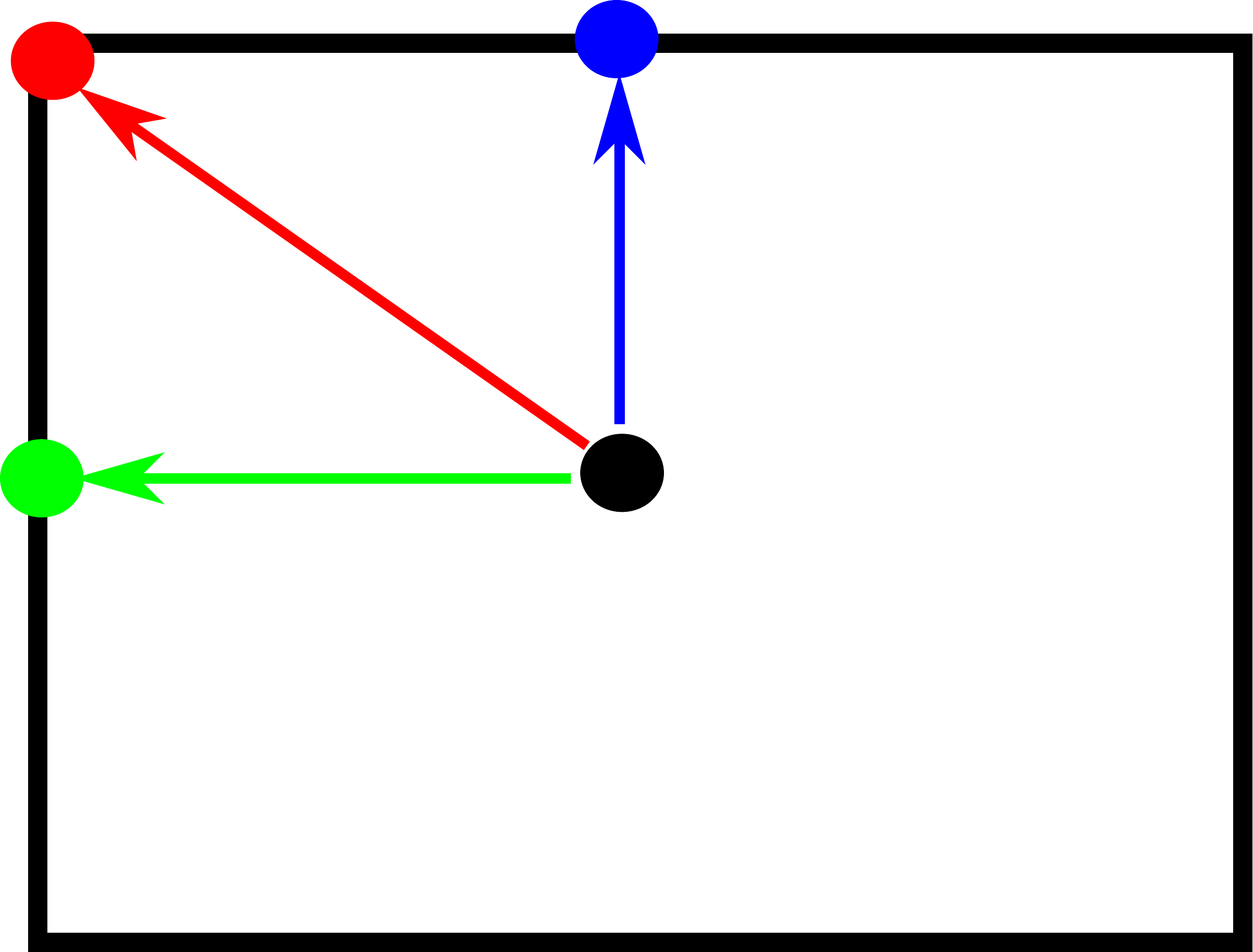}
 					\caption{We shift the weighting center based on the rotation that occurred. A rotation around the x-axis (blue) and a rotation
 					around the y-axis (green) will cause the center to be shifted to its final position (red).}
  					\label{fig_4_hotSpotIdea}
				\end{figure}

\begin{figure}[h!]
					\hspace*{-1.5cm}
					\subfigure[5 Pages per frame]
					{
						\includegraphics[scale=0.23]{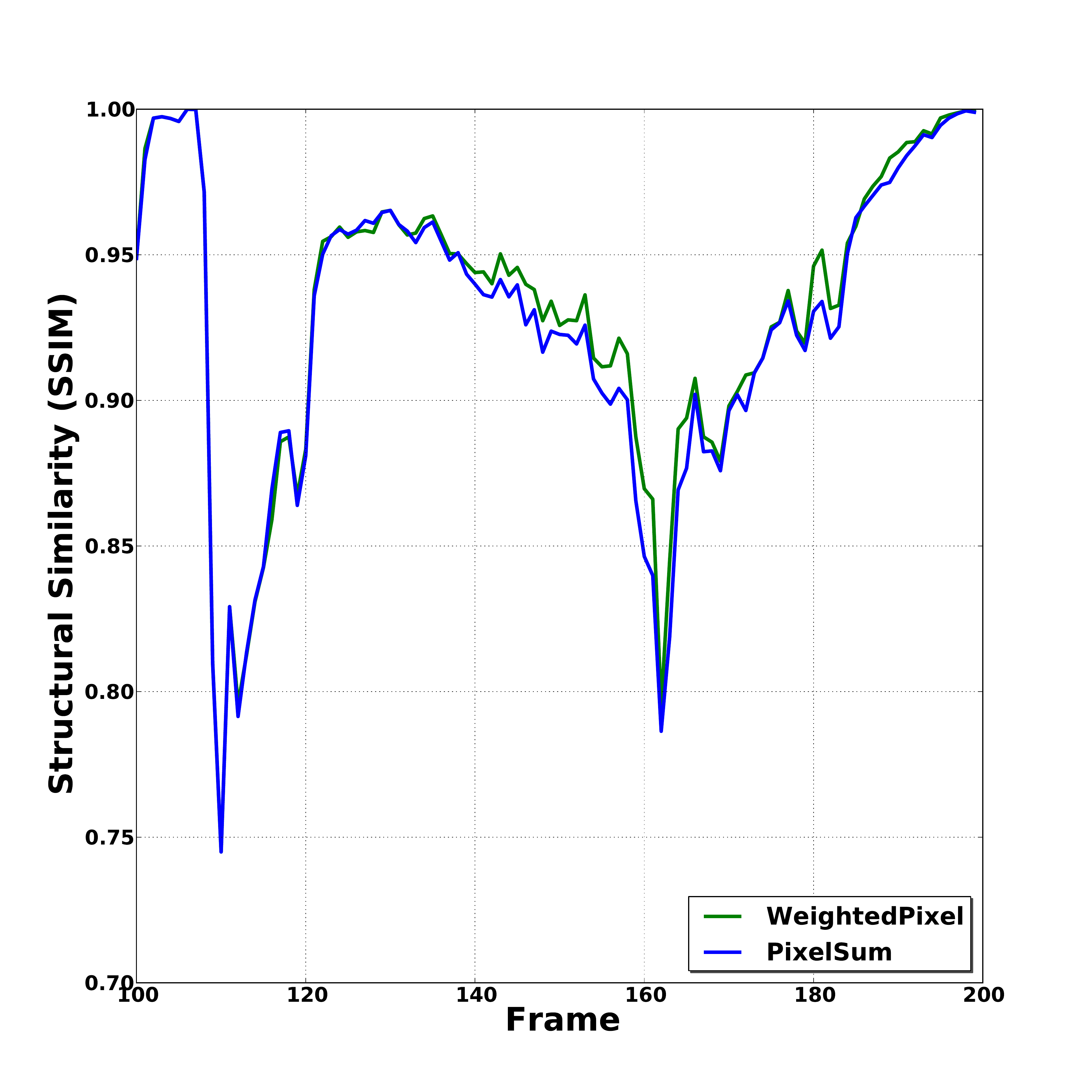}
					}	
					\hspace*{-1.0cm}
					\subfigure[5 Pages per frame]
					{
						\includegraphics[scale=0.23]{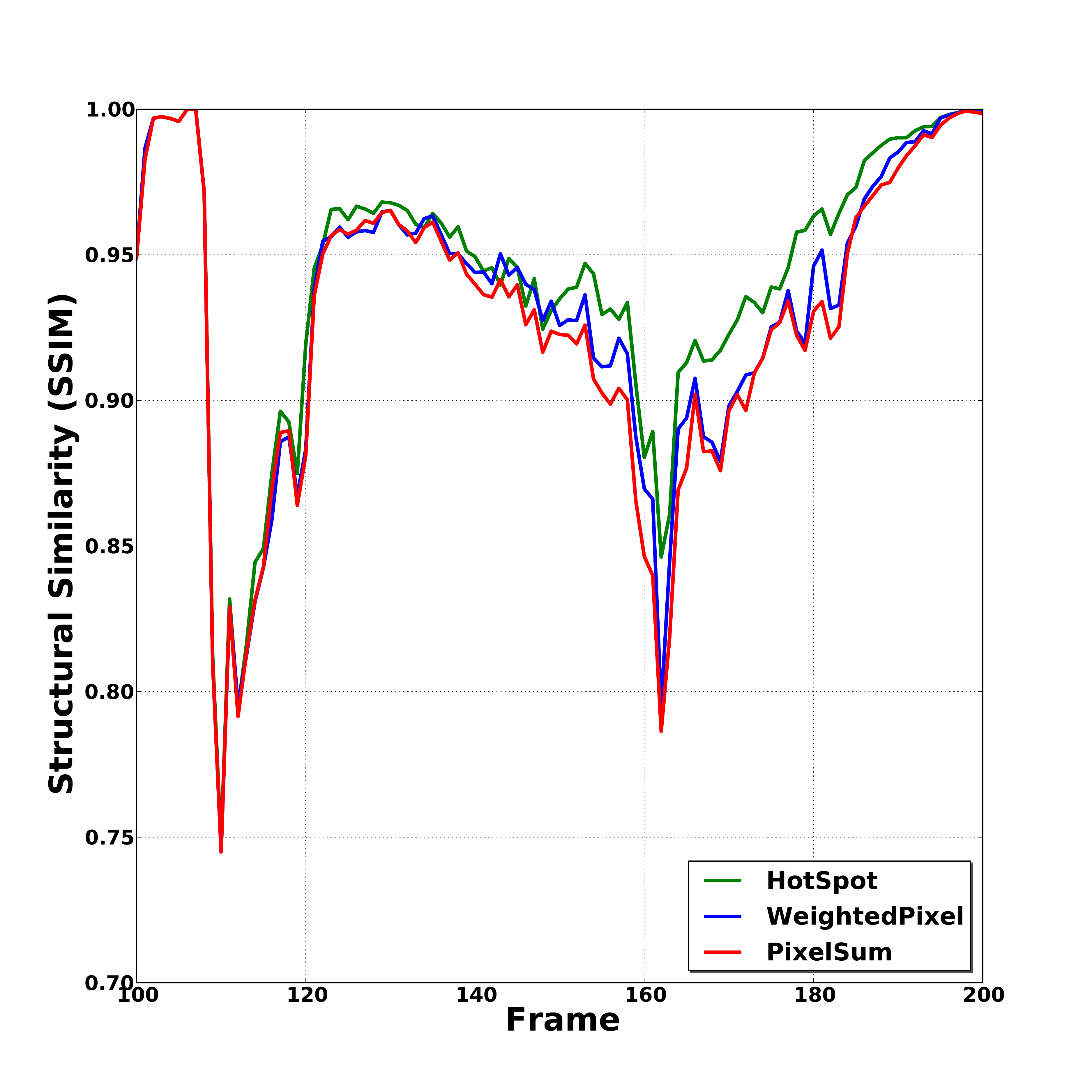}
					}
 				\caption{(a) WeightedPixel does not perform as good as we hoped for, but does a better job in the case of the translation through the corridor.
 				         (b) HotSpot exploits the strengths of WeightedPixel and does a better job in the case of slow rotations.}
  				\label{fig_4_advancedHeuristicsCenterWeight}
				\end{figure}

				Building on PixelSum we had the idea to use a center within the screen space to weight the number of pixels. 
				From this we derived the following two heuristics:

				\paragraph{WeightedPixel}
					This heuristic stems from the assumption, that the content at the center of the screen is more important
					than what is seen at the peripherals. The heuristic works quite similar to PixelSum, but each pixel that is
					accumulated gets weighted by its radial distance to the screen center.	
				\paragraph{HotSpot}
					The idea is to use the strength of WeightedPixel, but to get better results in the case of rotations.
					Rotations will cause pages to appear at specific sides of the screen, while other will leave it
					at the opposed ones. Due to this fact we take the rotation into account and shift the center based on the rotation
					to the borders of the screen. See Figure~\ref{fig_4_hotSpotIdea} for a schematic example of this idea.
				\\\\				
			\noindent Figure~\ref{fig_4_advancedHeuristicsCenterWeight} (a) shows the performance of WeightedPixel in
			comparison to PixelSum. Unfortunately WeightedPixel has not such a great impact on the quality as we hoped for, but it actually does a slightly 
			better job in the case where the camera moves down the corridor (motion 11 \& 12). 
			This is due to the nature of the heuristic itself, since it gives pages at the screen center a higher priority. 
			In such a motion through a corridor the pages at a high distance reside in the near of the screen center, while those on the wall on both
			sides of the camera span larger areas.
			WeightedPixel prefers those that reside at the center and so does exactly the right thing, because these pages will stay visible for a longer
			period of time.
				\\\\
			\noindent Figure~\ref{fig_4_advancedHeuristicsCenterWeight} (b) on the other hand compares Hotspot with both WeightedPixel and PixelSum.
			As we assumed: HotSpot uses the strength of WeightedPixel in the case of the corridor, but does also a better job in the case
			of slow rotations (see motion 13 up to 16). Unfortunately it suffers in the case of fast rotations from the same problem as all the
			other heuristics.\\
			\noindent We also made tests with a sliding HotSpot in which the amount of the shift depends on how fast the camera rotates. Although it seemed
			to be a more sophisticated idea, it did not lead to better results compared to WeightedPixel.
			
\begin{figure}[h!]
					\centering
						\includegraphics[scale=0.23]{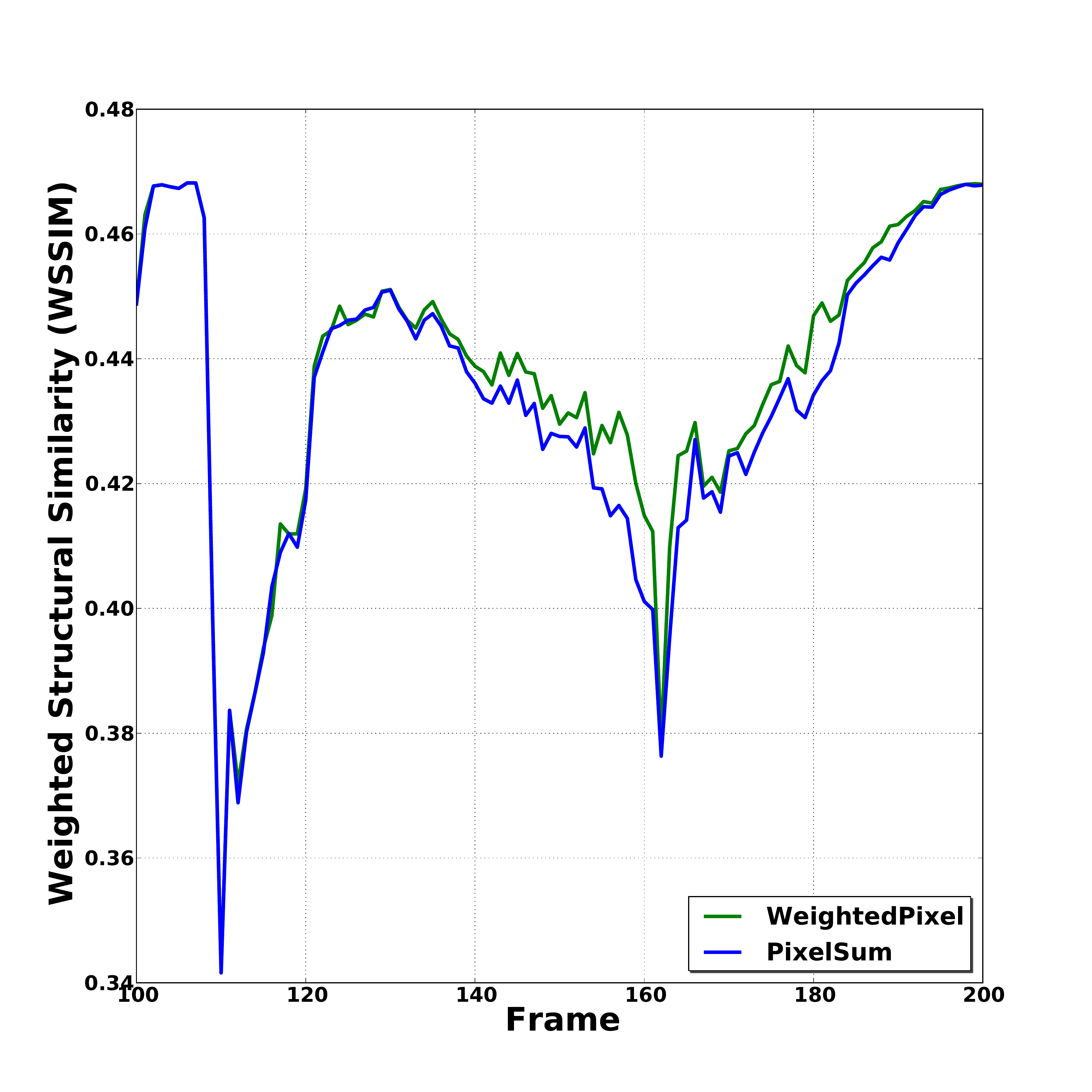}
 				\caption{Using WSSIM in order to take the importance of the screen center into account. The performance of PixelSum and WeightedPixel become
 					a bit more distinguishable.}
  				\label{fig_4_wssim}
				\end{figure}

			\paragraph{Weighted Structural Similarity} As we introduced WeightedPixel, we mentioned our assumption that the content at the screen center is
			more important than whats visible at the peripherals. 
			If the viewer looks most of the time at the center of the screen, then this assumption should be true.
			Unfortunately both SSIM and RMSE do not take this screen weighting directly into account. 
			Thus we extended structural similarity to weight the results of each compared window based on the distance of the windows midpoint to the screen center. 
			Figure~\ref{fig_4_wssim} shows again a comparison of PixelSum and WeightedPixel, but this time based on the results of our extension WSSIM.
			Although the scale of the results changes, we get curves that are quite similar to those that are yielded by SSIM.
			WeightedPixel again does not lead to overwhelmingly better results, but both heuristics become more distinguishable.
			In contrast to Figure~\ref{fig_4_advancedHeuristicsCenterWeight} (a) PixelSum performs with the exception of the fast rotation in motion 9
			overall better. 
			We mentioned this point for the reason of completeness. 
			Although we employed SSIM during our study and use it throughout this discussion, it could be very valuable to use a quality assessment that directly 
			takes the screen weighting into account.
				\\\\
				\noindent A common problem of the heuristics discussed so far remains: It can happen that they select pages that provide little or no 
				improvement to those that are currently available. The next section features a technique that tries to overcome this problem by taking the
				content of the pages into account.

			\subsubsection{NoiseValue} \label{04_noiseValue}				
				A perfectly white wall will stay white, no matter how close the viewer gets. 
				Even small amounts of noise will hardly be detectable and there will be little change between mip levels, whereas a very grainy rock 
				texture will exhibit a lot of change on every mip level.
				In order to take this fact into account, we had the idea, as previously stated in Section~\ref{03_noiseValueCalculation}, to provide
				the system with an indicator on how much the texture will change if it uses the child page instead of its parent during rendering.
				\begin{figure}[h!]
					\centering
					\includegraphics[scale=1.0]{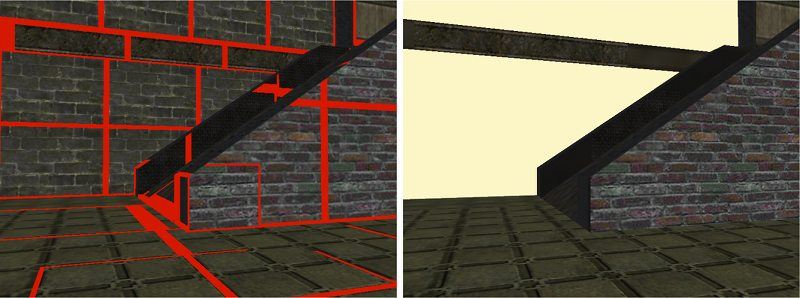}
 					\caption{We chose a source texture that is used by many pages and that spans large areas on different mip levels. We exchanged
 							 this texture so that the NoiseValues of the considered pages are zero.}
  					\label{fig_4_foollevel}
				\end{figure}\\\\		
				\noindent In order to test this idea we extended the aforementioned heuristics to consider this so called \emph{NoiseValue}:
				We accumulate all the NoiseValues between the considered page and the currently available ancestor. This sum is then used to
				scale the priority.\\
				\\
				We constructed a level that is based on the example given above and in which the NoiseValue should be relevant:
				We selected one often used source texture that spans large areas within the so far discussed level and exchanged it
				with another that does not contain any noise. See Figure~\ref{fig_4_foollevel} for screenshots of this idea.
				\begin{figure}[h!]
					\hspace*{-1.5cm}
					\subfigure[5 Pages prepared Level]
					{
						\includegraphics[scale=0.23]{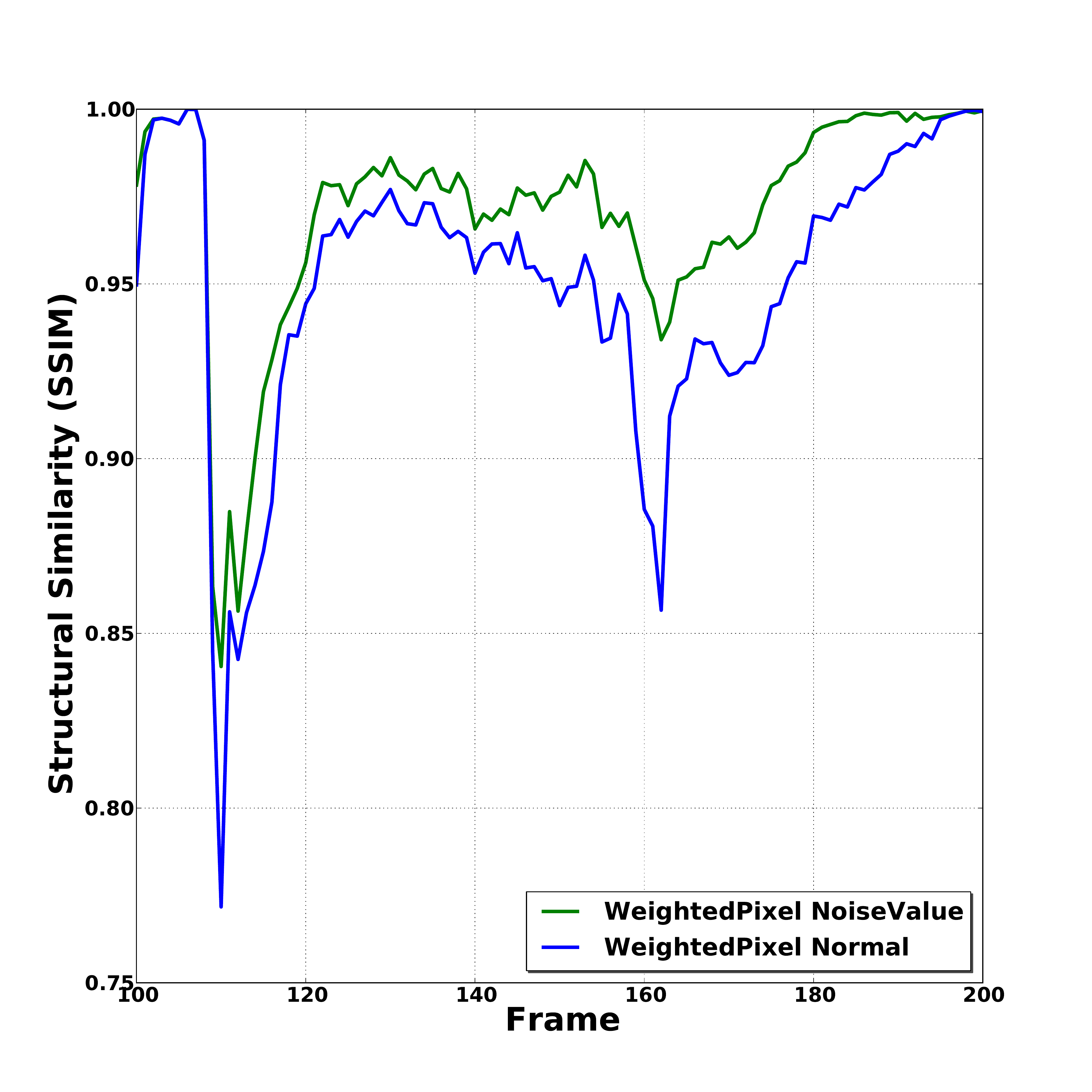}
					}
					\hspace*{-1.0cm}
					\subfigure[5 Pages prepared Level]
					{
						\includegraphics[scale=0.23]{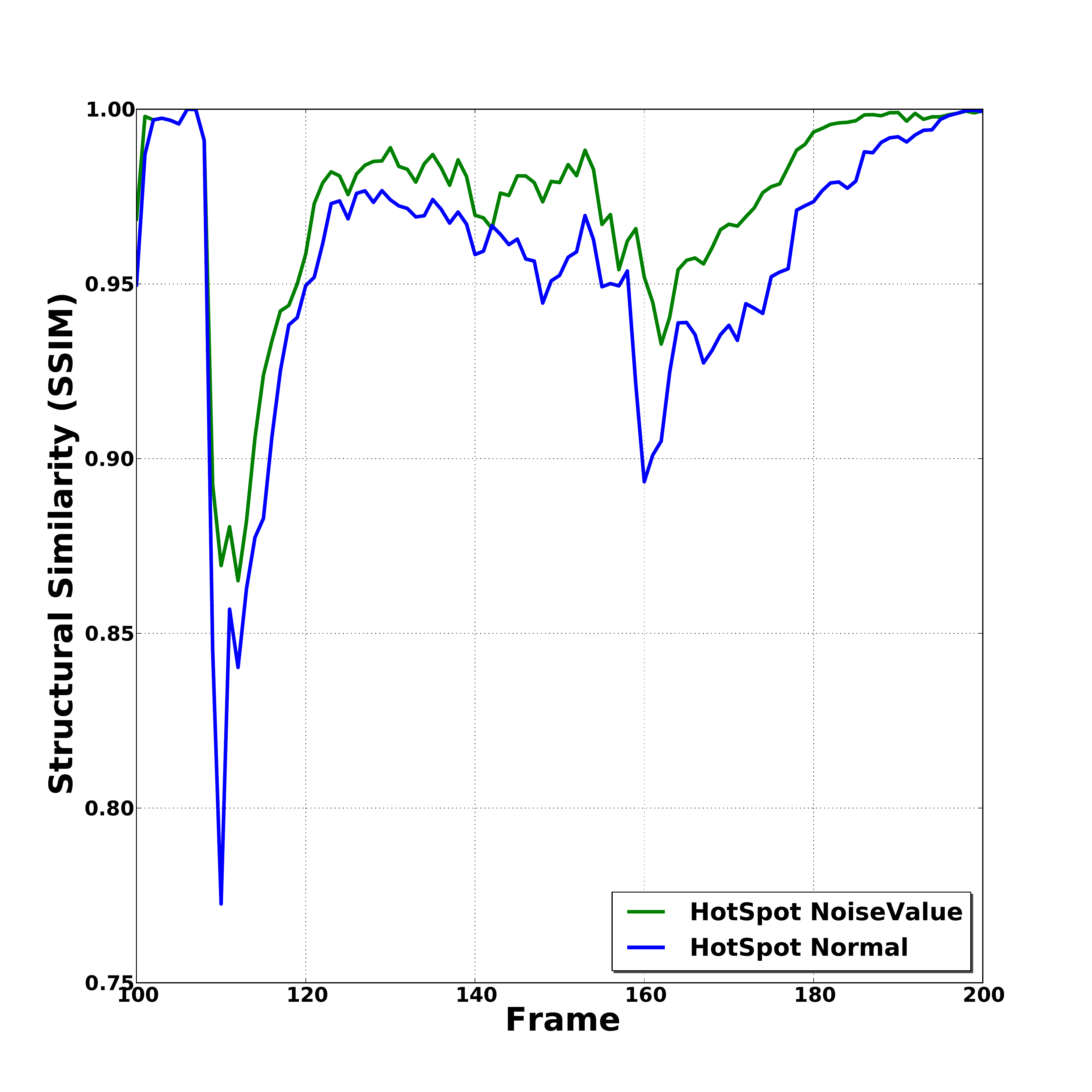}
					}
 				\caption{Considering the NoiseValue in the prepared level yields huge improvements.}
  				\label{fig_4_noiseValuePlots}
				\end{figure}\\\\				
				\noindent Figure~\ref{fig_4_noiseValuePlots} (a) and (b) show the results of WeightedPixel and HotSpot within this prepared level. 
				As we assumed: The overall quality increases, if we scale the computed priority by the accumulated NoiseValue.
				Furthermore we get much smaller peaks in the case of rotations, since it concentrates on the pages that actually result in a 
				quality improvement.
				\begin{figure}[h!]
					\hspace*{-1.5cm}
					\subfigure[5 Pages normal Level]
					{
						\includegraphics[scale=0.23]{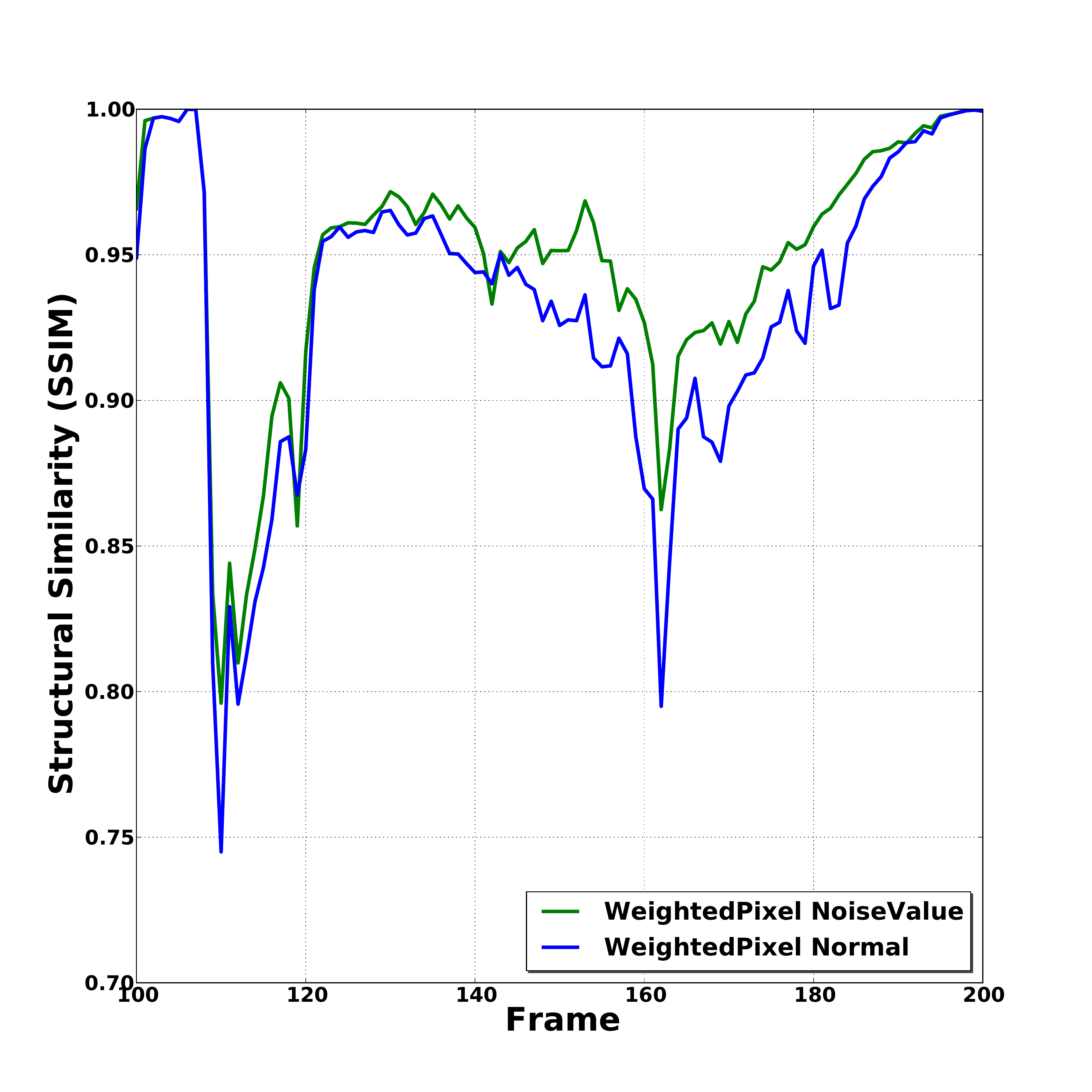}
					}
					\hspace*{-1.0cm}
					\subfigure[5 Pages normal Level]
					{
						\includegraphics[scale=0.23]{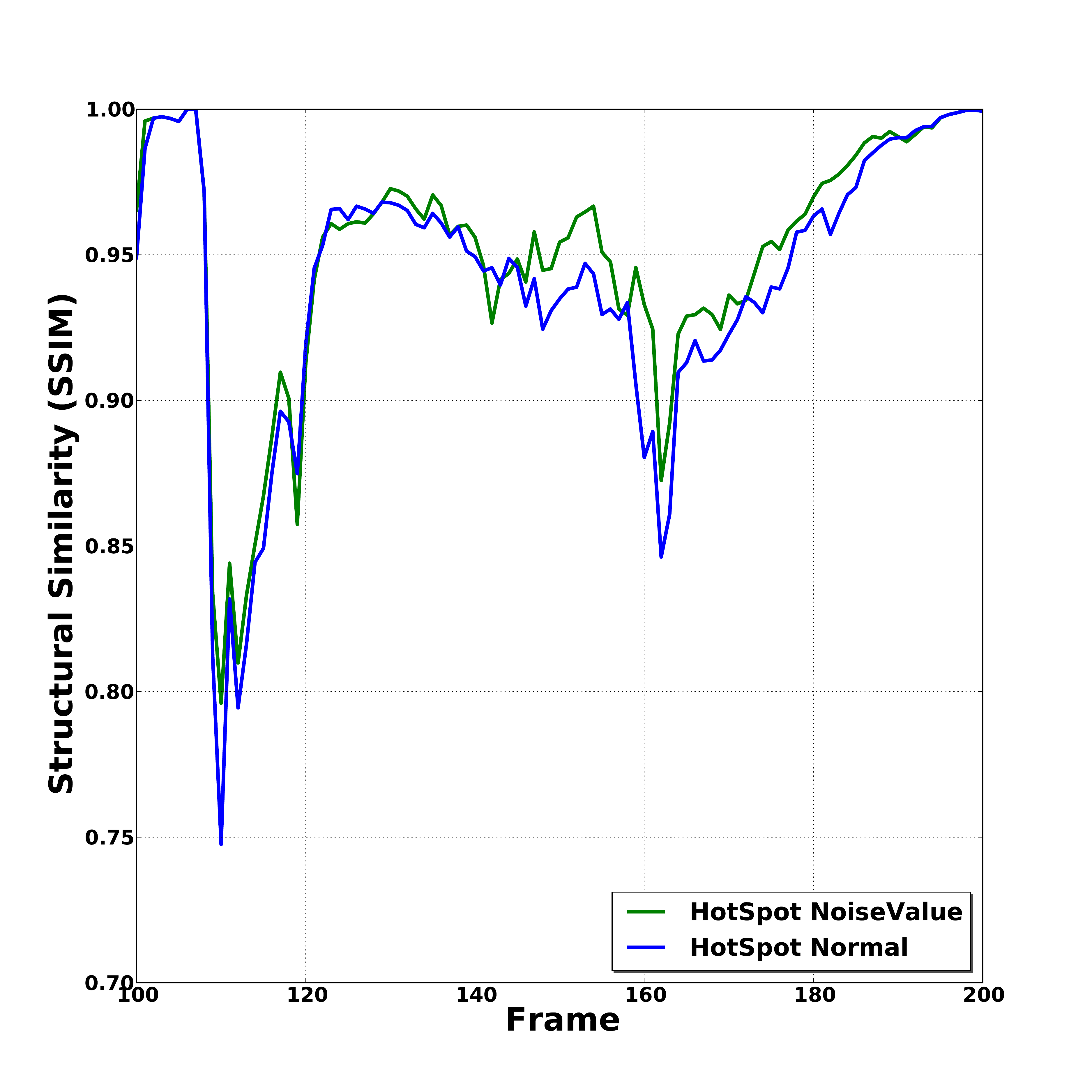}
					}
					\caption{Although not as strong as in the prepared level, we also get smaller Peaks in the case of rotations and better results overall.}
  				\label{fig_4_noiseValuePlotsNormalLevel}
				\end{figure}\\\\
				\noindent We also tested this idea in the original level, in order to see if the quality also increases. 
				The results are shown in Figure~\ref{fig_4_noiseValuePlotsNormalLevel}.
				Although the performance does not increase as strong as in the case of the prepared level, the trend is the same: 
				better results overall and smaller peaks in the case of rotations.
		
		\subsection{Page Prediction} \label{04_pagePrediction}
			The heuristics discussed so far suffer in their basic form from the fundamental problem that 
			they have to make their decision solely by using the information that is given within the current
			frame. This essentially means that they can not predict the future need of pages.
			HotSpot is the only exception, since it indicates what pages are more likely to stay
			visible within the next frame. But even HotSpot can not say which pages will become visible.  
		
			\subsubsection{LookAhead Camera} \label{04_lookAheadCam}
\begin{figure}[h!]
  				\centering
					\subfigure[Rotation around y-Axes]
					{
						\includegraphics[scale=1.0]{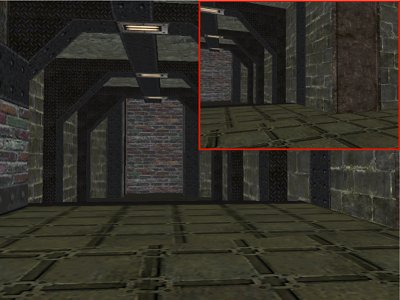}
					}
					\subfigure[Translation]
					{
						\includegraphics[scale=1.0]{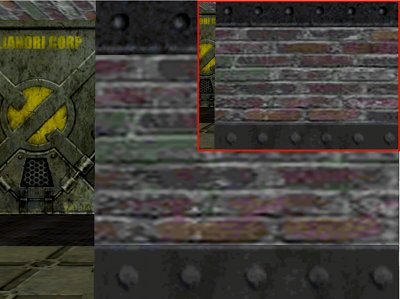}
					}
 				\caption{Visualization of the LookAhead Camera in our implementation. The red marked rectangle shows a scaled version
 				of what can be seen from the viewpoint of the LookAhead Camera in the current frame. 
 				If the user translates or rotates the camera, the LookAhead Camera will show the possible result of the next frame.}
  				\label{fig_4_lookAheadCamVisu}
				\end{figure}

\begin{figure}[h!]
					\hspace*{-1.5cm}
					\subfigure[5 Pages per frame]
					{
						\includegraphics[scale=0.23]{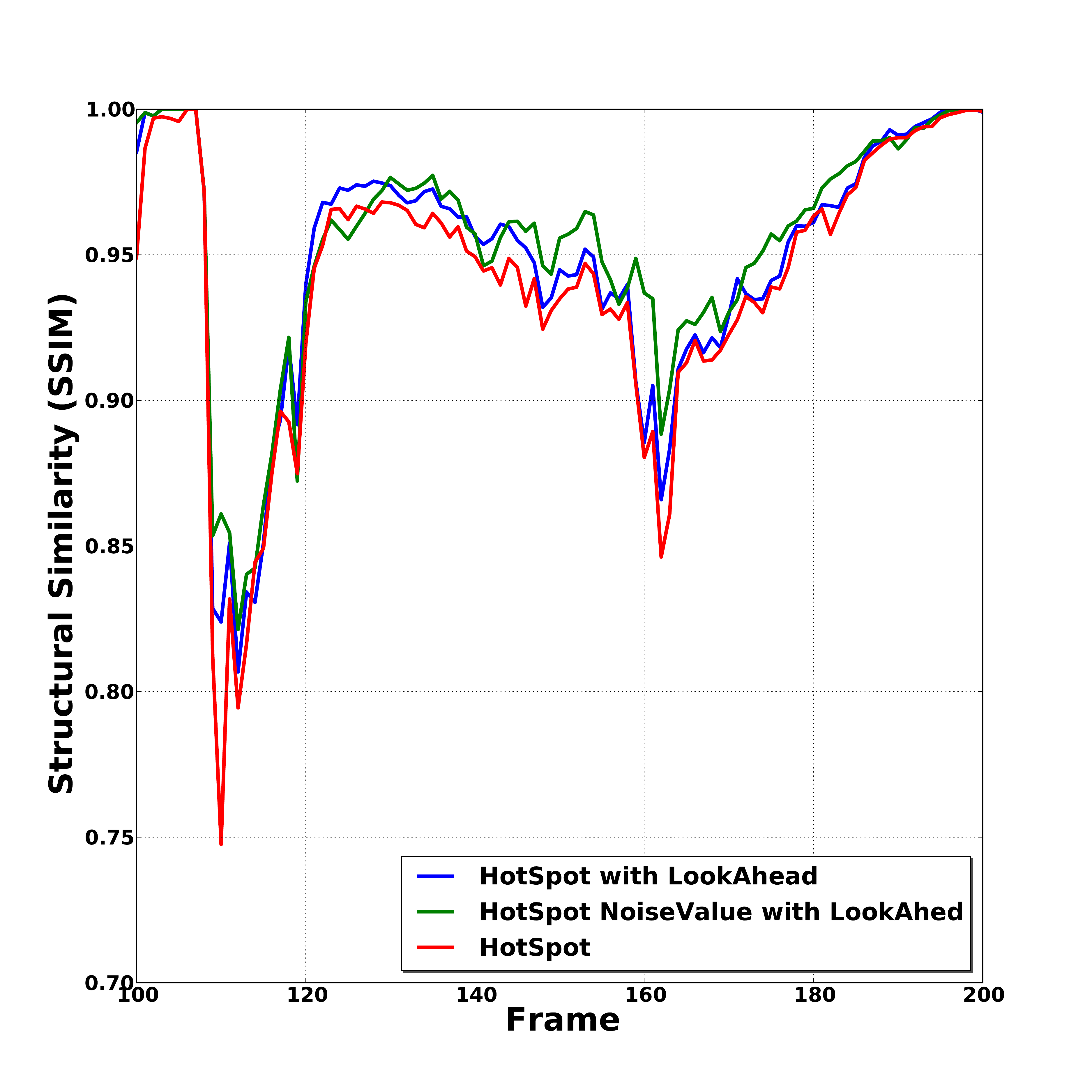}
					}
					\hspace*{-1.0cm}
					\subfigure[10 Pages per frame]
					{
						\includegraphics[scale=0.23]{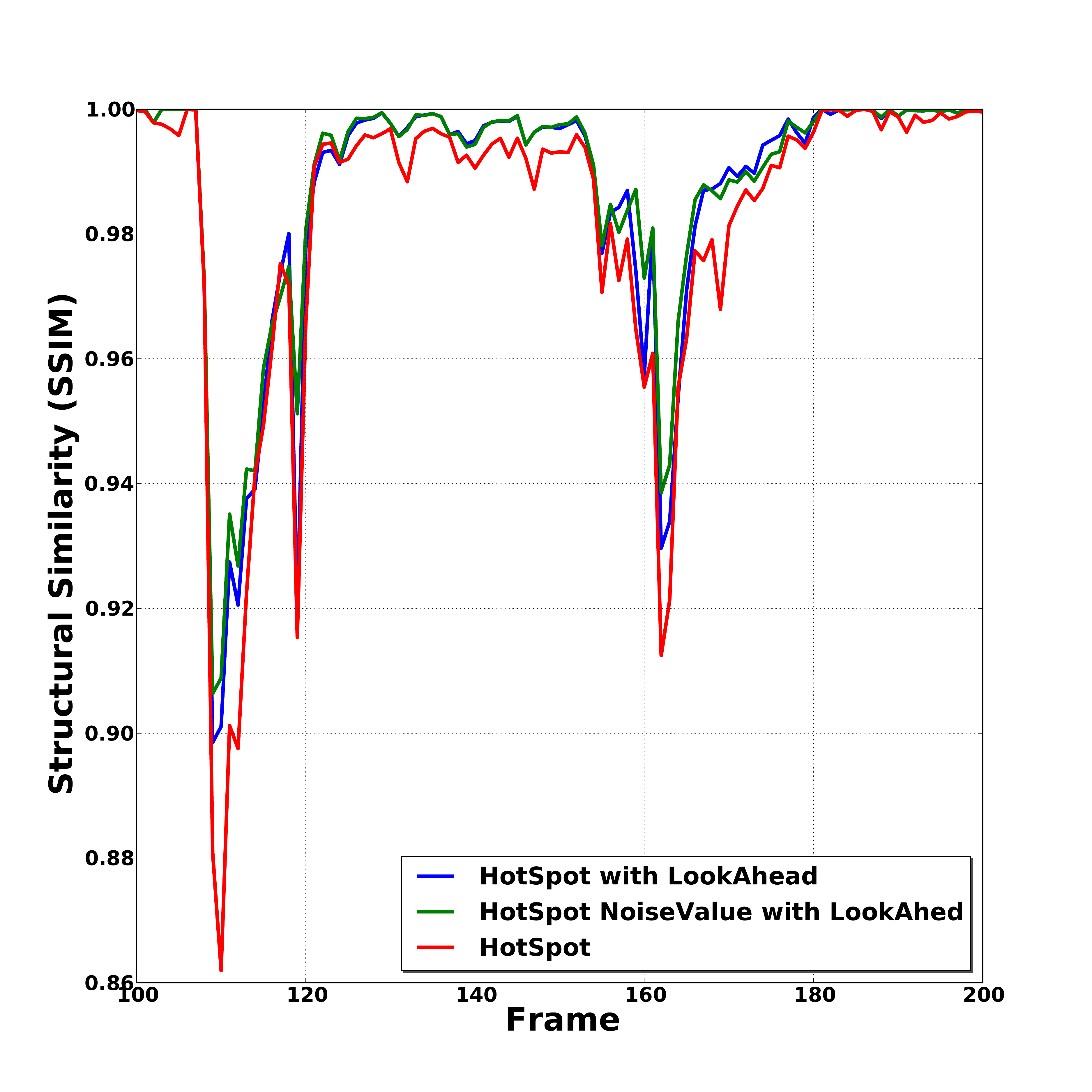}
					}
 				\caption{The LookAhead Camera leads to a overall better quality and smaller peaks in the case of rotations.}
  				\label{fig_4_lookAhead100_200}
				\end{figure}

\begin{figure}[h!]
					\hspace*{-1.5cm}
					\subfigure[5 pages per frame]                
					{
						\includegraphics[scale=0.23]{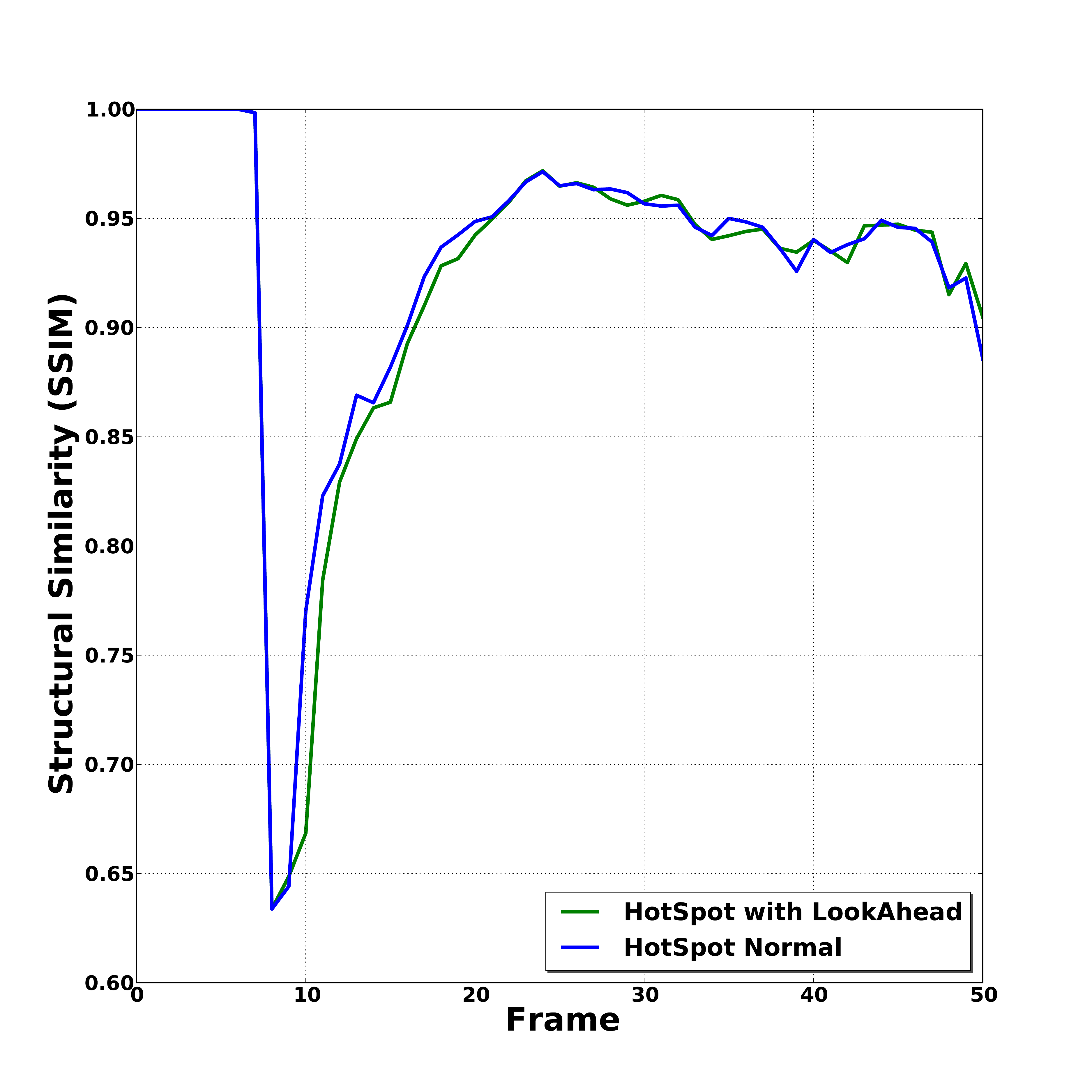}
					}
					\hspace*{-1.0cm}
					\subfigure[5 pages per frame]
					{
						\includegraphics[scale=0.23]{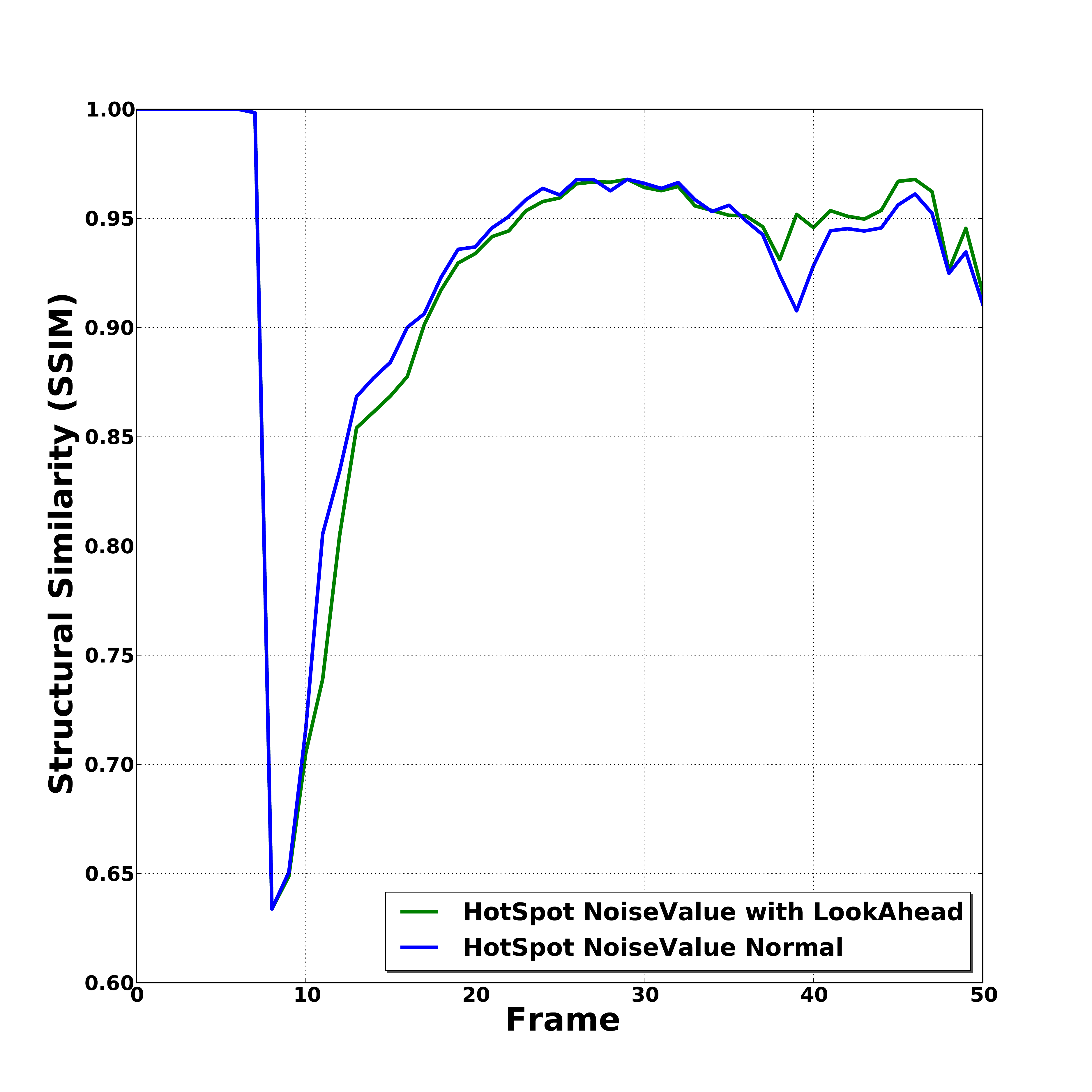}
					}
 				\caption{Problem of the LookAhead Camera. A fast rotation, that immediatly stops, fools the LookAhead Camera and hence a frame that
 				could have been used for streaming the correct pages is lost.}
  				\label{fig_4_lookAheadProblem}
				\end{figure}

				One method to predict the future need of pages is the idea that we call a \emph{LookAhead Camera}.
				It is based on the assumption that if we translated or rotated the camera in the last frame to a certain
				amount, we will probably do so again in the course of the next one.\\
				\\
				We track the motion that occurred during the current frame and build a second view matrix that simply 
				represents the transformation as if the motion occurred twice. We then do a second render pass and use
				the resulting needbuffer during the analysis.			
				A visualization we used for debugging purpose and that gives an example of the idea can be seen in 
				Figure~\ref{fig_4_lookAheadCamVisu}.				
				\\\\	
				\noindent Figure~\ref{fig_4_lookAhead100_200} (a) and (b) show a comparison of different HotSpot configurations.
				We see clearly that the overall performance gets better and that we can successfully diminish the effect of fast
				rotations, which was the main intent of using the LookAhead Camera.	
				\\\\	
				\noindent But using the LookAhead Camera is not without its problems. Figure~\ref{fig_4_lookAheadProblem}
				shows a peak that is the result of a very fast rotation, which immediately stops within the next frame.
				The LookAhead system is fooled by assuming that the next frame will feature a rotation of similar amount.
				So an immediate rotation of $45.0$ degrees for example will yield a $90.0$ degree rotation for the LookAhead Camera.
				This one frame that could have been used for streaming highly needed pages is actually lost.
				\begin{figure}[h!]
					\hspace*{-1.5cm}
					\subfigure[10 pages per frame]                
					{
						\includegraphics[scale=0.23]{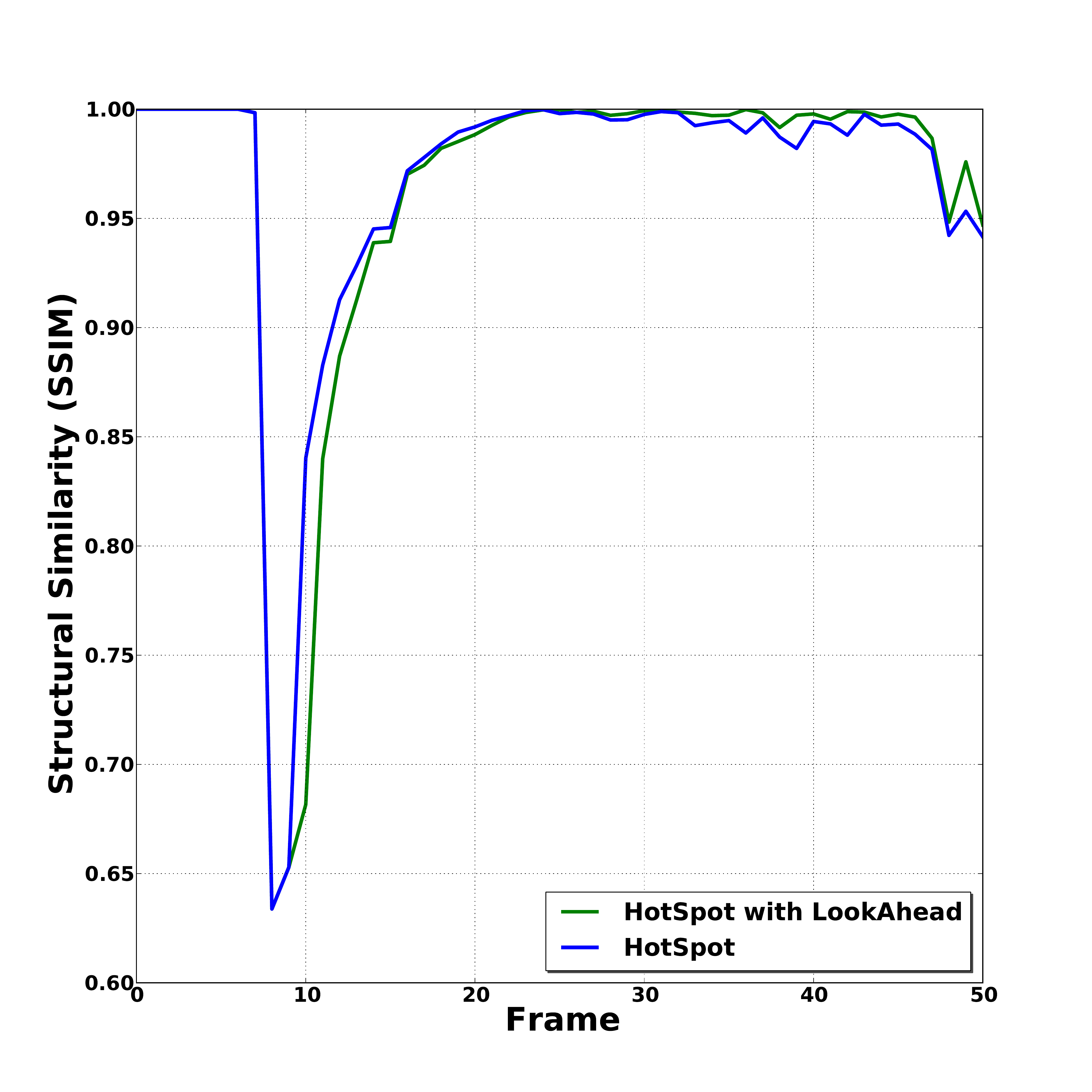}
					}
					\hspace*{-1.0cm}
					\subfigure[10 pages per frame]
					{
						\includegraphics[scale=0.23]{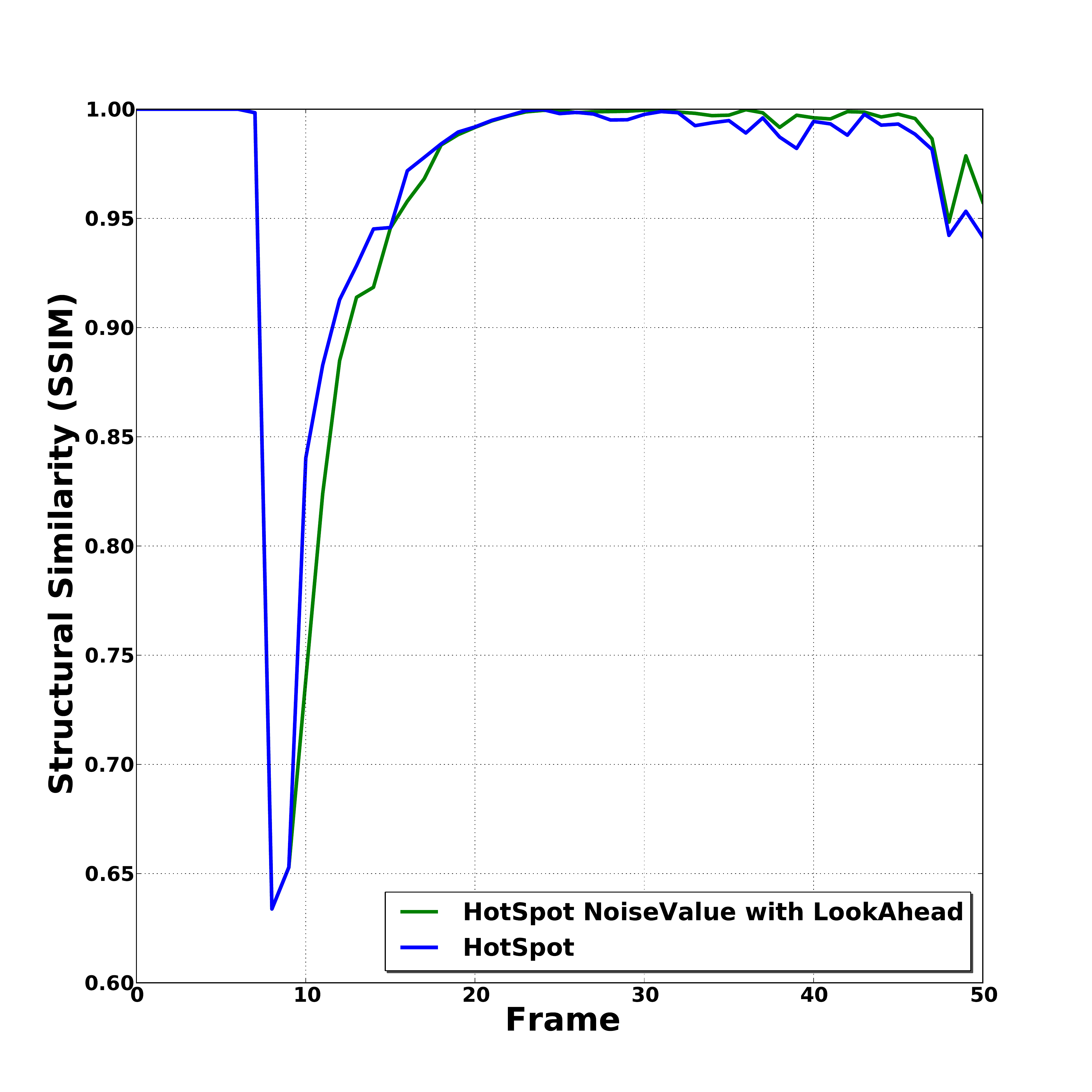}
					}
				\caption{The problem does also occur if the system is allowed to stream twice as much pages.}
  				\label{fig_4_lookAheadProblem10pages}
				\end{figure}	
				Figure~\ref{fig_4_lookAheadProblem} (a) and (b) show the results of this effect for HotSpot and HotSpot in combination with the NoiseValue.
				Looking at Figure~\ref{fig_4_lookAheadProblem10pages} reveals that this single frame is also lost in the case where the system is allowed
				to stream twice as much pages.\\
				\noindent We can conclude that it would probably be better to adjust the LookAhead Camera to coincide with the viewing 
				matrix of the current frame in the case that an immediate rotation occurs. 
				Using the second derivative of the rotation could be employed as an indicator for this adjustment.			

			\subsubsection{Prediction within the shader}
				At the end of our study we had an idea for a prediction within the fragment shader that could be used during
				a translation on the z-axis in order to indicate if a page is likely to stay visible. The basic idea is
				based on the assumption that if we could estimate the rate of change for the mip level of a page then we could
				store this value in the needbuffer and later incorporate it into a heuristic.
				In contrast to the LookAhead Camera this technique does not require a second render pass. 
				Unfortunately, we did not have enough time to implement it in the course of this thesis.
				Furthermore as we pondered about it, it became clear that fulfilling the task based on the information that is 
				available in a fragment shader becomes non trivial. 
				Due to the fact that it would only work for translations on the z-axis, one would have to weigh up costs and benefits. 
				
		\subsection{Result}
\begin{figure}[h!]
				\hspace*{-1.5cm}
				\subfigure[]
				{
					\includegraphics[scale=0.23]{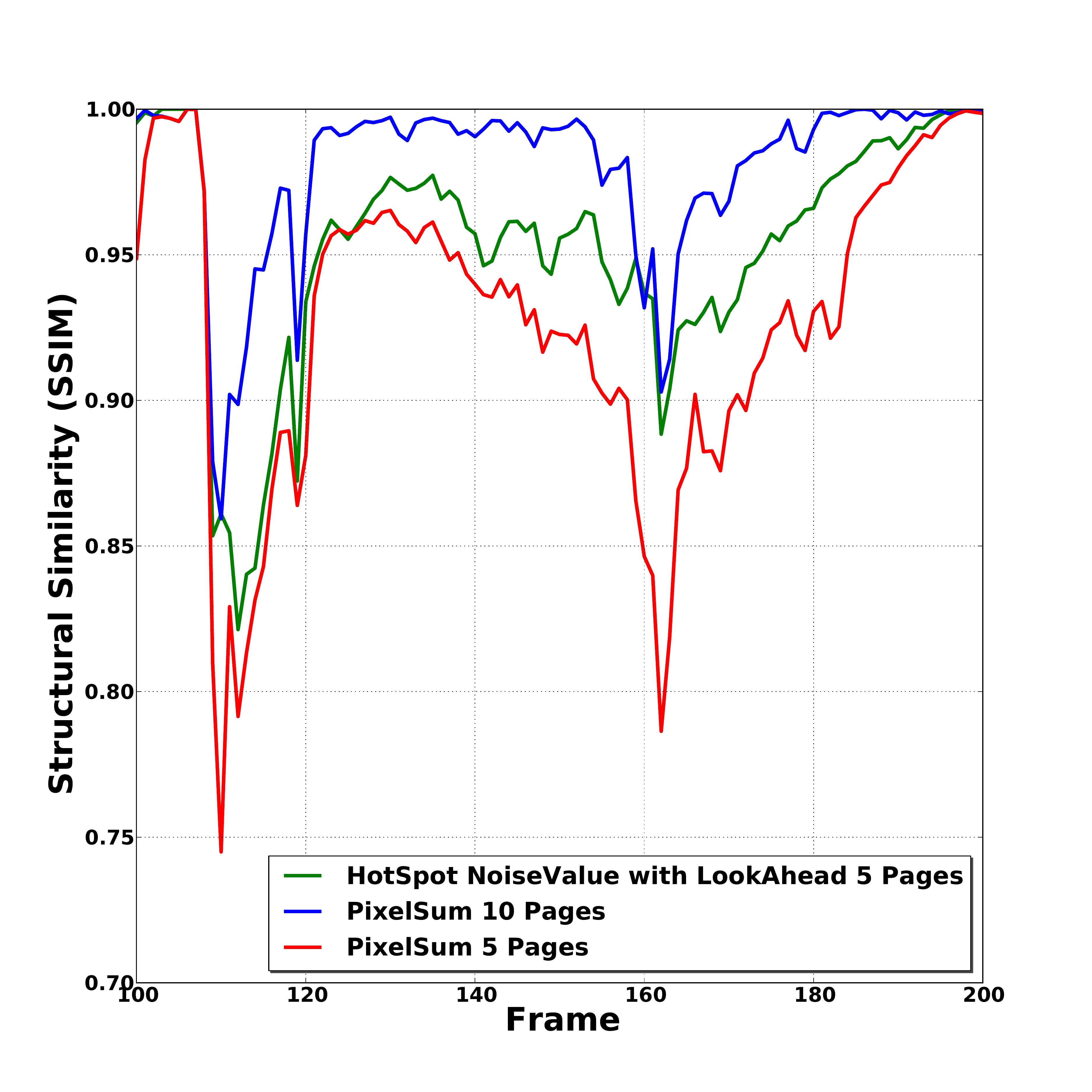}
				}
				\hspace*{-1.0cm}
				\subfigure[]
				{
					\includegraphics[scale=0.23]{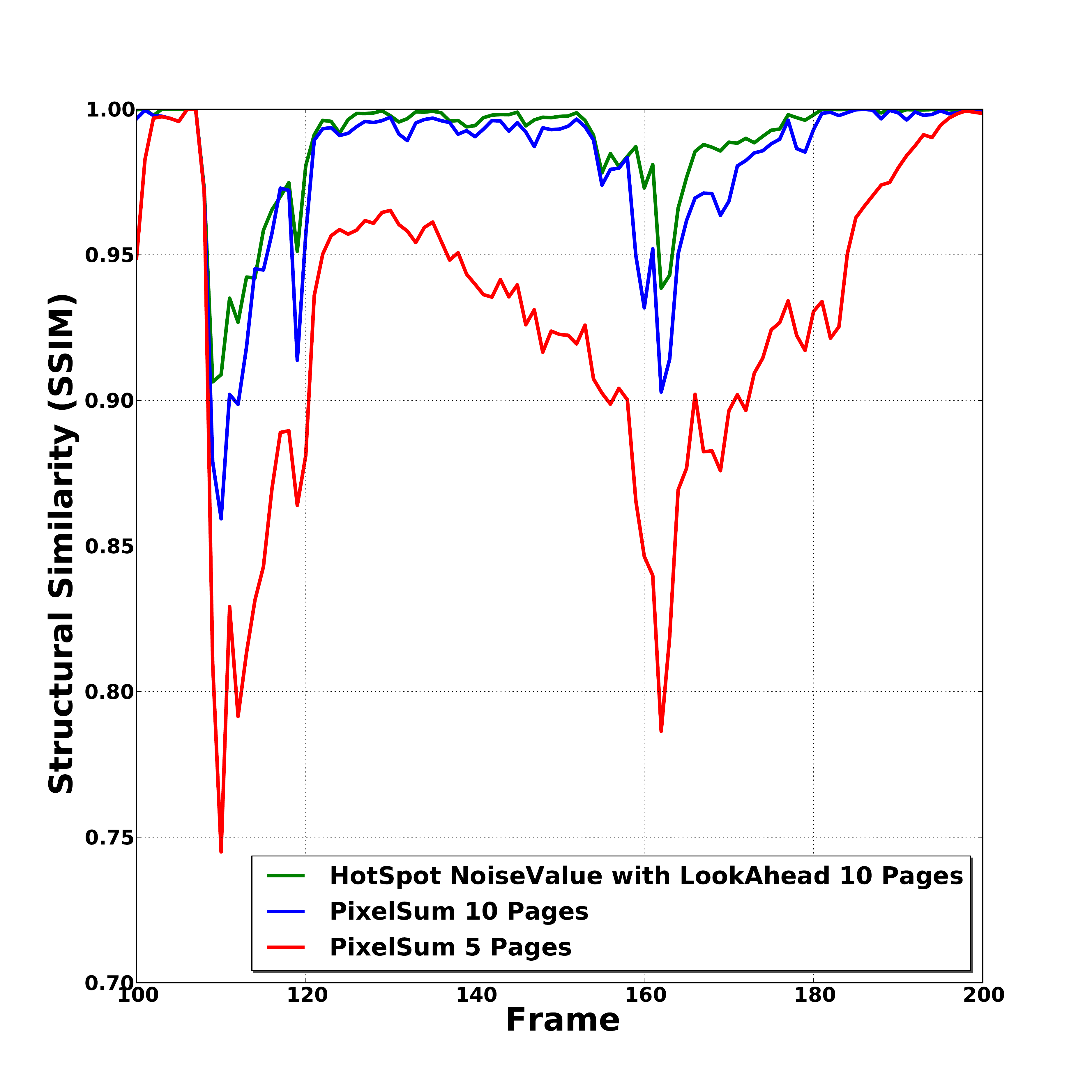}
				}		
 			\caption{Although we can not reach with just 5 pages per frame the same quality as in the case of 10 pages, we achieve quite distinguishable results
 			         in a setup that allows only a small number to be streamed. The difference in (b) is not as great as in (a), but this only underlines the
 			         importance of using such techniques within application areas that suffer from high latencies.}
  			\label{fig_4_heuristicResults}
			\end{figure}

			During this section we investigated several ideas that can be used to select the needed pages in an order that
			represents their importance for the current frame. Furthermore we tried to predict the future need of pages in
			order to diminish the effect that is caused by motions, most importantly rotations.\\
			\noindent We started with a set of basic heuristics and selected the best of them as a baseline for further extensions.
			Figure~\ref{fig_4_heuristicResults} shows a concluding comparison of the basic heuristic PixelSum and a combination of those 
			investigated techniques that yield very good results.
			While we achieve just little improvements over the basic heuristic in the case that we
			can stream many pages, we actually see that we can significantly do better in a setup that allows only a few number to 
			be streamed.\\
			\noindent It should be clear that it is not really possible to reach with just 5 pages per frame the same quality as a system that
			has the possibility to stream twice as much, but we can get closer to it.
			Our combination, HotSpot NoiseValue with LookAhead, performs much better in the case of rotations and yields overall better results compared to PixelSum.
			This is due to the different natures of these techniques:
			While the LookAhead Camera allows the system to select pages that will be probably visible within the next frame, HotSpot gives an indication
			on which of these are likely to stay visible. This is very important, since it filters those pages out that will probably become useless within the
			course of the next frames. Incorporating the NoiseValue into this combination narrows the set of useful pages even more, since it ignores those
			pages that will have only a small impact on the rendered image.
	\section{Ancestor Streaming} \label{04_ancestorStreaming}
		Despite of using a good page priority heuristic and employing some form of prediction,
		we still experience LOD Snaps when we stream only the pages that are really needed.
		As mentioned at the beginning of this chapter, we came to the conclusion that it makes sense
		to implicitly load the ancestors of the needed pages in advance to slowly fade into the higher resolution.
		As we investigated the problem, we started out with two strategies to stream the ancestors of a page.\\
		\\
		Let $P$ and $Q$ denote two pages, that have been identified as needed. This means, that they could be needed
		directly or passively by their children. In this case we could use one of the following strategies.
		\paragraph{InternMipMapOrder} We make sure, that we load page $P$ before $Q$, if $P$ is an ancestor of $Q$. In the case
		that $P$ is not an ancestor of $Q$, we simply let the priority decide on which page has to be streamed first. 
		\paragraph{ExternMipMapOrder} We make sure, that $P$ is loaded before $Q$, if $P$ is part of a lower resolution mipmap
		than $Q$.
		\\\\
		We define the priority of an ancestor page as the sum of the priorities of all its children.
		This makes sense, since it will yield a higher priority for ancestor pages that have multiple children with high priorities.
		In order to ensure the properties of the ExternMipMapOrder strategy, we simply modify the comparison function of the priority queue
		to check the level of detail before it compares the priority of two pages.\\
		\\	
		\begin{figure}[h!]
	 		\centering
  			\subfigure[InternMipMapOrder]{\includegraphics[scale=0.5]{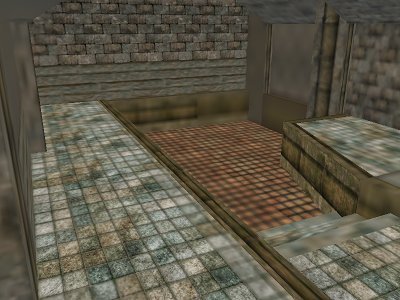}}
  			\subfigure[ExternMipMapOrder]{\includegraphics[scale=0.5]{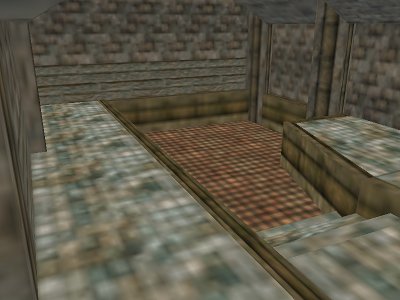}}
  			\caption{InternMipMapOrder streams the ancestors not uniformly, which can be more disturbing on slower systems than loading all ancestors in a uniform fashion.}
  			\label{fig_4_ancestorStreamingExample}
		\end{figure}
		
		\begin{figure}[h!]
    		\hspace*{-1.5cm}
			\subfigure[]
			{
				\includegraphics[scale=0.23]{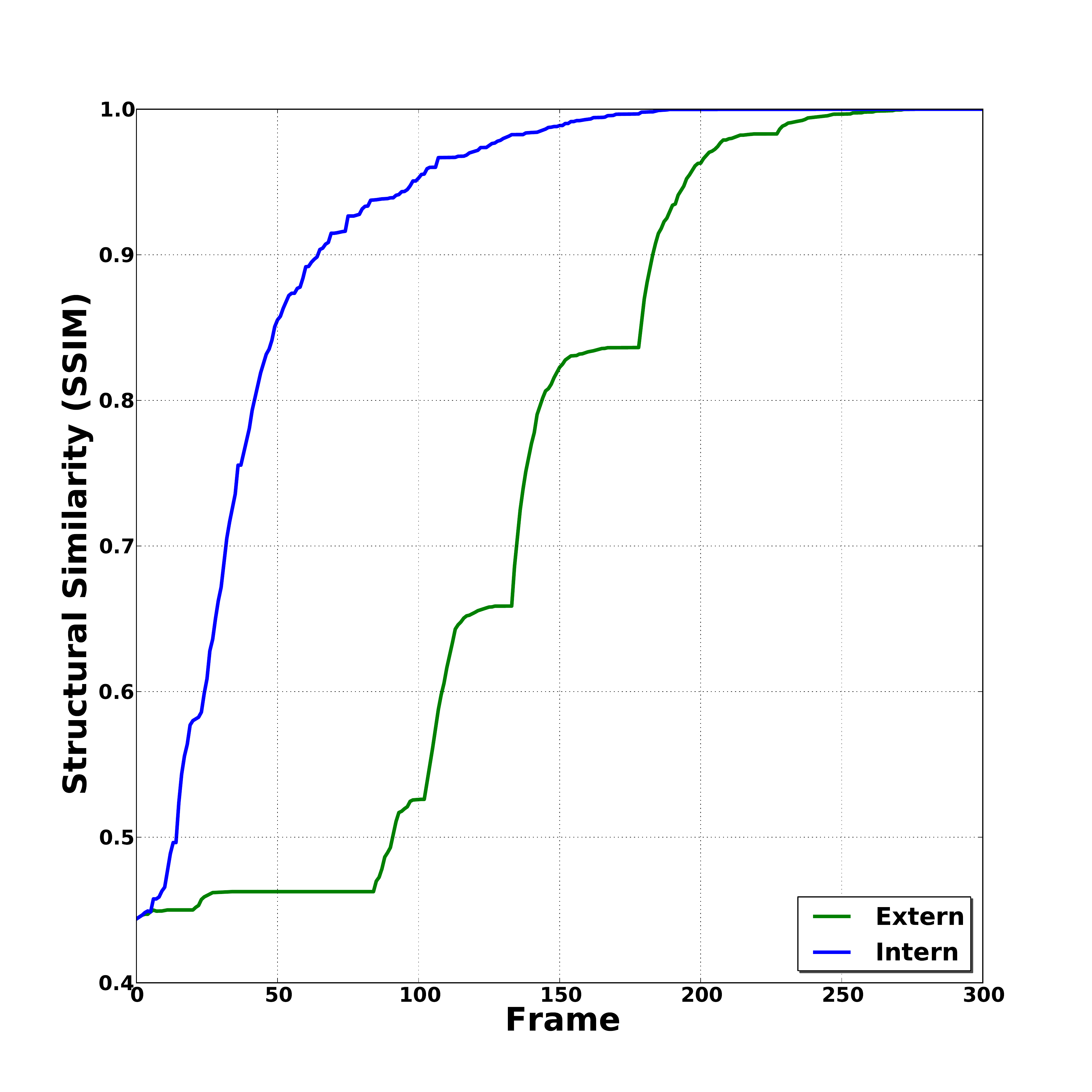}
			}
			\hspace*{-1.0cm}
			\subfigure[]
			{
				\includegraphics[scale=0.23]{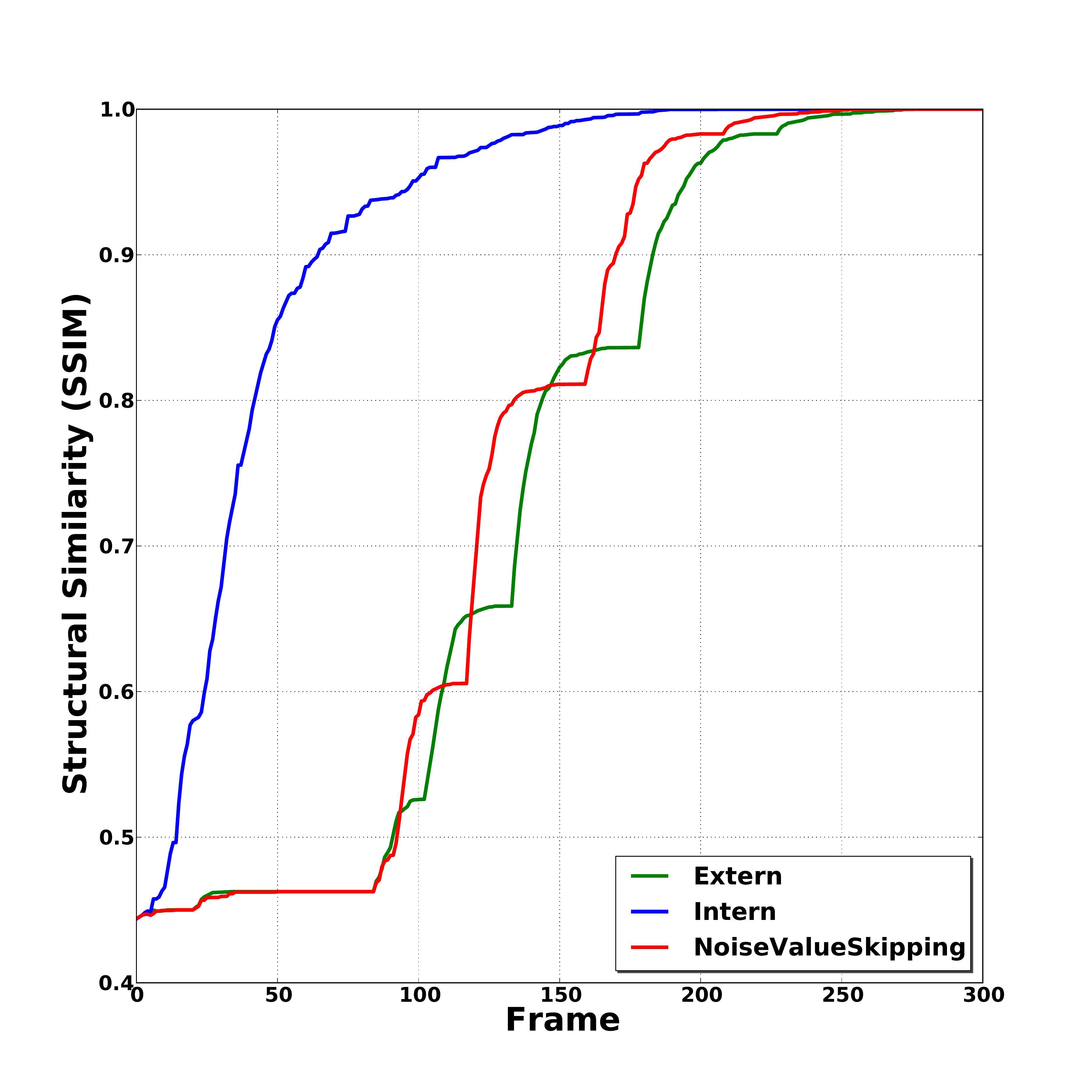}
			}
  			\caption{(a) Comparison of ExternMipMapOrder and InternMipMapOrder. 
  					 (b) NoiseValueSkipping reaches higher levels of quality faster.}
  			\label{fig_4_ancestorStreamingPlots}
			\end{figure}
			
		\noindent Figure~\ref{fig_4_ancestorStreamingPlots} (a) shows how both strategies behave in an indoor level test, that conforms to the
		setup we mentioned in Section~\ref{04_testCases}.
		If we would just decide based on how fast a strategy increases the quality measured by SSIM, we would choose InternMipMapOrder.
		Unfortunately we have to say, that this fast increase in quality is misleading. 
		Figure~\ref{fig_4_ancestorStreamingExample} shows an example for why this is the case. InternMipMapOrder will lead to rendered
		images in which some parts of the scene stay on a low resolution for a long time, while the rest of the scene is rendered
		in the anticipated quality.
		During our tests we had to admit that this result is as disturbing as the LOD Snaps, whose effect we wanted to diminish.
		ExternMipMapOrder in contrast increases the quality in a uniform fashion, since it ensures that all visible spots are on the same
		level of detail before it streams pages of the next higher mipmap.
		\begin{figure}[h!]
			\hspace*{-1.5cm}
			\subfigure[]
			{
				\includegraphics[scale=0.23]{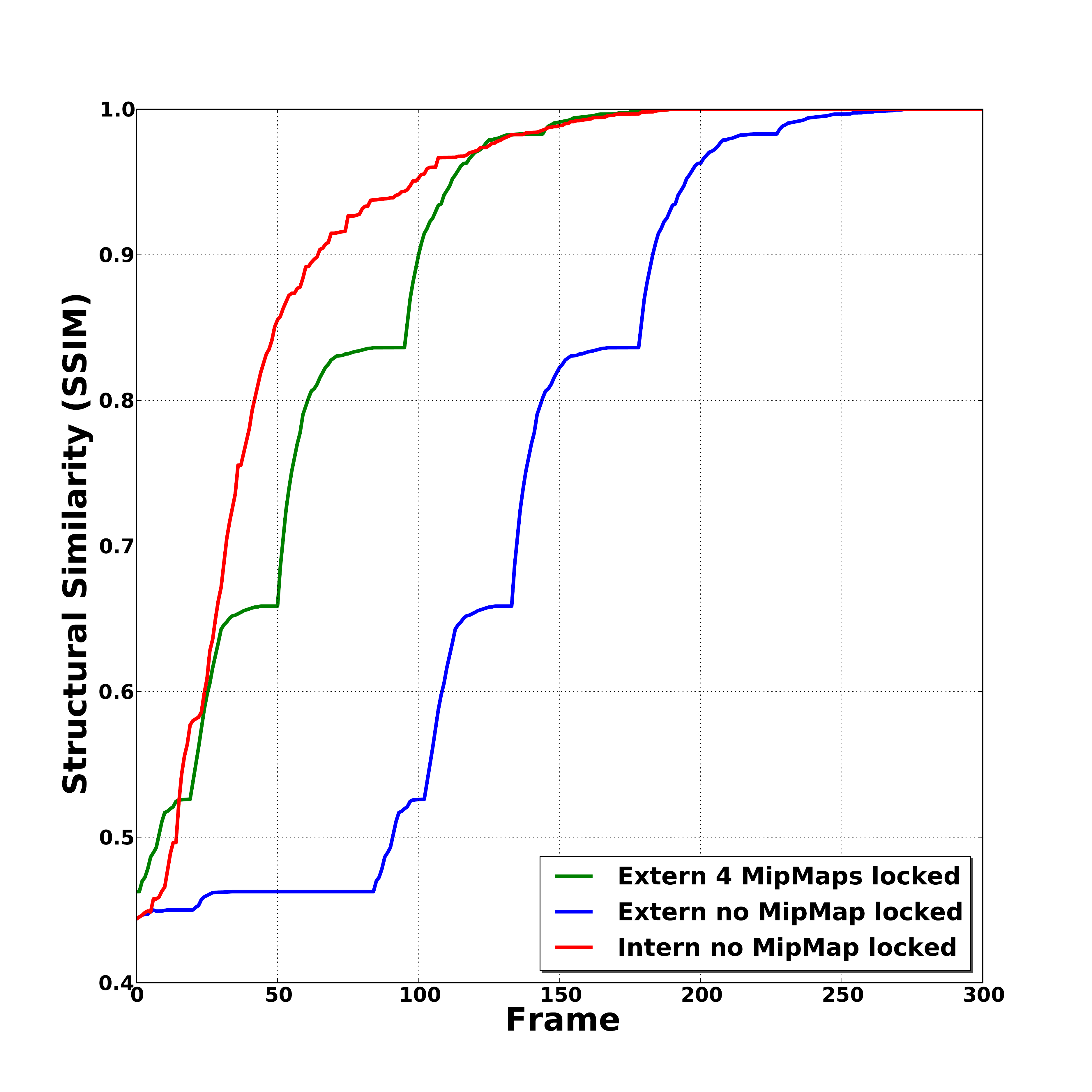}
			}
			\hspace*{-1.0cm}
			\subfigure[]
			{
				\includegraphics[scale=0.23]{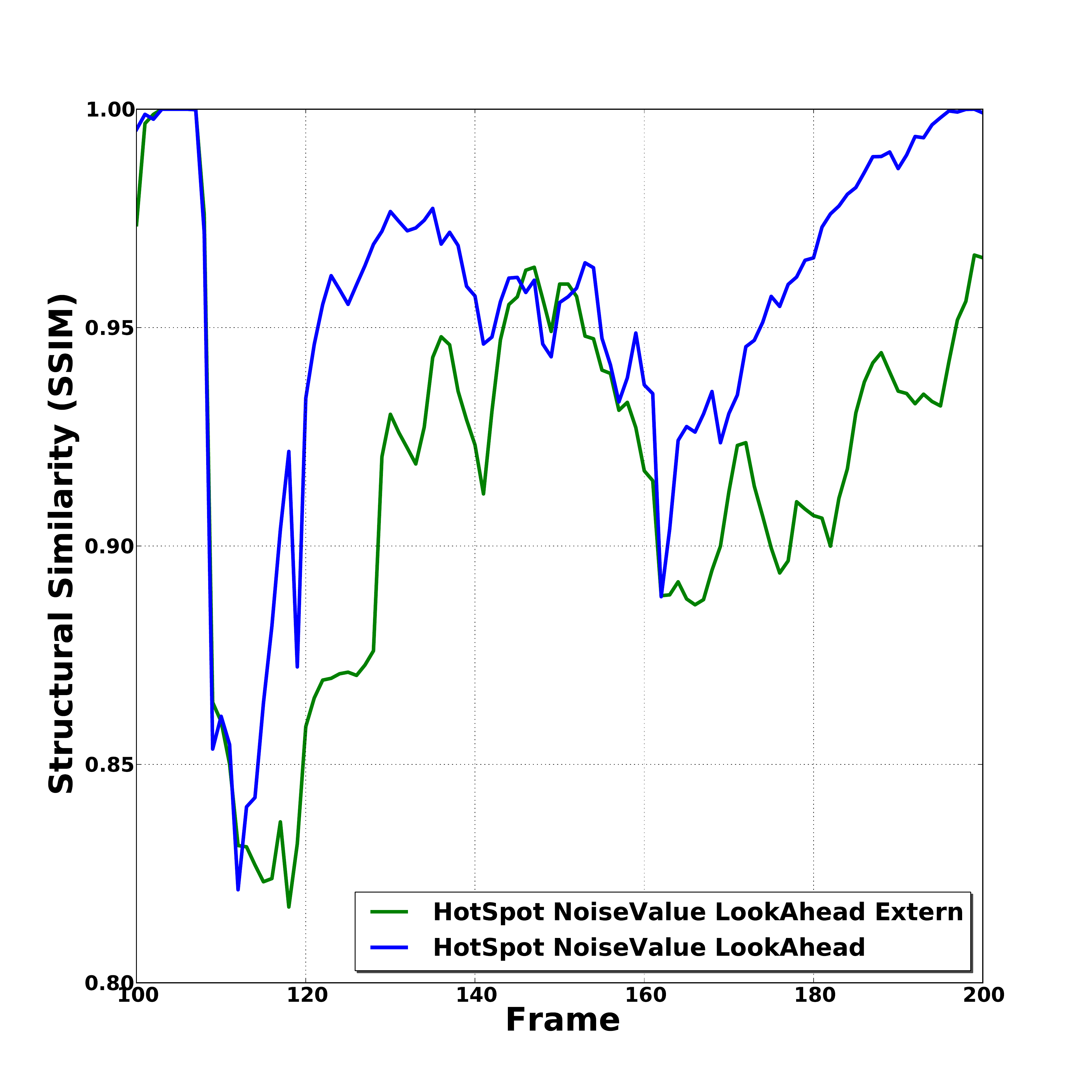}
			}
  			\caption{
  			         (a) Locking the first four mip levels makes sense in the case that a high page coherence can not be guaranteed.
  			         (b) We have to accept a overall lower quality in order to avoid LOD Snaps.}
  			\label{fig_4_ancestorStreamingPlots2}
			\end{figure}
			
		\subsection{A closer look on ExternMipMapOrder}
			Since we identified InternMipMapOrder to be no improvement over LOD Snaps, we took a closer look at the visually more
			satisfying ExternMipMapOrder.
			
		\paragraph{NoiseValueSkipping} In Section~\ref{04_noiseValue} we used the so called \emph{NoiseValue} to scale the page priority.
			Due to the nature of the NoiseValue we can incorporate this idea into ExternMipMapOrder to estimate if it is possible
			to skip the ancestor of a page. If the NoiseValue between a child and its parent page is lower than a certain threshold
			we can actually ignore the ancestor.
			We implemented this idea and Figure~\ref{fig_4_ancestorStreamingPlots} (b) shows that we can reach higher levels of quality faster. 
			We used the mean NoiseValue of all pages within the Virtual Texture as a threshold, but this
			could lead to false results if the arithmetic mean is large. Trial and error could possibly lead to a better
			threshold that is independent from the input NoiseValues.
		
		\paragraph{Locking mipmaps} Looking again at Figure~\ref{fig_4_ancestorStreamingPlots} reveals that ExternMipMapOrder 
			loses much time in this example by loading the first 85 pages. 
			The number 85 represents the fact that each page within the first four mipmaps is implicitly or directly needed.
			We stated in Section~\ref{03_embeddingProblems} that a high page coherence can not be guaranteed by our tool chain.
			This means that pages from all over the Virtual Texture could be visible from the current viewpoint. 
			If such a scene property is predominant, we advise to lock the first four mipmaps within the cache. 
			This means that the system loads the respective pages once at the initialization and will never swap them out of the cache. 
			Figure~\ref{fig_4_ancestorStreamingPlots2} (a) shows that doing so leads to the assumed result.
								
		\paragraph{Testing with motion} We also tested ExternMipMapOrder within the setup of Section~\ref{04_priorityHeuristics}.
			Figure~\ref{fig_4_ancestorStreamingPlots2} (b) shows an comparison of our best configuration in which we can see
			the impact of incorporating ExternMipMapOrder. Using ExternMipMapOrder leads to slower quality improvements,
			which is quite natural since the tests we made in Section~\ref{04_priorityHeuristics} did not take ancestor streaming
			into account. This means that we have to accept a overall lower quality in order to diminish the effect of LOD Snaps.

	\section{Terrains - Page Coherence} \label{04_pageCoherence}		
		Although we focused on indoor scenes during our study, we also did some of the tests in Section~\ref{04_priorityHeuristics}
		with terrains.\\
		\noindent We mentioned in Section~\ref{03_embeddingProblems} that we can not guarantee a page coherence between faces
		that are geometrical near to each other. 
		While valid for the indoor scenes, the terrains we generated do not suffer from the missing page coherence, 
		because we can simply overlay the triangle network that represents them with a continuous area within the Virtual Texture.
		Another difference between these landscapes and the indoor scenes is the fact, that the viewer is not completely surrounded
		by the level geometry. Instead he sees the horizon that is either represented by a skybox that is so far away, that it stays constantly
		on low resolution mip levels, or simply not rendered at all.\\
		\\
		Our tests yielded more favourable results compared to the indoor levels. This can actually be explained by
		the predominant page coherence, which makes it possible to improve the quality of neighbouring areas by streaming
		a smaller set of pages. Furthermore terrains feature a wide viewing range that allows the user to see areas far away, so
		that the system can stream higher resolution fallbacks in advance.
        Although not as large as in the case of indoor scenes, the rendering of terrains suffers from the same problems as soon as
        rotations occur.
         \begin{figure}[h!]
			\hspace*{-1.5cm}
			\subfigure[]                
			{
				\includegraphics[scale=0.23]{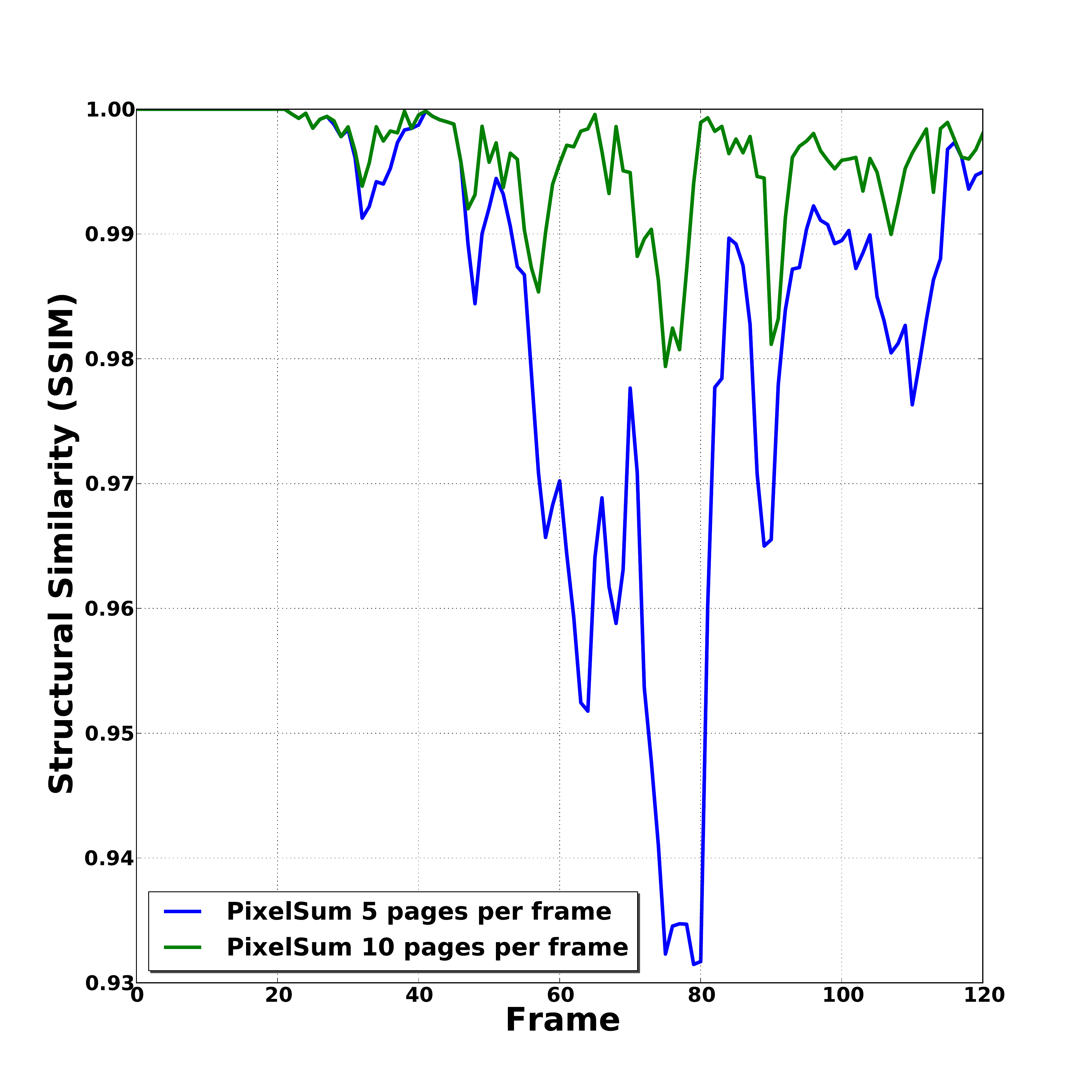}
			}
			\hspace*{-1.0cm}
			\subfigure[]
			{
				\includegraphics[scale=0.23]{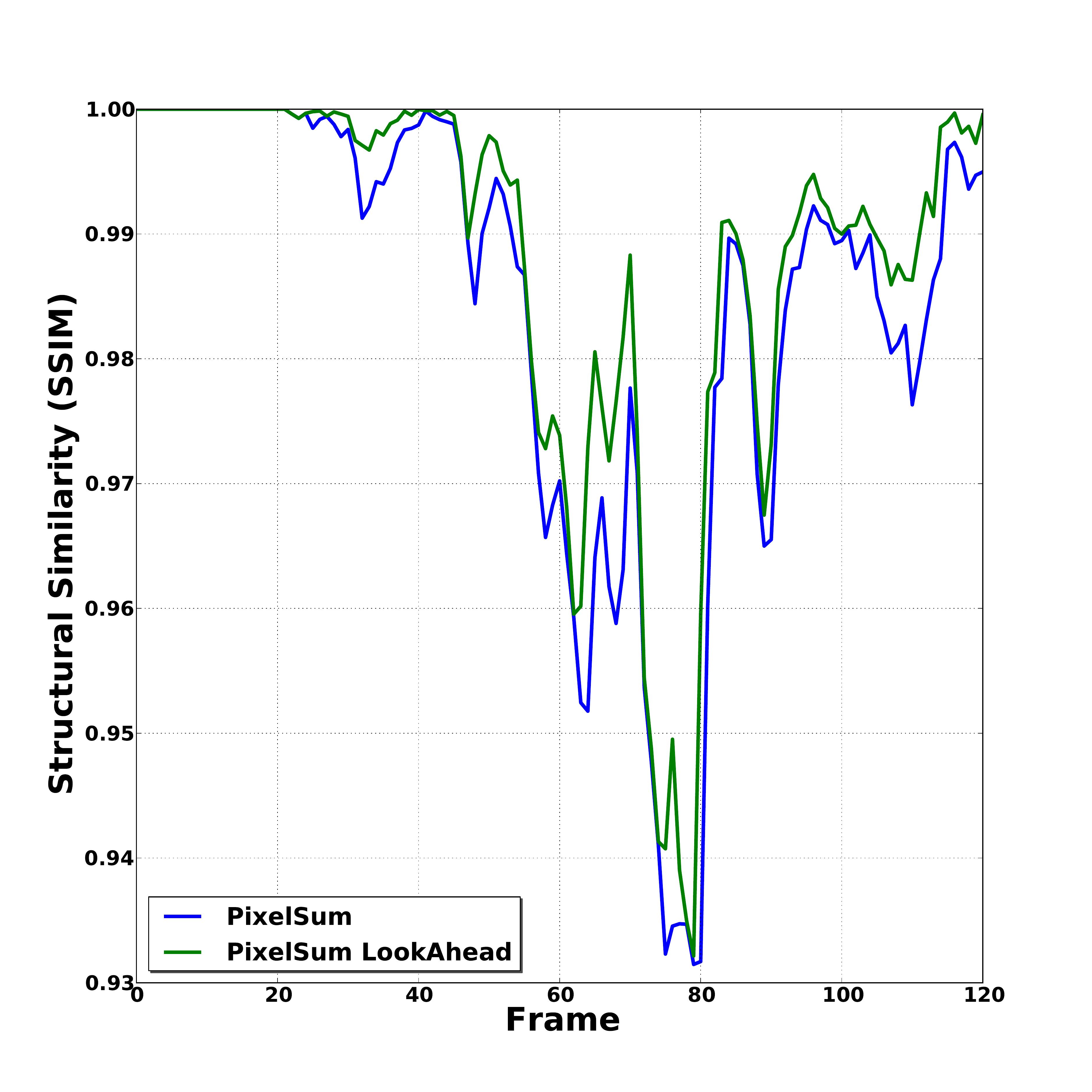}
			}
 			\caption{(a) The problem of small occlusions within the terrain gets more or less ignorable in the case that the system
 						 has the possibility to stream a large amount of pages.
 					 (b) The LookAhead Camera leads to non marginal improvements.}
  			\label{fig_4_terrainOcclusion}
		\end{figure}\\\\
        \noindent During our tests we came across a problem that is commonplace in terrains scenes, while it does not often occur
        in case of indoor levels. Small obstacles occlude parts of the scene that become visible as soon as the
        viewer can look behind such a obstacle. These obstacles are small hills in the case of landscape scenes.
        Figure~\ref{fig_4_terrainOcclusion} shows the results of one flythrough we made. Without performing any rotations, we
        just translated on the z-axis in order to fly over a sequence of small hills. 
        Instead of a more or less constant quality we actually see small
        peaks that are caused by the quality decrease that occurs as soon as areas that were occluded become visible.\\
        \noindent Figure~\ref{fig_4_terrainOcclusion} (a) shows that the impact of these occlusions becomes less
        relevant when the system can stream many pages per frame. We again tried the advanced techniques
        from Section~\ref{04_priorityHeuristics} in order to weaken the loss in quality for the case of a system that can only
        stream a small number of pages. Figure~\ref{fig_4_terrainOcclusion} (b) and~\ref{fig_4_terrainOcclusion2} reveal that only the LookAhead
        Camera has an non-marginal impact on the quality.
        \begin{figure}[h!]
			\hspace*{-1.5cm}
			\subfigure[]
			{
				\includegraphics[scale=0.23]{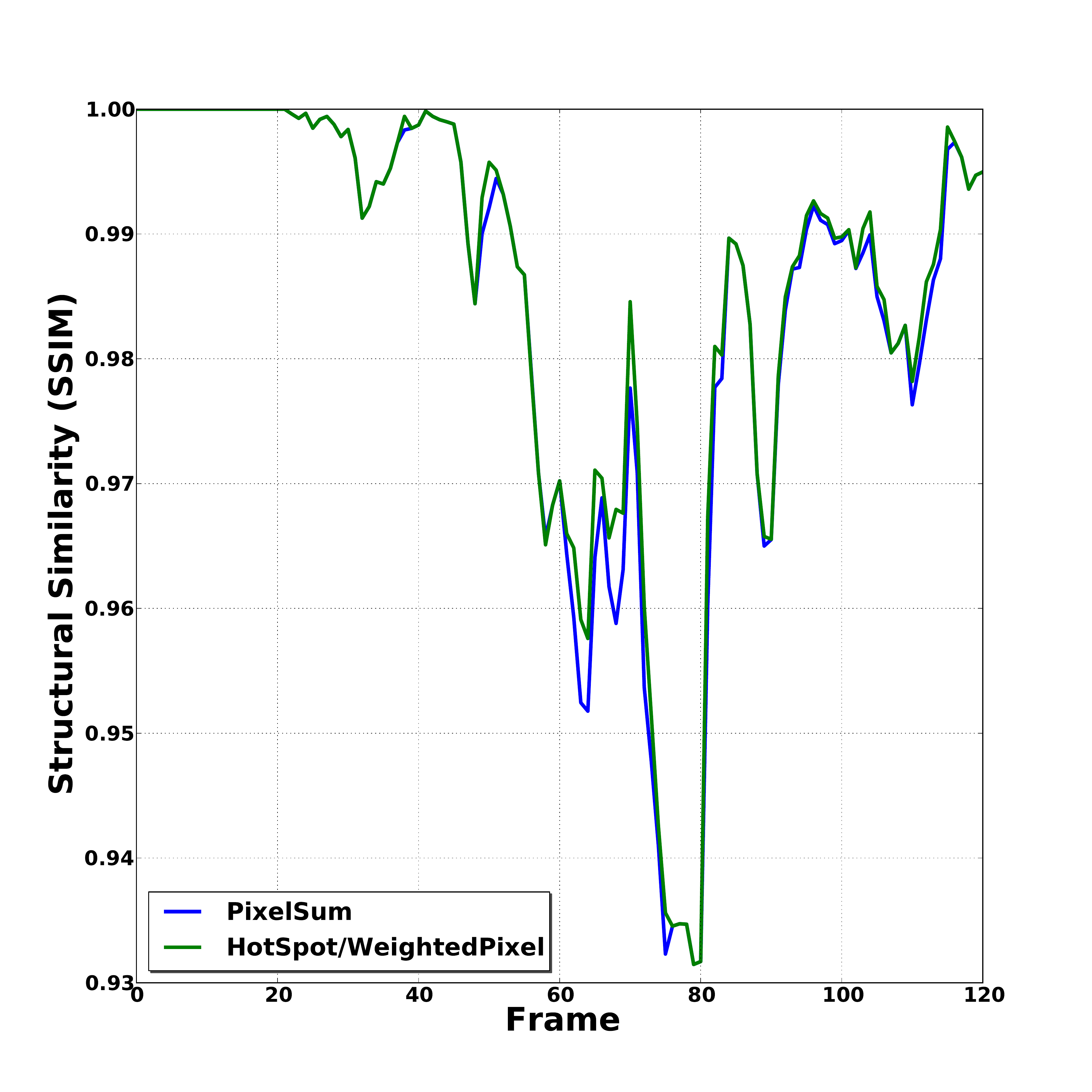}
			}
			\hspace*{-1.0cm}
			\subfigure[]
			{
				\includegraphics[scale=0.23]{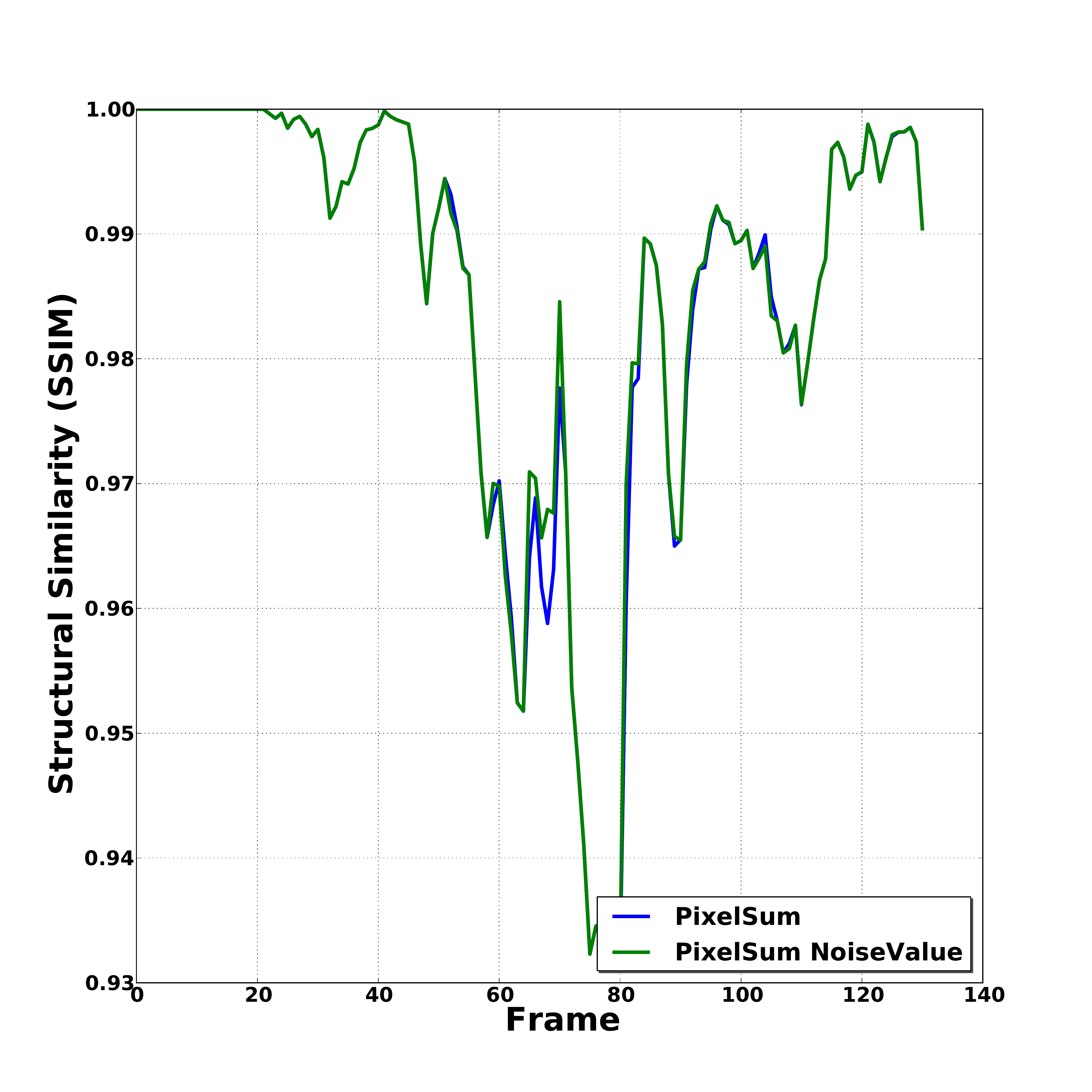}
			}
 			\caption{No huge improvements by using HotSpot (in this case WeightedPixel, since we do not rotate the camera) or incorporating the NoiseValue.}
  			\label{fig_4_terrainOcclusion2}
		\end{figure}        
        
	\section{Summary}
		In this chapter we investigated several techniques with the goal of improving the visual quality by collecting a set of different 
		information on the needed pages and have them streamed in a way that represents their importance for the current or next frame.
		In order to measure the performance of these ideas, we introduced a method that is based on the visual
		output. Provided with this method we were able to compare different system configurations and hence could conclude where specific techniques
		have a positive impact on the quality.
		Furthermore we took a look at how we can diminish the visual disturbing LOD Snap effect and in which way the differences of indoor- 
		and terrain scenes impact the quality.

\blankpage

\chapter{Conclusion \& Outlook}
During the course of this thesis we developed a renderer that employs Virtual Texturing in order
to store large texture sets within the limited size of available memory.
An accompanying tool chain has been implemented that allows the user to create textures of
multiple gigabytes and automatically retexture existing geometry.
Furthermore we analyzed several techniques that, if incorporated into the system, can improve the quality of rendering.\\
Looking at the ubiquitous usage of web-enabled mobile devices, we can imagine mobile applications
that employ Virtual Texturing in order to stream a vast amount of texture data from an internet server
and use it during rendering on those low spec machines.
Because of the limitations within this area of application, e.g. high latencies,
we feel confident that our investigation can be used as a starting point to make the visual output
for the user as good as possible.\\
\\
The Virtual Textures we used during the thesis were uncompressed and hence quite large. Although
we have large storage media available today, it would be interesting to incorporate
realtime decompression techniques~\cite{Waveren06} and analyze their impact on Virtual Texturing.
\\\\
Furthermore it would be worthwhile to delve more deeply into the development of interactive
tools purposely built for Virtual Texturing. The tool chain presented in Section~\ref{03_toolchain} was sufficient enough
for the course of this thesis, but in order to use this technique to its full potential one will have 
to offer artists a set of more intuitive applications.

\blankpage
\appendix

\chapter{Mathematical Background}
	\section{Full reference quality assessment} \label{app_frqa}
		To estimate the quality of an image, one can employ different methods that are classified
		by their dependency on reference images.
		Those of them that fall into the class of \emph{Full reference methods}, measure the
		difference in quality by comparing the distorted image directly to another that is considered
		to be perfect.
		\subsection{Mean Squared Error} \label{app_rmse}
			\emph{Mean squared error} is a widely adopted statistic to estimate the quality of an image.
			Its strength is the simplicity and rapidity with which it can be computed.\\
			\\
			Let us assume, that we want to compare two images $X$ and $Y$.
			Further let $x_i$ and $y_i$ denote the colors of two corresponding pixels within the images.
			The mean squared error between $X$ and $Y$ is then defined as
			\begin{center}
			\begin{math} 
				\displaystyle
				MSE = \frac{1}{n}\sum\limits_{i=1}^n(x_i - y_i)^2
			\end{math}
			\end{center}
			and obviously
			\begin{center}
			\begin{math} 
				\displaystyle
				RMSE = \sqrt[2]{MSE}
			\end{math}
			\end{center}
			for the \emph{rooted mean squared error}.
	
		\subsection{Structural Similarity} \label{app_ssim}
			\begin{figure}[h!]
 				\centering
				\subfigure[MSE = 0 \newline \hspace*{1.3em} SSIM = 1.0 ]
				{
					\includegraphics[scale=0.35]{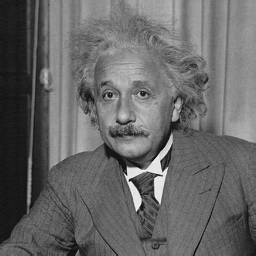}
				}
				\subfigure[MSE = 144 \newline \hspace*{1.3em} SSIM = 0.988]
				{
					\includegraphics[scale=0.35]{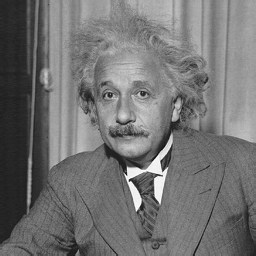}
				}
				\subfigure[MSE = 144 \newline \hspace*{1.3em} SSIM = 0.694]
				{
					\includegraphics[scale=0.35]{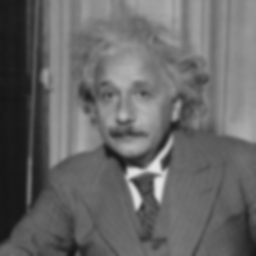}
				}
				\subfigure[MSE = 142 \newline \hspace*{1.3em} SSIM = 0.662]
				{
					\includegraphics[scale=0.35]{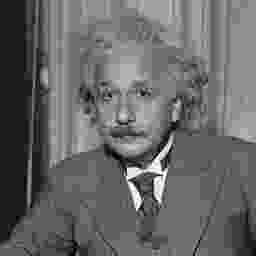}
				}
  
  				\caption{Example taken from~\cite{Wang04} that shows that SSIM classifies the images more differentiated than MSE.
  			         (a) Original image (b) meanshift (c) blur (d) jpeg compression.}
  				\label{figRMSESSIMDIFF}
			\end{figure}
			Using MSE and its variants for image quality estimation, suffers from the problem, that
			these metrics do not deliver results that are consistent with the image perception of
			humans~\cite{Wang98}.
			In order to overcome this problem, metrics like the \emph{universal image quality index}~\cite{Wang02} and its successor
			\emph{Structural Similarity}~\cite{Wang04} have been developed, which take the analysis of the \emph{Human Visual System} into account.
			Figure~\ref{figRMSESSIMDIFF} shows an example in which MSE would yield misleading results.\\
			\\
			Let $\mu_i$ denote the mean intensity of a block of pixels $i$ and $\sigma_i$ the corresponding
			variance.
			Furthermore let us assume, that $\sigma_{i,j}$ represents the covariance between two blocks
			$i$ and $j$.
			Then we can compute the SSIM Value between two pixel areas $x$ and $y$ by
			\begin{center}
			\begin{math} 
				\displaystyle
				Q(x,y) = \frac{(2\mu_x\mu_y + C_1)(2\sigma_{x,y}+C_2)}{(\mu_x^2+\mu_y^2+C_1)(\sigma_x^2+\sigma_y^2+C_2)}
			\end{math}
			\end{center}
			where $C_1 = (L \cdot K_1)^2$ and $C_2 = (L \cdot K_1)^2 $ are included to avoid instability.\\
			\noindent $L$ represents the range of color values, while $K_1$ and $K_2$ are constants $\ll 1$.
			\\\\
			To calculate the overall SSIM value between two complete images, a sliding window approach is used.
			A window of a constant size (e.g. 8 x 8 Pixels) loops over both images completely and compares
			the corresponding pixel areas.
			Let $m$ denote the number of areas that have been compared and $Q_i$ the SSIM value that resulted in the
			analysis of the i-th areas.
			Then the arithmetic mean of all SSIM values will yield the mean result between both images
			\begin{center}
			\begin{math} 
				\displaystyle
				Q_{mean} = \frac{1}{m} \sum\limits_{i=1}^m Q_i
			\end{math}
			\end{center}

	\section{Edge-compression based level of detail calculation} \label{app_lodCalculation}
		In order to calculate the mip level of each fragment that is processed by our shader (see~\ref{03_shader})
		we use the approach that has been proposed by Paul Heckbert~\cite{Heckbert83}.
		The basic idea is to measure the compression that occurs, when a parallelogram within the texture 
		gets mapped to the quadratic size of a fragment. This compression can be estimated by the maximal length
		of the parallelograms edges.
		By knowing the partial differentials of the texture coordinates at the currently considered fragment
		\begin{center}
		\begin{math} 
			\displaystyle
			s_x' = \frac{\partial s}{\partial x} 
			\hspace{1cm} 
			t'_x = \frac{\partial t}{\partial x}
			\hspace{1cm} 
			s'_y = \frac{\partial s}{\partial y}
			\hspace{1cm}
			t'_y = \frac{\partial t}{\partial y}
		\end{math}
		\end{center}
		we can compute the lengths of these edges.
		\begin{center}
		\begin{math} 
			e_x = \sqrt[2]{s^{'2}_x + t^{'2}_x} \hspace{1cm} e_y = \sqrt[2]{s^{'2}_y + t^{'2}_y}
		\end{math}
		\end{center}
		Choosing the maximal length
		\begin{center}
		\begin{math} 
			e_{max} = max(e_x,e_y)
		\end{math}
		\end{center}
		provides us with the maximal compression.
		We scale $e_{max} \in [0,1]$ with the dimension of the maximal mip level, for the reason of getting a useful result.
		By employing the logarithm we get the needed level within the pyramid.
		\begin{center}
		\begin{math} 
			d = \log_2{(e_{max} \cdot dim_{max})}
		\end{math}
		\end{center}		
		The OpenGL shading language GLSL~\cite{Kessenich09} provides everything we need to do a straight implementation
		of the described theory.\\
		\noindent Please note, that we return \textbf{uMaxMipMap - d} in our implementation, so that we can use the mip level enumeration
		we described in Section~\ref{03_virtualTexture}.
\begin{center}
\lstset{language=glsl}
\begin{lstlisting}
	uniform float uMaxDim;
	uniform float uMaxMipMap;

	float calculateMipMap()
	{
		vec2 dx = dFdx(gl_TexCoord[0].st);
		dx *= dx;
		vec2 dy = dFdy(gl_TexCoord[0].st);
		dy *= dy;

		float e_max = sqrt(max(dx.s+dx.t,dy.s+dy.t)); 
		
		float d = log2(e_max * uMaxDim);
	
		d = min(d,uMaxMipMap);
		d = max(d,0.0);
		  
		return uMaxMipMap-d;
	}
\end{lstlisting}
\end{center}

\backmatter

\bibliography{bachelorthesis}
\bibliographystyle{alpha}

\nocite{*}

\end{document}